\documentclass[prb,aps,twocolumn,superscriptaddress,showpacs,floatfix]{revtex4-1}

\usepackage{graphicx}
\usepackage{epstopdf}
\usepackage[colorlinks=true,linkcolor=black, citecolor=blue, urlcolor=blue, 
    unicode=true]{hyperref}

\usepackage{amsmath} 

\usepackage{amssymb}	
\usepackage{graphicx} 
\usepackage{color}
\usepackage{hyperref} 
\usepackage{anysize}
\usepackage{hhline,multirow,float}
\usepackage{makecell}
\usepackage{array}
\usepackage{dcolumn}
\usepackage{mathtools} 
\usepackage{multirow}
\usepackage[table]{xcolor}
\usepackage{capt-of}
\usepackage{wasysym}
\usepackage{placeins}

\marginsize{2.0 cm}{2.0 cm}{1 cm}{1 cm}

\setcitestyle{square}

\begin{document}
\newcommand\bbone{\ensuremath{\mathbbm{1}}}
\newcommand{\ul}{\underline}
\newcommand{\bp}{{\bf p}}
\newcommand{\vl}{v_{_L}}
\newcommand{\vc}{\mathbf}
\newcommand{\be}{\begin{equation}}
\newcommand{\ee}{\end{equation}}
\newcommand{\bk}{{{\bf{k}}}}
\newcommand{\bK}{{{\bf{K}}}}
\newcommand{\cE}{{{\cal E}}}
\newcommand{\bQ}{{{\bf{Q}}}}
\newcommand{\br}{{{\bf{r}}}}
\newcommand{\bg}{{{\bf{g}}}}
\newcommand{\bG}{{{\bf{G}}}}
\newcommand{\hbr}{{\hat{\bf{r}}}}
\newcommand{\bR}{{{\bf{R}}}}
\newcommand{\bq}{{\bf{q}}}
\newcommand{\hx}{{\hat{x}}}
\newcommand{\hy}{{\hat{y}}}
\newcommand{\hd}{{\hat{\delta}}}
\newcommand{\bea}{\begin{eqnarray}}
\newcommand{\eea}{\end{eqnarray}}
\newcommand{\ra}{\rangle}
\newcommand{\la}{\langle}
\renewcommand{\tt}{{\tilde{t}}}
\newcommand{\upa}{\uparrow}
\newcommand{\dna}{\downarrow}
\newcommand{\bS}{{\bf S}}
\newcommand{\vS}{\vec{S}}
\newcommand{\dg}{{\dagger}}
\newcommand{\pdg}{{\phantom\dagger}}
\newcommand{\tphi}{{\tilde\phi}}
\newcommand{\cf}{{\cal F}}
\newcommand{\ca}{{\cal A}}
\renewcommand{\ni}{\noindent}
\newcommand{\ct}{{\cal T}}
\newcommand{\brf}{\bar{F}}
\newcommand{\brg}{\bar{G}}
\newcommand{\jeff}{j_{\rm eff}}
\newcommand{\polarised}{$\left(\frac{d\sigma}{d\Omega}\right)_{\rm{mag}}$}
\newcommand{\yb}{Yb$^{3+}$}
\newcommand{\caruo}{Ca$_{2}$RuO$_{4}$}
\newcommand{\appropto}{\mathrel{\vcenter{
  \offinterlineskip\halign{\hfil$##$\cr
    \propto\cr\noalign{\kern2pt}\sim\cr\noalign{\kern-2pt}}}}}

\title{\texorpdfstring{Van Vleck excitons in \caruo}{}}

\author{P.~M.~Sarte}  
\affiliation{California NanoSystems Institute, University of California, Santa Barbara, CA 93106-6105, USA}
\affiliation{Materials Department, University of California, Santa Barbara, CA 93106-5050, USA} 
\author{C.~Stock}  
\affiliation{School of Physics and Astronomy, University of Edinburgh, Edinburgh EH9 3FJ, United Kingdom}
\author{B.~R.~Ortiz}
\affiliation{California NanoSystems Institute, University of California, Santa Barbara, CA 93106-6105, USA}
\affiliation{Materials Department, University of California, Santa Barbara, CA 93106-5050, USA} \author{K.~H.~Hong}
\affiliation{School of Chemistry, University of Edinburgh, Edinburgh EH9 3FJ, United Kingdom}
\affiliation{Centre for Science at Extreme Conditions, University of Edinburgh, Edinburgh EH9 3FD, United Kingdom}
\author{S.~D.~Wilson}
\affiliation{California NanoSystems Institute, University of California, Santa Barbara, CA 93106-6105, USA} 
\affiliation{Materials Department, University of California, Santa Barbara, CA 93106-5050, USA} 
\date{\today}

\begin{abstract}  

A framework is presented for modeling and understanding magnetic excitations in localized, intermediate coupling magnets where the interplay between spin-orbit coupling, magnetic exchange, and crystal field effects are known to create a complex landscape of unconventional magnetic behaviors and ground states. A spin-orbit exciton approach for modeling these excitations is developed based upon a Hamiltonian which explicitly incorporates single-ion crystalline electric field and spin exchange terms.  This framework is then leveraged to understand a canonical Van Vleck $j\rm{_{eff}}=0$ singlet ground state whose excitations are coupled spin and crystalline electric field levels.  Specifically, the anomalous Higgs mode [Jain \emph{et al.} Nat. Phys. \textbf{13}, 633 (2017)], spin-waves [S. Kunkem\"{o}ller \emph{et al.} Phys. Rev. Lett. \textbf{115}, 247201 (2015)], and orbital excitations [L. Das \emph{et al.} Phys. Rev. X \textbf{8}, 011048 (2018)] in the multiorbital Mott insulator Ca$_2$RuO$_4$ are captured and good agreement is found with previous neutron and inelastic x-ray spectroscopic measurements.  Furthermore, our results illustrate how a crystalline electric field-induced singlet ground state can support coherent longitudinal, or amplitude excitations, and transverse wavelike dynamics.  We use this description to discuss mechanisms for accessing a nearby critical point.
\end{abstract}
\maketitle

\renewcommand{\topfraction}{0.85}
\renewcommand{\bottomfraction}{0.85}
\renewcommand{\textfraction}{0.1}
\renewcommand{\floatpagefraction}{0.75}

\section{Introduction}

The magnetism inherent to Hund's metals~\cite{Georges17:4} and their parent multiorbital Mott states~\cite{Rau17:4} is believed to play a crucial role in many of their anomalous electronic properties,~\cite{Krempa14:5} ranging from unconventional superconductivity,~\cite{Yin11:10} to violations of Fermi liquid theory, to the recent unveiling of novel nonequilibrium states.~\cite{Zhao19:100}  This is perhaps most clearly illustrated in the range of emergent phenomena that appear in $4d$ transition metal oxides---where an intermediate coupling regime manifests.  In this regime, spin-orbit coupling, magnetic exchange, and crystalline electric field energies compete with one another on equal footing in determining a material's ground state properties.  

Due to the interplay of these competing energy scales, modeling the excitations out of the unusual ground states realized in this intermediate coupling regime and ultimately understanding their microscopic Hamiltonians is an enduring challenge. Magnetic excitations in this space become intertwined with other degrees of freedom as local orbital degeneracies are quenched via both the electrostatic crystal field potential and via spin-orbit coupling.  This often yields a complex excitation spectrum reflective of transitions between exchange-coupled single-ion states and an ``excitonic" energy landscape which is difficult to experimentally interpret.    

\begin{figure*}[htb!]
	\centering
	\includegraphics[width=1\linewidth]{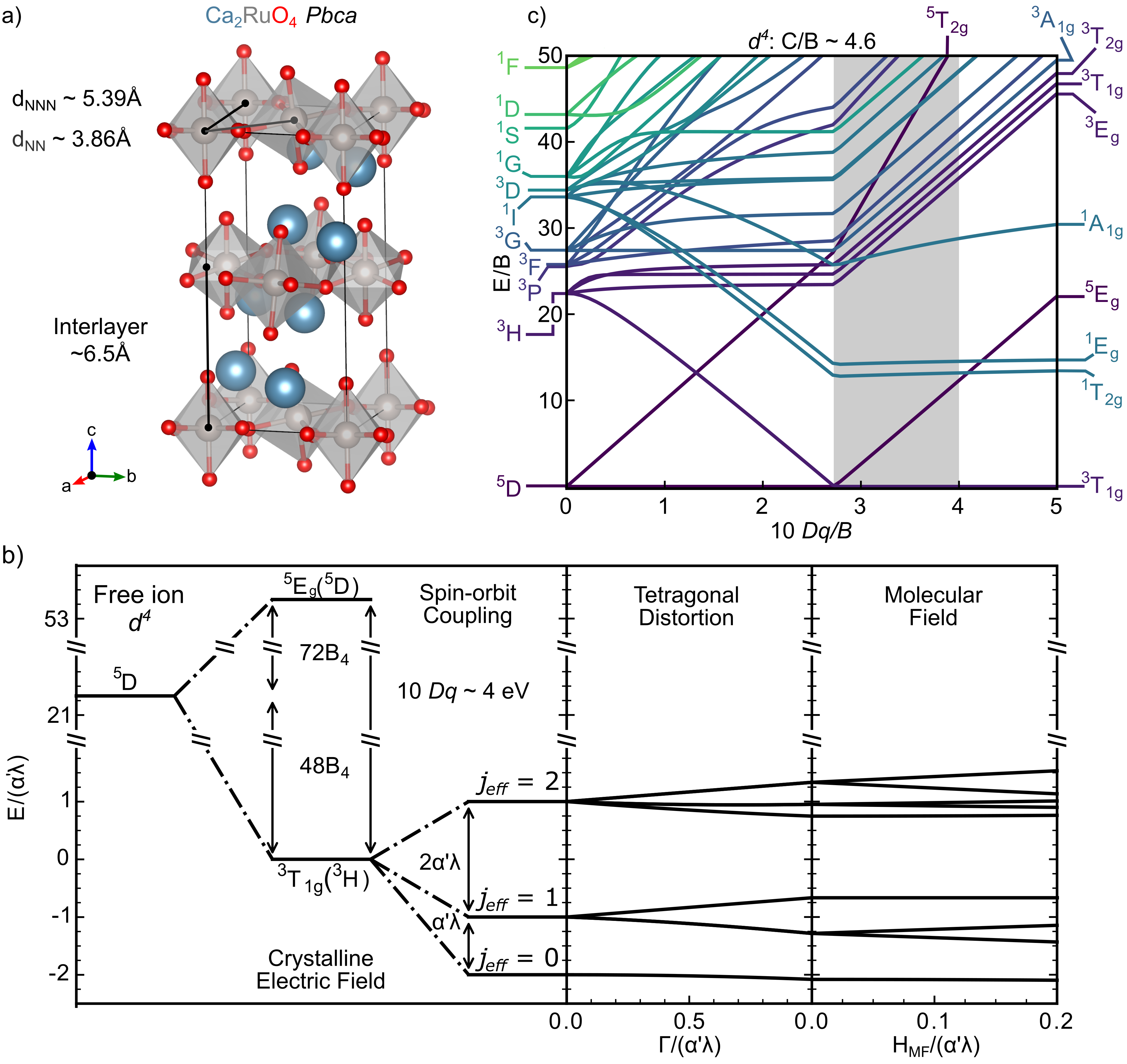}
	\caption{(a, top) Isometric view of the orthorhombic $Pbca$ (S.G.~\#61) unit cell of Ca$_{2}$RuO$_{4}$, consisting of (a, bottom) quasi-two-dimensional psuedo-square lattice layers of corner-sharing distorted RuO$_{6}$ octahedra.\cite{braden98:58} (b) Calculated energy eigenvalues as a function of the tetragonal distortion and the magnetic molecular field perturbations applied to the individual $j\rm{_{eff}}$ manifolds of the ground-state crystal-field triplet $^{3}$T$\rm{_{1g}}$ ($l$=1, $S$=1) of Ru$^{4+}$ in octahedral coordination. Both the eigenvalues and individual parameters: $\Gamma$ and $H\rm{_{MF}}$ have been normalized, and are presented to scale. (c) Calculated Tanabe-Sugano diagram for $d^{4}$ in perfect octahedral coordination with a $C/B$ ratio of 4.6. Shaded rectangle corresponds to the $10~Dq$/$B$ regime of interest for \caruo.\cite{Cao:book}}
	\label{fig:fig1}
\end{figure*}

One example of this richness appears in the unusual magnetic ground state of the multiorbital Mott insulator Ca$_{2}$RuO$_{4}$.~\cite{Cao97:56,nataksuji97:66,nakatsuji00:84,nakatsuji00:62,nataksuji03:90,Anisimov02:25,Nakamura02:65,Porter18:98}  
Ca$_{2}$RuO$_{4}$ possesses a distorted K$_{2}$NiF$_{4}$ structure~\cite{braden98:58,Steffens05:72,Pincini18:98,Friedt01:63} (Fig. \ref{fig:fig1}) with Ru$^{4+}$ cations in a $t^{4}_{2g}$ orbital configuration present in a strong crystalline electric field.  In this setting, Ru$^{4+}$ cations possess a spin angular momentum $S=1$ and an effective orbital angular momentum of $l$=1, yielding a nonmagnetic singlet $j\rm{_{eff}}=0$ ground state.  Neutron scattering measurements nevertheless observe long-range ordered antiferromagnetism, albeit with a reduced ordered moment $\sim$ 1~$\mu_{B}$. The result is an unusual manifestation of a spin-orbit induced mixing of higher energy crystal field levels \cite{Feldmaier19:xx} that stabilizes a static magnetic moment in a naively $j\rm{_{eff}}=0$ singlet ground state---a higher order state mixing leading to analogies with Van Vleck susceptibility.\cite{Khaliullin13:111}   The ground state in Ca$_{2}$RuO$_{4}$ is different from the weakly magnetic ground states observed in compounds based on Kramers ions ($e.g.$ CeRhSi$_{3}$~\cite{Pasztorova19:99} and YbRh$_{2}$Si$_{2}$~\cite{Stock12:109}) that result from the near cancellation of the elastic magnetic cross section from contributions from differing members of the ground state doublet.

Recent experiments have since confirmed this mixed level structure in Ca$_2$RuO$_4$,\cite{Gretarsson19:100} harking back to previous effects observed in rare earth intermetallic compounds.~\cite{Wang68:172} Compounds such as PrTl$_{3}$~\cite{Birgeneau71:27,Holden74:9} and TbSb~\cite{hsieh72:6} possess similar singlet magnetic non-Kramers ground states yet nevertheless also exhibit coherent sharp spin-waves~\cite{Birgeneau72:6} with ferromagnetically polarized ground states.\cite{Cooper71:59,pink68:1} Ca$_{2}$RuO$_{4}$ is analogous~\cite{Fang04:69} in this regard given the apparent contradiction of its non-magnetic $j\rm{_{eff}}=0$ ground state hosting weak antiferromagnetic order along with highly dispersive, coherent spin excitations.\cite{Kunkem15:115} Notably however, the spin dynamics arising from the weak magnetic order in Ca$_{2}$RuO$_{4}$ have been so far analyzed through conventional Holstein-Primakoff approaches~\cite{jain13:17} rather than employing the spin-orbit exciton framework~\cite{Peschel71:5} endemic to more strongly spin-orbit coupled rare earth systems.

Here we adapt the spin-orbit exciton framework to intermediate coupling oxides by modeling the low energy collective and single-ion magnetic excitations in Ca$_{2}$RuO$_{4}$. Comparison is made with previously reported spectroscopic data with the goal of understanding the origin of the anomalous transverse and longitudinal spin fluctuations reported in neutron scattering measurements.\cite{jain13:17,Kunkem17:95}  The model further accounts for the single-ion physics and crystal field levels reported with inelastic x-rays \cite{das18:8} by including the Heisenberg interaction between spins and the single-ion terms in the Hamiltonian equally.  This formalism implicitly includes the multiorbital nature of the Ru$^{4+}$ ion in Ca$_{2}$RuO$_{4}$'s Mott state where intermediate coupling requires orbital and spin degrees of freedom are necessarily linked in the Hamiltonian through the spin-orbit interaction term.  In using this minimal model and alternative approach, we are able to reproduce the anomalous collective transverse and longitudinal/amplitude excitations observed with neutrons as well as the higher energy spin-orbit transitions reported from inelastic x-ray experiments.  This approach and its ability to quantitatively parameterize the magnetic Hamiltonian of Ca$_{2}$RuO$_{4}$ suggests its broader utility to model other multiorbital Mott states in the intermediate coupling regime as well its broader relevance for understanding the parent magnetic instabilities governing the behaviors of Hunds' metals.  

This paper is divided into five sections including this introduction.  In section two, we first state the definitions of the problem in terms of response functions and the Hamiltonian under consideration.  In section three, we establish the theoretical framework of the spin-orbit exciton model.  In the fourth section, we then utilize the exciton model to account for previously reported inelastic neutron\cite{jain13:17} and x-ray spectroscopic data,\cite{das18:8} yielding optimized parameters in the model Hamiltonian whose values are directly compared to their corresponding physical quantities reported in literature. In the fifth and final section, we infer necessary conditions for a crystalline electric field-induced singlet ground state to host both longitudinal excitations and transverse wavelike dynamics, while providing possible mechanisms to achieve quantum criticality.


\section{Definitions:  Correlation, response functions, and scattering cross sections} 

\indent In this section, we present the theoretical framework and the definitions of the spin-orbit exciton model and its ability to directly parametrize the magnetic neutron scattering response in Ca$_{2}$RuO$_{4}$. 

\indent Previously utilized to address the temperature dependence of the low energy magnetic fluctuations in PrTl$_{3}$ by Buyers \emph{et al.}~\cite{buyers11:75} and more recently to understand the complex excitation spectra in CoO,\cite{sarte100:19} the spin-orbit exciton model employs the direct proportionality of the magnetic neutron cross section and the magnetic dynamic structural factor $S({\bf{Q}},\omega)$ given by  

\begin{equation}
S({\bf{Q}},\omega)=g_{L}^{2}f^2({\bf{Q}})\sum_{\alpha \beta} (\delta_{\alpha \beta}-\hat{Q}_{\alpha}\hat{Q}_{\beta}) S^{\alpha \beta}({\bf{Q}},\omega), 
\nonumber
\end{equation}

\noindent corresponding to a product of the Land\'{e} $g$ factor $g_{L}$, the magnetic form factor $f(\mathbf{Q})$, a polarization factor providing sensitivity to the component exclusively perpendicular to the momentum transfer $\mathbf{Q}$, and the dynamic spin structure factor $S^{\alpha \beta}({\bf{Q}},\omega)$. Corresponding to the Fourier transform of the spin-spin correlations

\begin{equation}
S^{\alpha\beta}({\bf{Q}},\omega)=\frac{1}{2\pi} \int dt e^{i\omega t} \langle \hat{S}^{\alpha} ({\bf{Q}},t) \hat{S}^{\beta}(-{\bf{Q}},0) \rangle,
\nonumber
\end{equation}

\noindent where $\alpha,\beta=x, y, z$, $S^{\alpha \beta}({\bf{Q}},\omega)$ as written above considers only the spin contribution to the neutron scattering cross section, a valid approximation given that the expectation value of the orbital angular momentum $\langle {\hat{\bf{L}}}\rangle \equiv$ 0 \emph{via} quenching for $d$-orbitals,\cite{Yosida:book}. In the next section, we will show that the orbital contribution to the scattering cross section does exist, and is enabled through a spin-orbit ($\hat{\mathbf{L}}\cdot\hat{\mathbf{S}}$) coupling term in the model Hamiltonian.

\indent The relation of the structure factor $S^{\alpha \beta}({\bf{Q}},\omega)$ to the response function is given by the fluctuation-dissipation theorem 

\begin{equation}
S^{\alpha \beta}({\bf{Q}},\omega)=-\frac{1}{\pi} \frac{1}{1-\exp(\omega/k{\rm{_{B}}}T)} \Im{G^{\alpha \beta} (\bf{Q},\omega)}, 
\label{eq:2}
\end{equation}

\noindent and allows the magnetic neutron cross section to be defined in terms of a Green's response function $G^{\alpha \beta}(\bf{Q},\omega)$.\cite{Zubarev60:3} Recognizing that the neutron response function is proportional to the temperature dependent Bose factor multiplied by the Fourier transform of the retarded Green's function

\begin{equation}
\begin{split}
G^{\alpha\beta}(ij, t)  & =  G(\hat{S}^{\alpha}(i,t),\hat{S}^{\beta}(j,0))  \\
 & =  -i\Theta(t)\langle[\hat{S}^{\alpha}(i,t),\hat{S}^{\beta}(j,0)]\rangle,
\nonumber
\end{split}
\end{equation}

\noindent where $\hat{S}$ here denotes a generic angular momentum operator, it can be shown that the application of appropriate boundary conditions onto the time Fourier transform of the first time derivative of $G^{\alpha\beta}(ij, t)$, yields an equation-of-motion of the general form  

\begin{equation}
\omega G(\hat{a},\hat{a}',\omega) = \langle [\hat{a},\hat{a}'] \rangle + G([\hat{a},\hat{\mathcal{H}}],\hat{a}',\omega), 
\label{eq:4}
\end{equation}

\noindent for a magnetic Hamiltonian $\hat{\mathcal{H}}$ and a generic component of a general angular momentum operator $\hat{a}$. 
The presence of the $[\hat{a},\hat{\mathcal{H}}]$ in Eq.~\ref{eq:4} demonstrates that the employment of response function $G^{\alpha \beta}(\bf{Q},\omega)$ allows for the magnetic neutron cross section to be directly parametrized \emph{via} the individual contributing terms to $\hat{\mathcal{H}}$. This equation-of-motion provides a direct connection between the model microscopic Hamiltonian of interest to the neutron scattering cross section that is measured experimentally.

\indent In the case of magnetic fluctuations that stem from localized magnetic moments on site $i$ with spin $\hat{\mathbf{S}}(i)$, each with a single-ion crystal field (CF) contribution, and coupled to each other by Heisenberg exchange between sites $i$ and $j$ defined by $J(ij)$, $\hat{\mathcal{H}}$ can be written as 

\begin{equation}
\begin{split}
\hat{\mathcal{H}} &= \hat{\mathcal{H}}_{CF} + \sum\limits_{ij}J(ij)\hat{\mathbf{S}}(i) \cdot \hat{\mathbf{S}}(j).
\label{eq:5} 
\end{split} 
\end{equation}

\noindent At temperatures $T < T\rm{_{N}}$ (or $T\rm{_{C}}$), a molecular field stemming from the assumption of long range magnetic order will be present at each given site $i$. This collective effect can be accounted for by a Zeeman-like term in the Hamiltonian given by

\begin{equation}
\hat{\mathcal{H}}_{MF}(i) = \sum\limits_{i}H_{MF}(i)\hat{S}_{z}(i),
\label{eq:6_1}
\end{equation}

\noindent where the molecular field is

\begin{equation}
H_{MF}(i) = 2\sum\limits_{i>j}J(ij)\langle \hat{S}_{z}(j)\rangle.
\label{eq:6} 
\end{equation}

\noindent Using these definitions, it can be shown~\cite{buyers11:75,sarte100:19} that $\hat{\mathcal{H}}$ can be divided into a sum of a single-ion ($\hat{\mathcal{H}}_{1}$) and an inter-ion ($\hat{\mathcal{H}}_{2}$) term given by

\begin{equation}
\hat{\mathcal{H}}_{1} = \sum\limits_{i}\hat{\mathcal{H}}_{CF}(i) + \sum\limits_{i}\hat{\mathcal{H}}_{MF}(i) ,
\label{eq:7}
\nonumber
\end{equation}

\noindent and

\begin{equation}
\begin{split}
\hat{\mathcal{H}}_{2} &= \sum\limits_{ij}J(ij)\hat{S}_{z}(i)[\hat{S}_{z}(j)-2\langle \hat{S}_{z}(j)\rangle] \\
&+ \frac{1}{2}\sum\limits_{ij}J(ij)[\hat{S}_{+}(i)\hat{S}_{-}(j) + \hat{S}_{-}(i)\hat{S}_{+}(j)],
\end{split}
\nonumber
\end{equation}

\noindent respectively. 

\indent In the spin-orbit exciton model for magnetic excitations, the single-ion term is first diagonalized for a given molecular field to provide the basis states $|n\rangle$. In second quantization formalism, the diagonalization of $\hat{\mathcal{H}}_{1}$ is written as

\begin{equation}
\hat{\mathcal{H}}_{1}|n\rangle = \omega_{n}|n\rangle,
\label{eq:9}
\end{equation}

\noindent where $\omega_{n}$ corresponds to the energy eigenvalue of the $|n\rangle$ Fock state. Such a diagonalization facilitates a redefinition of $\hat{\mathcal{H}}_{1}$ in terms of ladder operators $C(i)$ and $C^{\dagger}(i)$, such that 

\begin{equation}
\hat{\mathcal{H}}_{1}=\sum_{n} \sum_{i}\omega_{n} C^{\dagger}_{n}(i)C_{n}(i), 
\label{eq:10}
\end{equation}

\noindent where $C(i)$ and $C^{\dagger}(i)$ satisfy the commutation relations $[C_{n}(i),C^{\dagger}_{m}(j)]=\delta_{ij}\delta_{nm}$. It should be noted that for the purposes of conciseness, the circumflex accent $\hat{}$ in the case of operators have not been included, as will remain convention for the remainder of this section. 

\indent  Since the equation-of-motion (Eq.~\ref{eq:4}) involves the commutator of angular momentum operators with $\hat{\mathcal{H}}$ (Eq.~\ref{eq:5}), the evaluation of the response function requires a common basis. In the spin-orbit exciton model, the Fock states $\{ |n\rangle  \}$ accompanying the diagonalization of $\hat{\mathcal{H}}_{1}$ (Eqs.~\ref{eq:9}-\ref{eq:10}) provides a natural basis for such a task. By noting that the inter-ion term $\hat{\mathcal{H}}_{2}$ is itself a function of angular momentum operators, it becomes clear that the evaluation of the equation-of-motion requires the rotation of the components of the angular momentum operator $\hat{\mathbf{S}}$ onto the $\{ |n\rangle  \}$ basis. Such a coordinate rotation is given by 

\begin{equation} 
 	S_{(\pm,z)}  = \sum\limits_{mn}S_{(\pm,z)mn}C^{\dagger}_{m}C_{n},
 	\nonumber
 \end{equation}

\noindent utilizing the same ladder operators that were previously defined in Eq.~\ref{eq:10}. Such an approach is analogous to the bosonic approach of SU($N$) spin-wave theory previously discussed in the context of Kramers $j\rm{_{eff}}=1/2$ magnets.~\cite{Dong18:97,Muniz04:14} However, our excitonic description includes the crystalline electric field contribution explicitly in this analysis to calculate the uncoupled single-ion states.  In later sections, we will use this excitonic formalism to investigate how the crystalline electric field can be tuned to access nearby critical points and induce anomalous excitations.\\

\indent As shown by Buyers \emph{et al.}~\cite{buyers11:75} in the context of PrTl$_{3}$, and further applied to CoO by Sarte \emph{et al.},\cite{sarte100:19,Sarte20:32} the evaluation of the equation-of-motion begins by first defining an inter-level susceptibility $\tilde{G}$ 
 
\begin{equation}
G^{\alpha\beta}(i,j,\omega) = \sum\limits_{mn}S_{\alpha mn}\tilde{G}^{\beta}(m,n,i,j,\omega). 
\nonumber
\end{equation}

\noindent The use of $G^{\alpha\beta}(i,j,\omega)$, in combination with the projection of the full magnetic Hamiltonian $\hat{\mathcal{H}}= \hat{\mathcal{H}}_{1} + \hat{\mathcal{H}}_{2}$ onto the $\{ |n\rangle  \}$ basis, reduces the second term of the equation-of-motion (Eq.~\ref{eq:4}) to three sets of commutators, termed \textit{diagonal}, \textit{transverse}, and \textit{longitudinal}, with each commutator involving spin operators that are written in terms of ladder operators. When combined~\cite{sarte100:19} with the \emph{random phase decoupling method},~\cite{wolff60:120,cooke73:7,yamada66:21,yamada67:22} the evaluation of the three commutators in the $T\rightarrow0$~K limit reduces the Fourier transform of the equation-of-motion (Eq.~\ref{eq:4}) into a set of coupled linear and homogeneous equations given by:

\begin{widetext}
\begin{equation}
G^{\alpha\beta}(\mathbf{Q},\omega) = g^{\alpha\beta}(E) + g^{\alpha+}(\omega)J(\mathbf{Q})G^{-\beta} (\mathbf{Q},\omega) +  g^{\alpha-}(\omega)J(\mathbf{Q})G^{+\beta} (\mathbf{Q},\omega) + 2g^{\alpha z}(\omega)J(\mathbf{Q})G^{z\beta} (\mathbf{Q},\omega),
\label{eq:13} 
\end{equation}
\end{widetext}  

\noindent describing the coupling of the single-site response function   

\begin{equation}
g^{\alpha\beta}(\omega) = \sum\limits_{n}\left\{\frac{S_{\alpha 0n}S_{\beta n0}}{\omega  - \omega_{n0}} - \frac{S_{\alpha n0}S_{\beta 0n}}{\omega + \omega_{n0}} \right\},
\label{eq:14} 
\end{equation}

\noindent by the Fourier transform of the exchange interaction  

\begin{equation}
J(\mathbf{Q}) = \sum\limits_{i \neq j}J_{ij}e^{i\mathbf{Q}\cdot\mathbf{d}_{ij}}.
\label{eq:15}
\end{equation} 

\noindent By noting that the non-zero single-site response functions in a highly symmetric local octahedral coordination environment are restricted to $+-$, $-+$, or $zz$ combinations for $\alpha\beta$, Eq.~\ref{eq:13}, upon the inclusion of site indices, can be simplified to 

\begin{equation}
\begin{split}
& G^{\alpha\beta}_{ij}(\mathbf{Q},\omega) = \\
&\delta_{ij}g^{\alpha\beta}_{i}(\omega) + \sum\limits_{0\leq k\leq j-1}g^{\alpha\beta}_{i}(\omega)\Phi J_{i,i+k}(\mathbf{Q})G^{\alpha\beta}_{i+k,j}(\mathbf{Q},\omega),  
\label{eq:16}
\end{split}
\end{equation} 

\noindent where $\alpha\beta=\{+-,-+, zz\}$, and the prefactor $\Phi=2$ when $\alpha$ = $z$, and 1 otherwise. 

\indent Finally, the sum of Eq.~\ref{eq:16} over both $\alpha\beta$ and $ij$ combinations yields the total response function 

\begin{equation}
\begin{split}
& G(\mathbf{Q},\omega) = \sum\limits_{\alpha\beta}\sum\limits_{ij}G^{\alpha\beta}_{ij} \\
& = G^{+-}(\mathbf{Q},\omega) + G^{-+}(\mathbf{Q},\omega) + G^{zz}(\mathbf{Q},\omega), 
\end{split}
\label{eq:17}
\end{equation}

\noindent whose imaginary component, by the fluctuation-dissipation theorem (Eqs.~\ref{eq:2}), is proportional to the total dynamic structure factor, and thus the low temperature magnetic neutron cross section.

\section{Microscopic model: single-ion terms and Heisenberg spin exchange} 

\indent The simplification of the equation-of-motion (Eq.~\ref{eq:4}) to a set of coupled linear and homogeneous equations (Eq.~\ref{eq:13}, or equivalently Eq.~\ref{eq:16}), yields a model whose $\omega$- and $\mathbf{Q}$-dependence is explicitly parametrized by the single-site response function $g^{\alpha\beta}(\omega)$ coupled by the Fourier transform of the exchange interaction $J(\mathbf{Q})$, respectively. This explicit paramtrization corresponds to one of the main advantages of the excitonic approach in addressing the scattering cross section as the single-ion physics, corresponding to the effects of the crystalline electric field, is directly incorporated through the uncoupled single site response function $g(\omega)$.  In this section, we will discuss both the individual contributions to, and evaluation of, both these two terms in the equation-of-motion.  

\subsection{\texorpdfstring{Single site $g^{\alpha\beta}(\omega)$: Parameters \& Approximations}{}}

\indent With a sample temperature ($5$~K) much smaller than the energy transfers of interest ($\hbar\omega > 15$~meV), it is a valid approximation to restrict the discussion exclusively to the $T\rightarrow0$~K limit. In this limit, the single-site response function $g^{\alpha\beta}(\omega)$ (Eq.~\ref{eq:14}) is solely a function of the single-ion Hamiltonian $\hat{\mathcal{H}}_{1}$ (Eq.~\ref{eq:7}).  

\indent As illustrated in Fig.~\ref{fig:fig1}(b), there are four terms comprising the single-ion Hamiltonian for Ru$^{4+}$ in a local octahedral crystalline electric field

\begin{equation}
\begin{split}
\hat{\mathcal{H}}_{1} &= \hat{\mathcal{H}}_{CF}+ \hat{\mathcal{H}}_{MF}=\\
&(\hat{\mathcal{H}}_{CEF}+\hat{\mathcal{H}}_{SO} + \hat{\mathcal{H}}_{dis})+ \hat{\mathcal{H}}_{MF},
\nonumber
\end{split} 
\end{equation}

\noindent corresponding to the individual contributions from the octahedral crystalline electric field $\hat{\mathcal{H}}_{CEF}$, spin-orbit $\hat{\mathcal{H}}_{SO}$, the structural distortion $\hat{\mathcal{H}}_{dis}$ away from ideal octahedral coordination, and a mean molecular field $\hat{\mathcal{H}}_{MF}$ stemming from long range magnetic order. Additional contributions such as hyperfine nuclear splitting exhibit much weaker energy scales ($\sim\mu$eV), and are thus neglected in the current discussion.~\cite{Chatterji09:79}  We will now discuss each contributing term to this single-ion Hamiltonian, and thus the terms that ultimately parametrize the uncoupled single site susceptibilities $g(\omega)$.

\textit{The Octahedral Crystalline Electric Field}, $\hat{\mathcal{H}}_{CEF}$. In the case of the 4$d^{4}$ ion Ru$^{4+}$ in an octahedral environment surrounded by six oxygens, with a reported crystal field strength (10$Dq$) of $\sim$~4~eV,~\cite{Cao:book,Gretarsson19:100} it cannot be assumed that the $d$-orbital splitting induced by the crystalline electric field is small in comparison to the energy cost of violating the Pauli principle and pairing electrons in individual $d$-orbitals.  This contrasts with the case of $3d$ ions such as Co$^{2+}$ where a weak crystalline electric field is present, allowing both Hund's rules and the Pauli exclusion principle to be applied amongst the five degenerate $d$ orbitals.~\cite{Cowley13:88} With a strong crystalline electric field, the basis is taken to be $\{ |t\rangle, |e\rangle \}$, corresponding to the $d$ orbital states that have been split electrostatically into a low energy triplet $|t\rangle$ and higher energy doublet $|e\rangle$ states.~\cite{McClure59:9,Tanabe54:9,Tanabe54:9_2}, separated in energy by a value of $\Delta=10Dq$.

\indent The electronic configuration of the ground state can be determined by Hund's rules combined with the Pauli principle, albeit in the context of a large gap $\Delta\equiv10Dq$. Whereas in the case of weak crystal field theory, where the values of both $S$ and $L$ are those for the free-ion states that are obtained from the direct application of Hund's rules and Pauli principle on 5 degenerate $d$-orbitals, in the case of Hund's first rule and the spin angular momentum quantum number $S$, its value is still maximized, but bound by the restriction that the $|t\rangle$ manifold must be fully populated before proceeding to populating the higher energy $|e\rangle$ manifold. Based on such a restriction, Hund's first rule yields two unpaired electrons for a $d^{4}$ free-ion configuration, corresponding to a total spin angular momentum quantum number $S$=$\sum m_{S}$=1.  This gives a $|t^{4}\rangle|e^{0}\rangle$ configuration.

\indent In the case of the total orbital angular momentum number $L$, it is not clear on what this value should be simply based on Hund's second rule since the $d_{xy}$, $d_{xz}$, and $d_{yz}$ orbitals which make up the $|t\rangle$ manifold are mixtures of uncoupled $|m \rangle$ states which are eigenstates of the $L_{z}$.  Given the application of the Pauli principle discussed in the previous paragraph, we expect the orbital ground state to be triply degenerate given the $|t^{4}\rangle|e^{0}\rangle$ electron configuration and only one of the $| t\rangle$ orbitals are fully filled.  This is confirmed by the Sugano-Tanabe diagram~\cite{Tanabe54:9,Tanabe54:9_2} presented in Fig.~\ref{fig:fig1}(c), that was calculated using the electrostatic matrices and character tables supplied in Refs. \onlinecite{McClure59:9} in the strong crystal field basis with the assumption that the ratio of the Racah parameters $\frac{C}{B}\sim4.6$.  In the strong crystal field limit ($10Dq>B$) the ground state is a $^{3}T_{1g}$ orbital triplet, itself stemming from the $^{3}H$ ($L$=5, $S$=1) free-ion state ($10Dq\rightarrow0$).  


\textit{Spin-Orbit Coupling}, $\hat{\mathcal{H}}_{SO}$. Corresponding to the relativistic interaction between the spin and orbital degrees of freedom, spin-orbit coupling is given by

\begin{equation}
\hat{\mathcal{H}}_{SO} = \lambda\hat{\mathbf{L}}\hat{\cdot\mathbf{S}},
\label{eq:SO} 
\end{equation}

\noindent where $\lambda$ is the spin-orbit coupling constant and expected to increase with atomic number as $\sim Z^{2}$.~\cite{Landau:book} The inclusion of spin-orbit coupling in the magnetic Hamiltonian yields a non-zero $[S_{z}, \mathcal{H}]$, and as a result, the expectation value of $S_{z}$ is not conserved. This lack of conservation is in stark contrast with an exclusively Heisenberg magnetic Hamiltonian, and enables the possibility of an amplitude, or longitudinal $zz$, mode to exist.~\cite{pekker15:36}  In the case of Ca$_{2}$RuO$_{4}$, $|\lambda|\sim0.1$~eV is an order of magnitude smaller than $10Dq$~\cite{das18:8,mizokawa01:87,fatuzzo15:91}, and thus spin-orbit coupling can be considered as a perturbation to the crystal field states defined by $\hat{\mathcal{H}}_{CEF}$ in the $\{ |t\rangle, |e\rangle \}$ basis. 

\indent The treatment of this particular perturbation can be simplified by noting that the magnetic fluctuations of interest originate exclusively from the $^{3}T_{1g}$ crystal field ground state presented in Fig. \ref{fig:fig1} $(c)$, being already accessible with neutron incident energies $E_{i}\lesssim|\lambda|$ incident on a sample at 5~K.  This exclusivity in the determination of the magnetic properties of Ca$_{2}$RuO$_{4}$, allows one to confine the current discussion to the triply degenerate crystal field ground state. Requiring a projection from the original $\{|t\rangle,|e\rangle\}$ basis onto the smaller $|l=1,m_{l}\rangle$ basis that defines the subspace spanned by the $^{3}T_{1g}$ orbital triplet, the spin-orbit Hamiltonian (Eq.~\ref{eq:SO}) can be rewritten as  

\begin{equation}
\hat{\mathcal{H}}_{SO} = \alpha'\lambda\hat{\mathbf{l}}\hat{\cdot\mathbf{S}}, 
\label{eq:SO2} 
\end{equation}

\noindent consisting of new orbital angular momentum operators that act on the new $|l=1,m_{l}\rangle$ basis, accompanied by a scalar projection factor $\alpha'$.   
\indent In contrast to magnets located in the \emph{weak-intermediate field} regime, the determination of the scalar $\alpha'$ for Ca$_{2}$RuO$_{4}$ is not particularly straightforward. In the case of the weak crystal field limit, the projection is between bases with good quantum numbers $L$ ($m_{L}$) or $l$ (m$_{l}$), and $S$ ($m_{S}$). Possessing fixed orbital and spin angular momentum values, the projection is amenable to methods based on the matrix representation of the angular momentum operators, significantly simplifying the process for determining $\alpha'$ as was done for CoO.~\cite{sarte100:19} For magnetic ions located in the \emph{strong crystal field} regime such as is the case for Ca$_{2}$RuO$_{4}$, the value of $L$ ($m_{L}$)  was addressed by Griffith~\cite{Griffith60:56} using the $T$-$P$ equivalence relation 

\begin{equation}
\hat{\mathbf{L}}(t) = -\hat{\mathbf{L}}(p).
\nonumber
\end{equation}

\noindent Valid in the $\Delta \gg \lambda$ regime where minimal mixing between $t$ and $e$ states occur, the projection of the $L=2$ $t$ onto the $L=1$ $p$ states greatly simplifies the process of calculating $\alpha'$ directly using representation theory. In the case of the orbital triplet ground state of the $d^{4}$ configuration, Griffith~\cite{Griffith60:56} and Moffitt~\cite{moffitt59:2} determined a value of $-\frac{1}{2}$ for $\alpha'$ in the pure $L$-$S$ Russell-Saunders coupling scheme. 

\indent Having projected $\hat{\mathbf{L}}$ onto a fictitious operator $\hat{\mathbf{l}}$, the basis of the new spin-orbit Hamiltonian $\hat{\mathcal{H}}_{SO}$ (Eq.~\ref{eq:SO2})  is now comprised of the 9 $|l=1,m_{l};S=1,m_{S}\rangle$ states. Based on both the Land\'{e} interval rule and the addition theorem, $\hat{\mathcal{H}}_{SO}$ yields three unique effective total angular momentum $\hat{\mathbf{j}}=\hat{\mathbf{l}} + \hat{\mathbf{S}}$ manifolds, corresponding to $j\rm{_{eff}}=0$, 1, and 2, with energy eigenvalues 

\begin{equation}
E = \left(\frac{\alpha'\lambda}{2}\right)[j(j+1)-S(S+1)-l(l+1)]. 
\label{eq:diff}
\end{equation}

\indent By employing both the projection constant $\alpha'=-\frac{1}{2}$ previously determined by Griffith, and the reported value of 75~meV for $|\lambda|$,~\cite{das18:8} the diagonalization of $\hat{\mathcal{H}}_{SO}$
 
\begin{equation}
\begin{split}
&diag(\hat{\mathcal{H}}_{SO}) =	\\
			\setlength{\arraycolsep}{3pt}
				\def\arraystretch{0.9}
&\left[\begin{array}{ccccccccc}
		-75&0&0&0&0&0&0&0&0 \\
		0&-37.5&0&0&0&0&0&0&0 \\ 
		0&0&-37.5&0&0&0&0&0&0 \\
		0&0&0&-37.5&0&0&0&0&0 \\
		0&0&0&0&37.5&0&0&0&0 \\
		0&0&0&0&0&37.5&0&0&0\\ 
		0&0&0&0&0&0&37.5&0&0 \\
		0&0&0&0&0&0&0&37.5&0 \\
		0&0&0&0&0&0&0&0&37.5\\
		\end{array}	\right], 
		\nonumber
\end{split}
\end{equation}

\noindent confirms a singlet ground state, separated from triplet and pentet excited manifolds by $\Delta$(singlet$\rightarrow$triplet) = $37.5$~meV $\equiv\alpha'\lambda$, and  $\Delta$(singlet$\rightarrow$pentet) = $112.5$~meV $\equiv3\alpha'\lambda$ (Fig.~\ref{fig:fig1}(b)), in agreement with Eq.~\ref{eq:diff}. Confirmation of the assignment of $j\rm{_{eff}}=0$, 1, and 2 to the singlet, triplet, and pentet manifolds, respectively, is accomplished by the projection of the components of the effective total angular momentum operator $\hat{\mathbf{j}}=\hat{\mathbf{l}} + \hat{\mathbf{S}}$  onto the subspaces that are spanned by the three individual manifolds defined by $\hat{\mathcal{H}}_{SO}$. Such a projection corresponds to a rotation of the individual angular momentum operators from the original $|l,m_{l},S,m_{S}\rangle$ basis to a basis $|\phi{\rm{_{SO}}}\rangle$, consisting of the individual eigenvectors of $\hat{\mathcal{H}}_{SO}$. In the matter of a generic angular momentum operator $\hat{\mathcal{O}}$, this particular rotation is achieved by 

\begin{equation}
\hat{\mathcal{O}}_{|\phi_{SO}\rangle} = \mathcal{C}^{-1}\hat{\mathcal{O}}_{|l,m_{l}\rangle}\mathcal{C},
\label{eq:rotation}
\end{equation}

\noindent corresponding to the matrix multiplication of $\hat{\mathcal{O}}$ by a transformation matrix $\mathcal{C}$ (and its inverse) which consists of the individual eigenvectors $\phi\rm{_{SO}}$ that are arranged in order of increasing energy. In the case of $z$-component $\hat{j_{z}}$, the rotation given by Eq.~\ref{eq:rotation} yields  

\begin{equation}
\mathcal{C}^{-1}\hat{j_{z}}\mathcal{C}=
					\setlength{\arraycolsep}{3pt}
						\def\arraystretch{1}
\left[\begin{array}{c|ccc|ccccc}
\mathbf{0}& 	0 &	0 &	0&	0& 	0&	0&	0&	0 \\  \hline
0&	\mathbf{-1}&	\mathbf{0}&	\mathbf{0}&	0&	0&	0&	0&	0\\
0&	\mathbf{0}&	\mathbf{0}&	\mathbf{0}&	0&	0&	0&	0&	0\\
0&	\mathbf{0}&	\mathbf{0}&	\mathbf{1}&	0&	0&	0&	0&	0\\ \hline 
0&	0&	0&	0&	\mathbf{-2}&	\mathbf{0}&	\mathbf{0}&	\mathbf{0}&\mathbf{0}\\
0&	0&	0&	0&	\mathbf{0}&	\mathbf{-1}&	\mathbf{0}&	\mathbf{0}&	\mathbf{0}\\ 
0&	0&	0&	0&	\mathbf{0}&	\mathbf{0}&	\mathbf{0}&	\mathbf{0}&	\mathbf{0} \\
0&	0&	0&	0&	\mathbf{0}&	\mathbf{0}& \mathbf{0}&	\mathbf{1}&	\mathbf{0} \\
0&	0&	0&	0&	\mathbf{0}&	\mathbf{0}&	\mathbf{0}&	\mathbf{0}&	\mathbf{2}\\
		\end{array}	\right], 
		\nonumber 
\end{equation}

\noindent corresponding to a matrix whose top 1 $\times$ 1, middle 3 $\times$ 3, and bottom 5 $\times$ 5 block matrices are identical to the $\hat{J_{z}}$ operator in $|j{\rm{_{eff}}}=0,m_{j,{\rm{eff}}}\rangle$, $|j{\rm{_{eff}}}=1,m_{j,{\rm{eff}}}\rangle$, and $|j{\rm{_{eff}}}=2,m_{j,{\rm{eff}}}\rangle$ bases, respectively. By performing the same projection for the $\hat{j_{x}}$ and $\hat{j_{y}}$ operators, it can be shown that these block matrices satisfy the canonical commutation relations of angular momentum $\hat{\mathbf{j}}=i\hat{\mathbf{j}}\times\hat{\mathbf{j}}$, thus confirming that these block matrices do indeed correspond to valid angular momentum operators.


\textit{The Distortion Hamiltonian, $\hat{\mathcal{H}}_{dis}$:} Employing single crystal neutron diffraction, Braden \emph{et al.}~\cite{braden98:58} were the first to identify that the $d^{4}$ Ca$_{2}$RuO$_{4}$ exhibits a strong cooperative Jahn-Teller distortion away from an ideal octahedral environment. Over the next two decades, a plethora of extensive studies would establish that the distortion accompanies orbital ordering, corresponding to the driving mechanism for the metal-to-insulator transition at $T\rm{_{MI}}$=357~K~\cite{Alexander99:60,Zegkinoglou05:95,Qi10:105,Lee02:89} and the presence of the ``Higgs" mode.~\cite{Zhang20:101}

\indent To account for such a distortion in our model of the single-ion eigenstates, deviations of the crystalline electric field away from ideal local octahedral coordination are considered. By noting that the triplet-doublet gap $\Delta$ induced by a crystalline electric field in close proximity to the orbital cross-over value $10Dq/B\sim$2.7  (Fig.~\ref{fig:fig1}(b)) is identical (in magnitude) to the splitting observed near the free-ion limit, it is a valid assumption that in the case of Ca$_{2}$RuO$_{4}$, the undistorted crystalline electric field $\hat{\mathcal{H}}_{CEF}$ can be written in terms of Stevens operators $\hat{\mathcal{O}}^{0}_{4}$ and $\hat{\mathcal{O}}^{4}_{4}$ as 

\begin{equation}
\hat{\mathcal{H}}_{CEF} = B^{0}_{4}\left(\hat{\mathcal{O}}^{0}_{4} + 5\hat{\mathcal{O}}^{4}_{4}\right).
\nonumber
\end{equation} 

\noindent The Stevens parameter $B^{m}_{l}$ prefactor is a numerical coefficient given by~\cite{Hutchings64:16,Bauer:book} 
\begin{equation}
B^{m}_{l} = -|e| p^{m}_{l}\gamma^{m}_{l}\langle r^{l}\rangle\Theta_{l},
\label{eq:stevens}
\end{equation}

\noindent where $\Theta_{l}$ corresponds to projection constants accompanying the conversion from Cartesian coordinates to angular momentum operators \emph{via} the Wigner-Eckart theorem, and is commonly denoted as $\alpha$ and $\beta$ for $l=2$ and $l=4$, respectively. $\langle r^{l}\rangle$ denotes the expectation values of the radial wavefunction and correspond to 3.319$a^{2}_{0}$ and 20.22$a^{4}_{0}$ for $l=2$ and $l=4$, respectively.~\cite{Abragam:book,Hotta06:69} $p^{m}_{l}$ are the scalar coefficients of the corresponding Tesseral functions $Z^{m}_{l}$. In the point charge approximation, where the charge density $\rho(\mathbf{r}) \propto \delta(\mathbf{r})$, the Tesseral functions are incoroporated into the $\gamma^{m}_{l}$ term given by 

\begin{equation}
\gamma^{m}_{l} = \frac{1}{2l+1}\sum\limits_{i}\frac{q_{i}Z^{m}_{l}(x_{i},y_{i},z_{i})}{\epsilon_{0}r^{l+1}_{i}},
\nonumber
\end{equation}   

\noindent where $q_{i}$ denotes the $i\rm{^{th}}$ charge. 

\indent As a first approximation, we have considered the simplest case of a uniaxial distortion along $z$. Given the definition above of the crystalline electric field in terms of the Stevens operator equivalents, an equivalent expression for a tetragonal distortion is given by~\cite{Walter87:59}

\begin{equation}
\hat{\mathcal{H}}_{dis}=B^{0}_{2}\hat{\mathcal{O}}^{0}_{2}=\Gamma  \left(\hat{l}^{2}_{z}-\frac{2}{3} \right),
\label{eq:distortion}
\nonumber
\end{equation}

\noindent and parametrized by $\Gamma$, whose sign and magnitude are determined by $B^{0}_{2}$.

\begin{table}[htb!]
\caption{Atomic coordinates of ions constituting a single distorted octahedron about a central magnetic Ru$^{4+}$ used in the determination of the Stevens parameters for \caruo~in the point charge limit.}
{\renewcommand{\arraystretch}{1.8}
\begin{tabular}{|c|c|c|c|}
\hline
~~Ion~~&~~$x$~~&~~$y$~~&~~$z$\\ 
\hline
Ru$^{4+}$ & 0  & 0.5 & 0.5  \\ \hline
O$^{2-}$(1) & 0.3058  & 0.7004 &  0.5225\\ \hline
O$^{2-}$(2) & 0.1942& 0.2004 & 0.5225\\ \hline
O$^{2-}$(3) & $-$0.3058 & 0.2996 & 0.4775 \\ \hline
O$^{2-}$(4) & $-$0.1942 & 0.7996  & 0.4775  \\ \hline
O$^{2-}$(5) & $-$0.0611 & 0.5152  & 0.6667  \\ \hline
O$^{2-}$(6) & 0.0611 &  0.4848 & 0.3333  \\ \hline
\end{tabular}}
\label{tab:1}
\end{table}

\indent To obtain an estimate for the energy scales and the signs of the Stevens coefficients to guide the analysis below, we have considered the simplest case for a point charge model. Corresponding to a single electron in a $d$-orbital ($L$=2), where $\alpha=-\frac{2}{21}$ and $\beta=+\frac{2}{63}$,~\cite{Hutchings64:16} Eq.~\ref{eq:stevens} for a single distorted octahedra about a magnetic Ru$^{4+}$ (Tab.~\ref{tab:1}) in the point charge limit produces Stevens parameters $B^{0}_{4}$ = $+1.32$~meV$>0$ and $B^{0}_{2}$=$+0.52$~meV$>0$, consistent with what is expected for the ground state orbital triplet configuration $^{3}T_{1g}$ of Ru$^{4+}$. Consisting of one fully filled and two partially filled $|t \rangle$ orbitals, the electronic configuration of Ru$^{4+}$ is analogous to that of $d^{2}$ V$^{3+}$ placed in a weak octahedral crystalline electric field, where the sign for $B^{0}_{4}$ and $B^{0}_{2}$ are both positive.~\cite{Abragam:book}  


\textit{The Molecular Field Hamiltonian, $\hat{\mathcal{H}}_{MF}$:} Corresponding to the final perturbative term to $\hat{\mathcal{H}}\rm{_{CEF}}$ in the single-ion Hamiltonian,  $\hat{\mathcal{H}}_{MF}$ (Eq.~\ref{eq:6_1}) addresses the effect of the mean molecular field stemming from the assumption of long range magnetic order by the Ru$^{4+}$ moments below $T\rm{_{N}}=110$~K. In the simplest case where a single dominant Ru$^{4+}$-O$^{2-}$-Ru$^{4+}$ superexchange pathway with an isotropic magnetic exchange constant $J_{1}$ between $z_{1}$ nearest neighbors is considered, $\hat{\mathcal{H}}_{MF}$ (Eq.~\ref{eq:6_1}) reduces to

\begin{equation}
\hat{\mathcal{H}}_{MF}=\sum_{i}H_{MF}(i)\hat{S}_{z} =2z_{1}J_{1}\langle \hat{S_{z}}\rangle \hat{S}_{z}.
\nonumber
\end{equation}

\noindent As illustrated in Fig.~\ref{fig:fig1}(b), $\hat{\mathcal{H}}_{MF}$ corresponds to a Zeeman-like term that removes time-reversal symmetry, splitting originally degenerate $j\rm{_{eff}}$ levels. In the case where $|J|\rightarrow|\lambda|$ (or equivalently $|H{\rm{_{MF}}}|/|\alpha'\lambda|\rightarrow1$), the corresponding increase in splitting induced by $\hat{\mathcal{H}}_{MF}$ results in the significant entanglement between individual $j\rm{_{eff}}$ levels.\cite{Sarte18:98_2,sarte98:18} The presence of such a strong admixture renders the modeling of the these systems difficult using psuedo-bosonic (Holstein-Primakoff) approaches that are based on conventional linear spin-wave theory, ultimately making it necessary to employ alternative methods such as the multi-level spin-orbit exciton model discussed here.  

\begin{table*}[htb!]
	\caption{Displacement vectors $\mathbf{d}_{m,ij}$ for each Ru$^{4+}$ constituting the $s$- and $d$-sublattices within the first two coordination shells $m$ of \caruo. All vectors and their magnitudes were calculated using the VESTA visualization software package,~\cite{Momma11:44} employing the reported unit cell parameters at $T=11$~K.~\cite{braden98:58} Numbers in parentheses indicate errors.}	\centering
	{\renewcommand{\arraystretch}{1.25}
		\begin{tabular}{|c|c|c|c|c|}
			\hline
			~~~$m$~~~&~~~$|\mathbf{d}_{m,ij}|$ (\AA)~~~&~~~Number of Neighbors~~~&~~~$\mathbf{d}_{m,ij}$ in $s$-Sublattice ($a$)~~~&~~~$\mathbf{d}_{m,ij}$ in $d$-Sublattice ($a$)~~~\\  
			\hline 
			\multirow{4}{*}{1} 	&   \multirow{4}{*}{3.8618(5)} & \multirow{4}{*}{4} & & $\left(\frac{1}{2}~\frac{1}{2}~0\right)$\\  
			&   &  & &  $\left(\frac{1}{2}~-\frac{1}{2}~0\right)$ \\ 
			&   &  & & $\left(-\frac{1}{2}~\frac{1}{2}~0\right)$ \\ 
			&   & & & $\left(-\frac{1}{2}~-\frac{1}{2}~0\right)$  \\ 
			\hline
			\multirow{4}{*}{2} 	&   \multirow{4}{*}{ 5.5150(11)} & \multirow{4}{*}{4} &   (1~0~0) & \\  
			&   &   & ($-1$~0~0) & \\
			&   &     &(0~1~0) &\\ 
			&   &     &(0~$-1$~0) & \\  \hline
	\end{tabular}}
\label{tab:tab2}
\end{table*}

\begin{figure*}[htb!]
	\centering
	\includegraphics[width=1.0\linewidth]{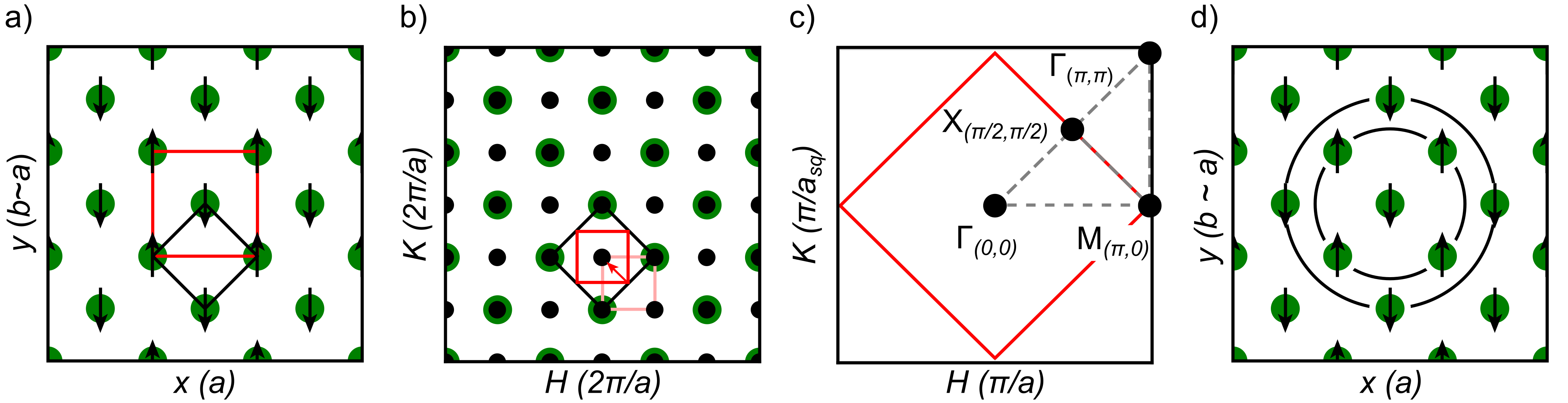}
	\caption{Non-primitive pseudo-tetragonal (red) and primitive square (black) (a) direct and (b) reciprocal space unit cells of Ca$_{2}$RuO$_{4}$ viewed in the $(a,b)$ and $(H,K)$ planes, respectively. The antiferromagnetic unit cell in the $(a,b)$ basal plane coincides with the non-primitive pseudo-tetragonal unit cell. (c) High symmetry directions between high symmetry points of the antiferromagnetic reciprocal unit cell (red) that has been symmetrized with respect to the square lattice (black).}
	\label{fig:fig2}
\end{figure*}

\subsection{\emph{J}(\textbf{Q}): Parameters \& Approximations} 

\indent Having already discussed the single-ion Hamiltonian and its role in determining both the eigenstates basis and the single site susceptibility $g(\omega)$, the discussion now shifts to addressing the coupling of these individual sites.  By enabling the coupling of $g^{\alpha\beta}(\omega)$, the Fourier transform of the exchange interaction $J(\mathbf{Q})$ uniquely specifies the $\mathbf{Q}$-dependence of the response function $G(\mathbf{Q},\omega$). Being itself parametrized by both $J_{ij}$ and $\mathbf{d}_{ij}$, corresponding to the magnetic exchange constant and displacement vector between moments located at sites $i$ and $j$, respectively, an analytical expression for $J(\mathbf{Q})$ requires detailed knowledge of both the nuclear and magnetic structures of the system under investigation. 

\indent In the case of Ca$_{2}$RuO$_{4}$,\cite{braden98:58,nakatsuji97:282} its orthorhombic $Pbca$ unit cell ($a$=5.4074~\AA, $b$=5.5150~\AA, $c$=11.90520~\AA) is a result of the reduction of symmetry of a $I4/mmm$ unit cell through a combination of the rotation and tilting of the compressed $\left[{\rm{RuO}}_{6}\right]^{8-}$ octahedra about the $c$ axis and $(a,b)$ plane, respectively. Corresponding to the ideal K$_{2}$NiF$_{4}$ structure (Fig. \ref{fig:fig1}), a structure type commonly observed among the cuprates and other high $T_{c}$ superconductors, the derivation of the Ca$_{2}$RuO$_{4}$ $Pbca$ unit cell from $I4/mmm$ suggests that the spin-orbit exciton model may utilize certain approximations commonly employed with these superconductors.~\cite{Zhou18:30}  

\begin{figure*}[htb!]
	\centering
	\includegraphics[width=1\linewidth]{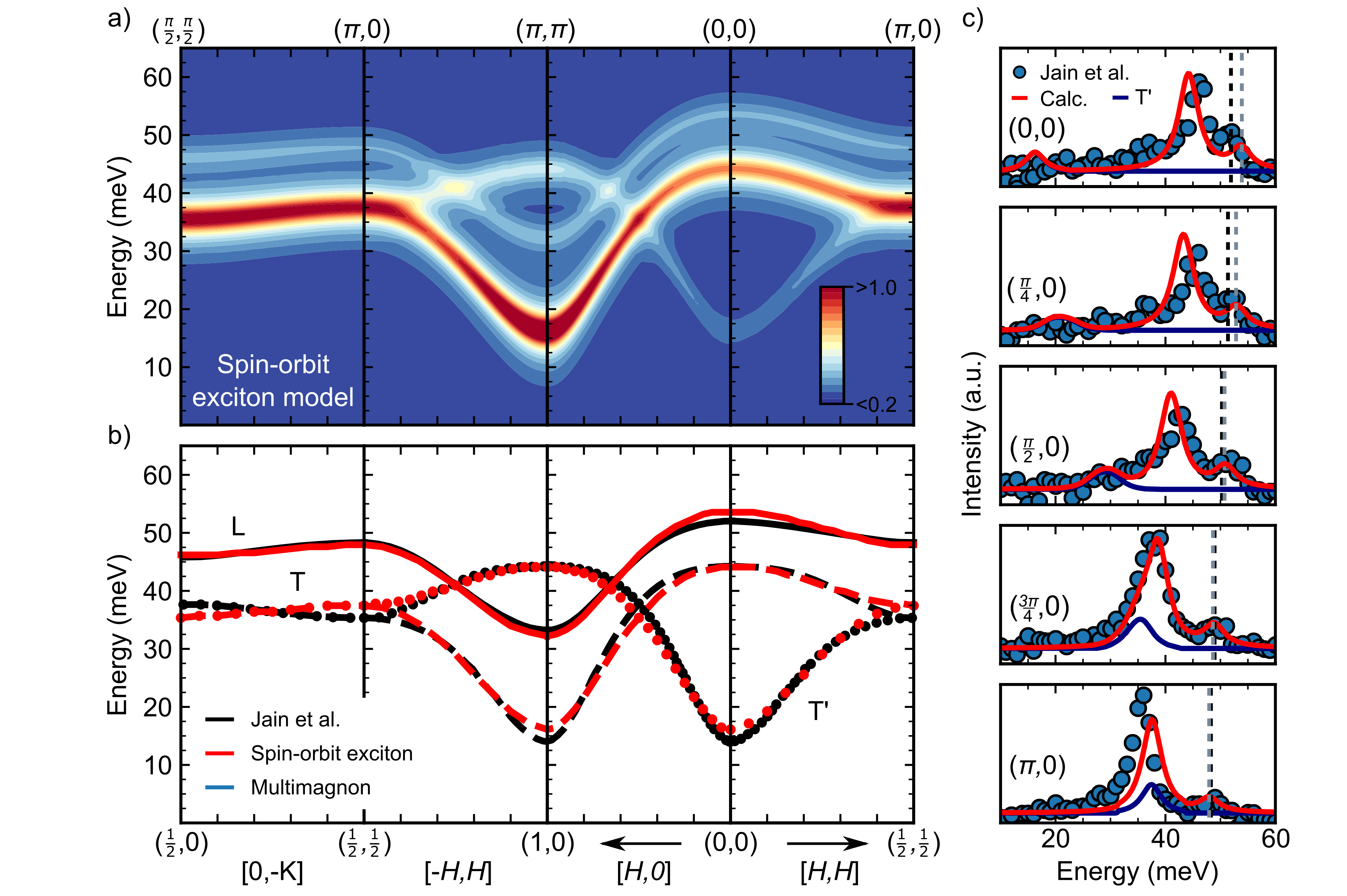}
	\caption{(a) Magnetic dynamic structure factor $S(\mathbf{Q},E)$ of Ca$_{2}$RuO$_{4}$ calculated in the $T\rightarrow0$~K limit using a mean-field multilevel spin-orbit exciton model with refined parameters summarized in Tab.~\ref{tab:3}. (b) Comparison of the dispersion relation for the longitudinal ($L$) and transverse modes ($T$, $T'$) calculated using a spin-orbit exciton model (red) and spin-wave theory by Jain \emph{et al.}~\cite{jain13:17} employing a phenomenological Hamiltonian (black). (c) Comparison of $\mathbf{Q}$-integrated cuts of data measured at 5~K by Jain \emph{et al.}\cite{jain13:17} and calculated using the spin-orbit exciton model for various $\mathbf{Q}$ $\in$ [(0,0), ($\pi$,$0$)] along [$H$,$H$]. For the purposes of comparison, the energy transfer for the maximum of the longitudinal mode that was previously calculated by spin-wave theory\cite{jain13:17} and spin-orbit exciton model have been both labeled explicitly for each $\mathbf{Q}$-integrated cut.}
	\label{fig:fig4}
\end{figure*}

\indent The large inter-plane distance illustrated in Fig.~\ref{fig:fig1}(a), combined with a N\'{e}el state consisting of magnetic moments that lie almost exclusively in the $(a,b)$ plane, suggests the restriction of the spin-orbit exciton model in the case of \caruo~to a single layer in the $(a,b)$ basal plane defined by a pseudo-tetragonal unit cell ($a\simeq b \neq c$) that is illustrated in Fig.~\ref{fig:fig2}. Such quasi-two dimensionality ($J_{c}\approx0$) has been experimentally validated with reported inelastic spectra (Fig.~\ref{fig:fig4}) exhibiting an $XY$-like dispersion with a maximum at $(H,K)=(0,0)$.~\cite{Kunkem17:95,jain13:17,Kunkem15:115} 

\indent As illustrated in Fig.~\ref{fig:fig2}(d) and summarized in Tab.~\ref{tab:tab2}, a $2d$ collinear antiferromagnet such as \caruo~can be reduced to two unique site indices corresponding to sublattices consisting of moments that are aligned anti-parallel and parallel relative to a reference moment. By exclusively considering the antiferromagnetically ordered Ru$^{4+}$ moments contained in the first two coordination shells within the $(a,b)$ basal plane that are coupled with isotropic magnetic exchange constants $J_{1}$ and $J_{2}$, Eq.~\ref{eq:15} yields two unique expressions for the Fourier transform of the exchange constants: 

\begin{equation}
J_{d} ({\bf{Q}}) = 2J_{1}(\cos(\pi(H+K))+\cos(\pi(H-K))),
\label{eq:jd}
\end{equation} 

\noindent and 

\begin{equation}
 J_{s} ({\bf{Q}}) = 2J_{2}(\cos(2\pi(H))+\cos(2\pi(K))),
\label{eq:js}
\end{equation}

\noindent for $J_{i,i+k}(\mathbf{Q})$ in Eq.~\ref{eq:17}, where labels $\mathbf{d}$ and $\mathbf{s}$ denote \textbf{d}ifferent $i\neq i+k$ and \textbf{s}ame $i=i+k$ site indices, respectively. 


\subsection{Model: Numerical Details} 

\indent In the case of the $2d$ collinear antiferromagnet \caruo, the restriction of the site indices $i,j$ to two unique values reduces Eq.~\ref{eq:17} to a series of four coupled linear equations given by:

\begin{align} 
G^{+-}_{11}(\mathbf{Q},E) &= g^{+-}_{1}(E) + g^{+-}_{1}(E) J_{s}(\mathbf{Q})G^{+-}_{11}(\mathbf{Q},E) \nonumber \\
&+ g^{+-}_{1}(E)J_{d}(\mathbf{Q})G^{+-}_{21}(\mathbf{Q},E)   \nonumber \\ 
G^{+-}_{21}(\mathbf{Q},E) &= g^{+-}_{2}(E)J_{s}(\mathbf{Q})G^{+-}_{21}(\mathbf{Q},E) \nonumber \\
&+ g^{+-}_{2}(E)J_{d}(\mathbf{Q})G^{+-}_{11}(\mathbf{Q},E)    \nonumber \\ 
G^{+-}_{12}(\mathbf{Q},E) &= g^{+-}_{1}(E)J_{s}(\mathbf{Q})G^{+-}_{12}(\mathbf{Q},E) \nonumber \\
&+ g^{+-}_{1}(E)J_{d}(\mathbf{Q})G^{+-}_{22}(\mathbf{Q},E)    \nonumber \\ 
G^{+-}_{22}(\mathbf{Q},E) &= g^{+-}_{2}(E) + g^{+-}_{2}(E)J_{s}(\mathbf{Q})G^{+-}_{22}(\mathbf{Q},E) \nonumber \\
&+ g^{+-}_{2}(E)J_{d}(\mathbf{Q})G^{+-}_{12}(\mathbf{Q},E) \nonumber 
\end{align}

\noindent and

\begin{align} 
G^{zz}_{11}(\mathbf{Q},E) &= g^{zz}_{1}(E) + 2g^{zz}_{1}(E) J_{s}(\mathbf{Q})G^{zz}_{11}(\mathbf{Q},E) \nonumber \\
&+ 2g^{zz}_{1}(E)J_{d}(\mathbf{Q})G^{zz}_{21}(\mathbf{Q},E)  \nonumber \\ 
G^{zz}_{21}(\mathbf{Q},E) &= 2g^{zz}_{2}(E)J_{s}(\mathbf{Q})G^{zz}_{21}(\mathbf{Q},E) \nonumber \\
&+ 2g^{zz}_{2}(E)J_{d}(\mathbf{Q})G^{zz}_{11}(\mathbf{Q},E)    \nonumber \\ 
G^{zz}_{12}(\mathbf{Q},E) &= 2g^{zz}_{1}(E)J_{s}(\mathbf{Q})G^{zz}_{12}(\mathbf{Q},E) \nonumber \\
&+ 2g^{zz}_{1}(E)J_{d}(\mathbf{Q})G^{zz}_{22}(\mathbf{Q},E)    \nonumber \\ 
G^{zz}_{22}(\mathbf{Q},E) &= g^{zz}_{2}(E) + 2g^{zz}_{2}(E)J_{s}(\mathbf{Q})G^{zz}_{22}(\mathbf{Q},E) \nonumber \\
&+ 2g^{zz}_{2}(E)J_{d}(\mathbf{Q})G^{zz}_{12}(\mathbf{Q},E) \nonumber 
\end{align}

\noindent where $\hbar$ has been set to 1, and thus $\omega$ being relabeled as $E=\hbar\omega$. We note here that we have assumed the different single-ion levels are coupled with the same $J(\mathbf{Q})$.  We make this assumption for simplicity and test this below against data in the next section. Solving these four coupled equations yields: 

\begin{widetext}
\begin{align}
\begin{split}
G^{+-}(\mathbf{Q},E) \equiv \sum\limits_{ij}G^{+-}_{ij}(\mathbf{Q},E)  = 
\frac{g^{+-}_{1}(E)+g^{+-}_{2}(E)+2g^{+-}_{1}(E)g^{+-}_{2}(E)[J_{d}(\mathbf{Q}) -J_{s}(\mathbf{Q})]} {[1-g^{+-}_{1}(E)J_{s}(\mathbf{Q})]\cdot [1  -g^{+-}_{2}(E)J_{s}(\mathbf{Q})]-g^{+-}_{1}(E)g^{+-}_{2}(E)[J_{d}(\mathbf{Q})]^{2}},\\ 
\\
G^{zz}(\mathbf{Q},E) \equiv \sum\limits_{ij}G^{zz}_{ij}(\mathbf{Q},E)  = 
\frac{g^{zz}_{1}(E)+g^{zz}_{2}(E)+4g^{zz}_{1}(E)g^{zz}_{2}(E)[J_{d}(\mathbf{Q}) -J_{s}(\mathbf{Q})]} {[1-2g^{zz}_{1}(E)J_{s}(\mathbf{Q})]\cdot [1 -2g^{zz}_{2}(E)J_{s}(\mathbf{Q})]-4g^{zz}_{1}(E)g^{zz}_{2}(E)[J_{d}(\mathbf{Q})]^{2}},
\end{split}
\label{eq:long} 
\end{align}

\end{widetext}

\noindent where $G^{-+}(\mathbf{Q},E)$ has the same form as $G^{+-}(\mathbf{Q},E)$ with indices $+$ $\longleftrightarrow$ $-$. Here, the energy $E$ was redefined as $E+i\delta$, where $\delta$ is a positive infinitesimal to ensure analyticity of $g^{\alpha\beta}$ (Eq.~\ref{eq:14}). Its value was set to 50\% of the experimental elastic resolution width (HWHM) on the ARCS time-of-flight neutron spectrometer (SNS, ORNL) set to the experimental parameters employed by Jain \emph{et al}.~\cite{jain13:17}

\indent In the $T\rightarrow0$~K limit where the Bose factor $n(E)\simeq 1$, the fluctuation-dissipation theorem (Eq.~\ref{eq:2}) is reduced to 

\begin{equation}
    S(\mathbf{Q},E) \propto -\Im G(\mathbf{Q},E), 
    \nonumber
\end{equation}

\noindent where the magnetic dynamic structure factor is directly proportional to the imaginary component of the total response function $G(\mathbf{Q},E)$. Since the dynamic structure factor is directly proportional to the magnetic neutron cross section $\frac{d^{2}\sigma}{d\Omega dE}$ in the $T\rightarrow 0$~K limit, plus the addition of the square of the magnetic form factor, an expression for the unnormalized raw inelastic neutron scattering intensity is given by   

\begin{equation}
    I(\mathbf{Q},E) \simeq -\mathcal{A}f^{2}(\mathbf{Q})\Im G(\mathbf{Q},E), 
\label{eq:raw} 
\end{equation}

\noindent where the magnetic form factor has been approximated by the isotropic magnetic form factor $f(\mathbf{Q})$ and the pre-factor $\mathcal{A}$ is a scalar corresponding to a combination of conversion and scale constants. The combination of the definition of $G(\mathbf{Q},E)$ (Eq.~\ref{eq:17}) with its individual components $G^{\alpha\beta}(\mathbf{Q},E)$ (Eq.~\ref{eq:long}) reduces the inelastic neutron scattering intensity given in Eq.~\ref{eq:raw} to a closed-form analytic expression describing the coupling of $g^{\alpha\beta}(E)$ by $J(\mathbf{Q})$. Since $g^{\alpha\beta}(E)$ (Eq.~\ref{eq:14}) is function of $\lambda$, $H_{MF}$, $\Gamma$, while $J(\mathbf{Q})$ (Eq.~\ref{eq:15}) is a function of $J_{1}$, and $J_{2}$, the analytic expression for the scattering intensity, for a fixed $\mathbf{Q}$ and $E$, is itself a closed-form function of five distinct parameters, making both parametrization and the subsequent optimization readily amenable to numerical calculation methods.~\cite{Sarte20:32}  

\begin{table}[htb!]
\caption{Refined parameter values of the spin-orbit exciton model for \caruo~obtained from a least squares minimization (Eq.~\ref{eq:least_squares}) performed throughout a 5$d$-hyperdimensional parameter space whose individual axes were defined by a range of values roughly centered about each parameter's initial value. All values are reported in meV and numbers in parentheses indicate calculated errors.}
{\renewcommand{\arraystretch}{1.8}
\begin{tabular}{|c|c|c|c|}
\hline
Parameter&Initial Value&~~Range~~&Refined Value\\ 
\hline
$\alpha'\lambda$ & 37.5 & [0,70] & 39(4) \\ \hline
$\Gamma$ & 14.9 & [0,25] &  19(2) \\ \hline
$J_{1}$ & 1.88 & [0,5]& 2.1(2) \\ \hline
$J_{2}$ & $-0.425$ & [$-2$,2] & $-0.37(6)$ \\ \hline
$H_{MF}$ & 11.6 & [0,25] & $0.1(1)$  \\ \hline
\end{tabular}}
\label{tab:3}
\end{table}
\indent The determination of optimal parameter values for: $\lambda$, $H_{MF}$, $\Gamma$, $J_{1}$, and $J_{2}$ was accomplished numerically in \texttt{MATLAB} by solving the least squares minimization problem    
\begin{equation}
    \sum\limits_{\beta=T,L}\sum\limits_{\mathbf{Q}}\left[E(\lambda,{\rm{H}}_{MF},\Gamma,J_{1},J_{2})_{\mathbf{Q},\beta}-E_{data,\mathbf{Q},\beta}\right]^{2},
    \label{eq:least_squares} 
\end{equation}

\noindent where $E(\dots)_{\mathbf{Q},\beta}$ and $E_{data,\mathbf{Q},\beta}$ denote the calculated and measured energy transfer possessing maximum intensity for a fixed $\mathbf{Q}$ and branch $\beta$, respectively. 
\indent The methods used for the determination of the initial values varied significantly from parameter to parameter. In the case of $\alpha'\lambda$ and $J_{1}$, the values of 37.5~meV and 1.88~meV simply correspond to their respective values reported in literature.~\cite{das18:8,mizokawa01:87,Veenstra14:112,fatuzzo15:91,Cao:book,Rho03:68,Rho05:71,Sugai90:42} An initial estimate for the mean molecular field $H_{MF}$ was determined by first extracting the value for $\sum\limits_{i>j}z_{ij}J_{ij}$ from the experimentally determined~\cite{nakatsuji00:84,nakatsuji00:62} Curie-Weiss temperature $\theta{\rm{_{CW}}}\sim-90$~K ($-7.76$~meV) \emph{via} its mean field definition

\begin{equation}
\theta{\rm{_{CW}}} = -\frac{2}{3}S(S + 1)\sum\limits_{i>j}z_{ij}J_{ij}.
\nonumber
\end{equation}   
\begin{figure*}[htb!]
	\centering
	\includegraphics[width=1\linewidth]{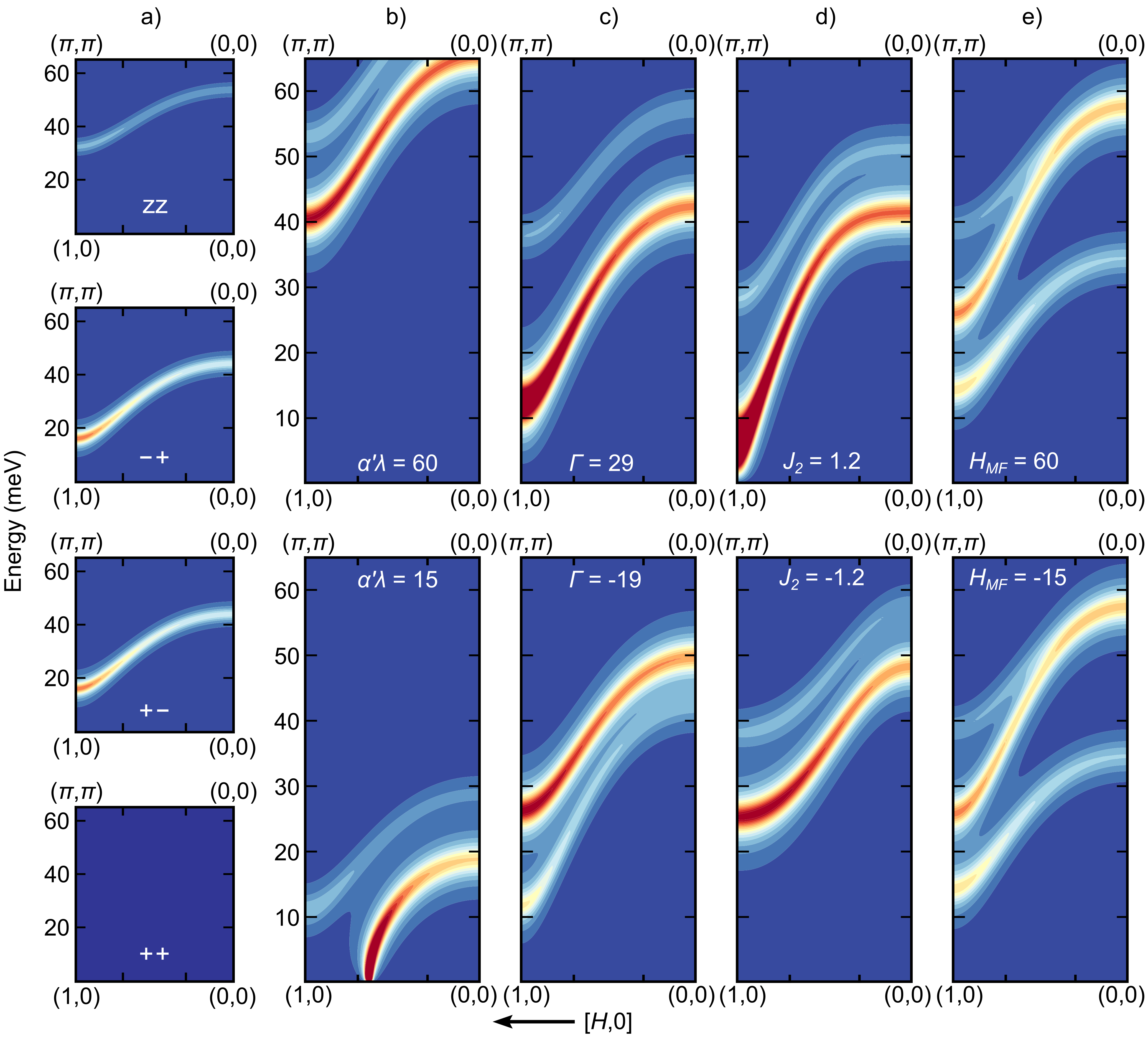}
	\caption{(a) Individual $\alpha\beta$ components of the magnetic dynamic structure factor $S$($\mathbf{Q}$,$E$) along [$H$,$0$] calculated using the spin-orbit exciton model. (b)-(e) Illustration of the effects of spin-orbit coupling $\alpha'\lambda$, tetragonal distortion $\Gamma$, next nearest neighbor magnetic exchange $J_{2}$, and the mean molecular field $H\rm{_{MF}}$ parameters, respectively, on the total calculated magnetic dynamic structure factor. Unless otherwise stated, the spin-orbit exciton model's parameters were fixed to their refined values that are listed in Tab.~\ref{tab:3}.}
	\label{fig:fig8}
\end{figure*}

\noindent The insertion of the extracted value of 5.82~meV for $\sum\limits_{i>j}z_{ij}J_{ij}$ into the definition of $H_{MF}$ (Eq.~\ref{eq:6}) yields an initial estimate of 11.6~meV for $H_{MF}$. Furthermore, since the spin-orbit exciton model considered here is restricted to the first two coordination shells, $\sum\limits_{i>j}z_{ij}J_{ij}$ is reduced to $4J_{1}+4J_{2}$. By combining the extracted value of 5.82~meV for $\sum\limits_{i>j}z_{ij}J_{ij}$ with the initial estimate of $1.88$~meV for $J_{1}$, the value of $-0.425$~meV is obtained for an initial estimate of $J_{2}$. A negative value whose magnitude is significantly smaller than $J_{1}$, consistent with a system that assumes long range antiferromagnetic order in the mean field limit. Finally, the initial estimate of 14.9~meV for $\Gamma$ was determined by the scaling of the distortion parameter reported for KCoF$_{3}$~\cite{Buyers71:4} by an empirical factor of 0.02/0.00197=10.15 corresponding to the ratio of their respective tetragonal distortions $\delta c/c$, while its positive sign corresponds to the restriction previously established in the point charge calculation that $\Gamma>0$.

\indent As is the case for all derivative-free methods, including the simplex search method specifically employed by \texttt{MATLAB}, the parameter values determined by the minimization algorithm do not necessarily yield the global minimum, or even a local minimum. This is particularly true when the initial values are too far removed from the true optimal values. As an attempt to address such a concern, the least squares minimization problem given by Eq.~\ref{eq:least_squares} was solved for various initial test values for the five parameters. These test values define a hyperdimensional $10^{d}$, $d=5$ parameter space, where each axis corresponds to a linear distribution of 10 values over a specified range that is roughly centered about the specific parameter's initial value. The set of optimized values for these five parameters that yield the minimum of the $10^{5}$ solutions to the least squares minimization problem was defined as parameters' refined values. The initial values, distribution ranges in the $5$ dimensional parameter space, and the final refined values of the five parameters are summarized in Tab.~\ref{tab:3}.

\section{Calculated Results} 

\indent Having established the underlying theoretical foundation and the corresponding physical parameters that constitute our model, we now present a direct comparison of the experimental data reported by Jain \emph{et al.}~\cite{jain13:17} to our calculated parametrization based on the spin-orbit exciton approach. 

\begin{figure*}[htb!]
	\centering
	\includegraphics[width=1\linewidth]{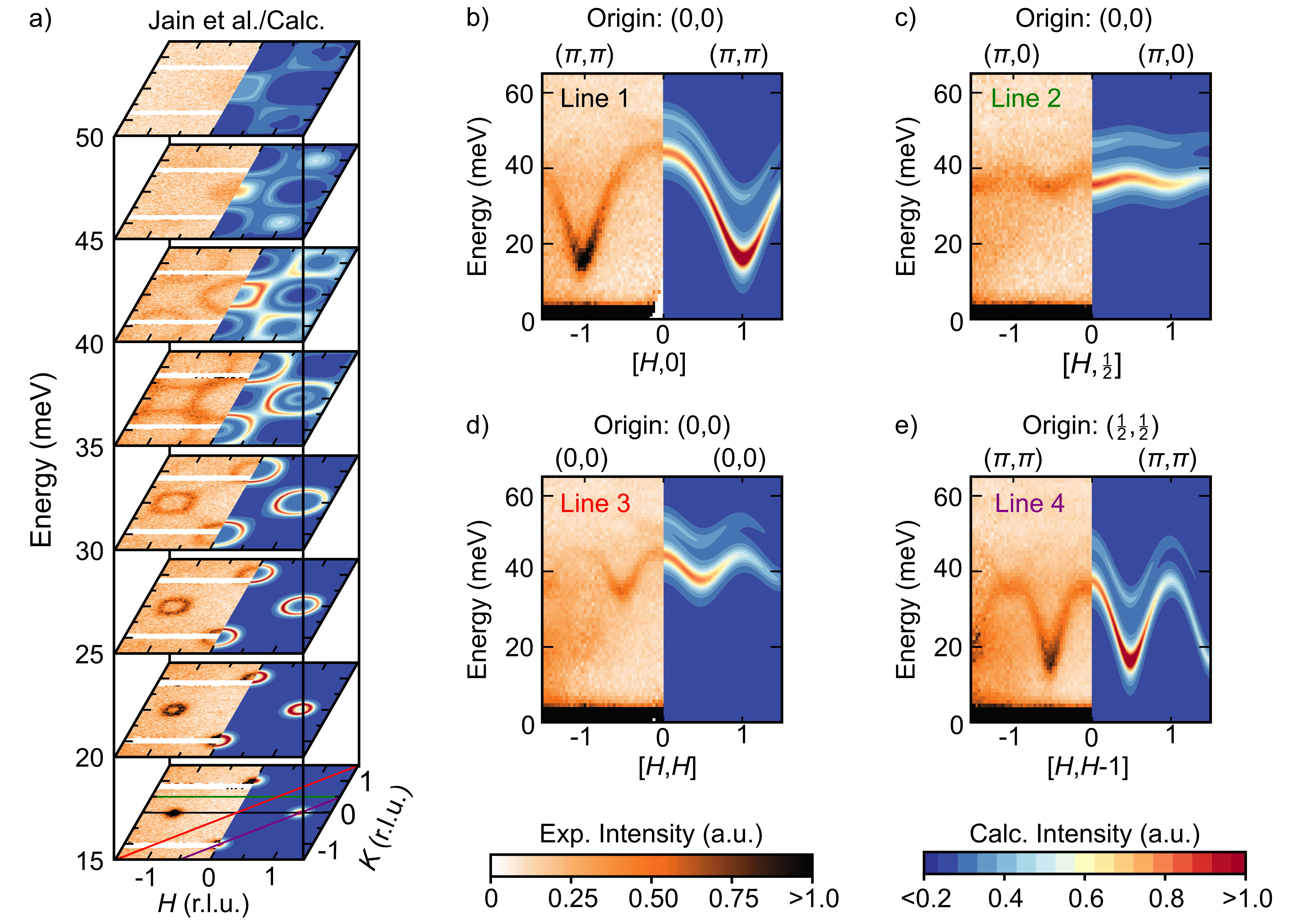}
	\caption{(a) Comparison of constant energy slices in the $(H,K)$ plane measured at 5~K by Jain \emph{et al.}~\cite{jain13:17} and calculated using the spin-orbit exciton model with corresponding ($\mathbf{Q}$,$E$)-slices along the (b) [$H$,0], (c) [$H$,$\frac{1}{2}$], (d) [$H$,$H$], and (e) [$H$,$H$-$1$] high symmetry directions. For the purposes of clarity, the origin in reciprocal space for each $(\mathbf{Q},E)$ slice has been stated explicitly.}
	\label{fig:fig7}
\end{figure*}

\indent As illustrated in Fig.~\ref{fig:fig4}(a), by employing the refined parameters in Tab.~\ref{tab:3} that were obtained through a least squares minimization, the spin-orbit exciton model yields two distinct modes (denoted as $L$ and $T$). We note that we will address the $T'$ mode present in Fig. \ref{fig:fig4} at the end of this section.  Fig.~\ref{fig:fig8}(a) illustrates that the strongly dispersive low energy $T$ mode corresponds to transverse fluctuations along the $(a,c)$ plane ($\alpha\beta$~=~$+-$ and $-+$), while the longitudinal $zz$ fluctuations along the $b$ axis uniquely constitute the second $L$ mode located at higher energy transfers. The dispersion relation for both modes throughout the Brillouin zone presented in Fig.~\ref{fig:fig4}(b) are in excellent agreement with their respective counterparts previously calculated by Jain \emph{et al.}~\cite{jain13:17} applying linear spin-wave theory to a phenomenological Hamiltonian in the $\Gamma \ll \lambda$ limit. As illustrated in Fig.~\ref{fig:fig7}, constant energy and $(\mathbf{Q},E)$ slices along select high symmetry directions identified the presence of minor discrepancies between the calculated model and experimental data that are predominately limited to a region in the Brillouin zone between $(\pi,0)$ and $\left(\frac{\pi}{2},\frac{\pi}{2}\right)$ along $[0,-K]$ at energy transfers $\sim$35-40~meV. Such discrepancies were previously noted by Jain \emph{et al.}\cite{jain13:17} and it is suspected that in the case of the current model, this particular discrepancy may stem from
a complex further neighbor exchange that has not been accounted for in our $J({\bf{Q}})$ (Eqs.~\ref{eq:jd} and~\ref{eq:js}) which has been restricted to employing isotropic magnetic exchange constants that span only over the first two coordination shells in the $(H,K)$ plane of the pseudo-tetragonal unit cell. 

\indent The influences of the spin-orbit exciton model's individual parameters, each corresponding to physically measurable quantities, on both the $T$ and $L$ modes are summarized in Figs.~\ref{fig:fig8}(b)-(e). By defining the splitting in energy between different $j\rm{_{eff}}$ manifolds of $\hat{\mathcal{H}}\rm{_{SO}}$ (Fig.~\ref{fig:fig1}(b)), $\alpha'\lambda$ determines the energy scale of interest. In the case of a fixed value for $\alpha'\lambda$, the magnetic exchange constants $J_{1}$ and $J_{2}$ both determine the modes' dispersion relation and bandwidth, while the value of the tetragonal distortion $\Gamma$ dictates the gap in energy between the $T$ and $L$ modes. In the case of $\Gamma$, a positive value yields an $L$ mode higher in energy relative to $T$, while a negative value simply reverses the order. Finally, for fixed values of $\alpha'\lambda$, $\Gamma$, $J_{1}$, and $J_{2}$, the magnitude and sign of the mean molecular field $H_{MF}$ determines the separation and relative order in energy, respectively, between the individual components ($+-$ and $-+$) that constitute the transverse $T$ mode.          
\begin{figure}[htb!]
	\centering
	\includegraphics[width=1\linewidth]{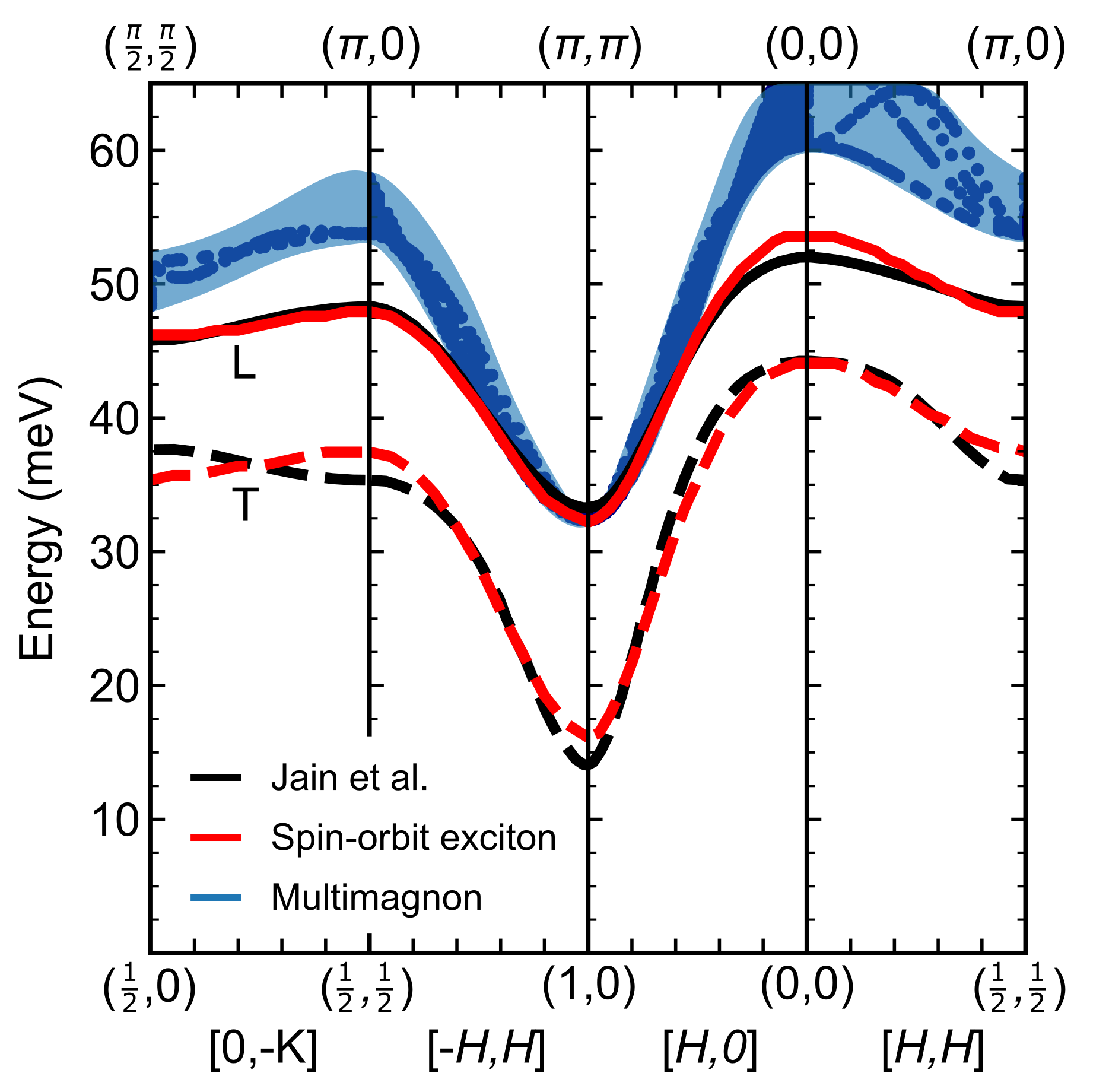}
	\caption{Comparison of the dispersion relation calculated using the spin-orbit exciton model (red) and spin-wave theory~\cite{jain13:17} employing $\hat{\mathcal{H}}\rm{_{phen}}$ (black) with the kinematically permissable ($\mathbf{Q}$,$E$) region for the two-magnon continuum.}
	\label{fig:fig9}
\end{figure}

\indent As summarized in Tab.~\ref{tab:3}, the refined values for four (out of the five) parameters: $\alpha'\lambda$, $\Gamma$ , $J_{1}$, and $J_{2}$ are in good agreement with their initial values. While the refined value of 39(4)~meV for $\alpha'\lambda$ agrees within error with the value previously deduced from RIXS, the slight deviation of the refined tetragonal distortion parameter $\Gamma$ of 19(2)~meV from its initial value of 14.9~meV may be attributed to the oversimplification of the reported\cite{braden98:58} monoclinic distortion into one that is bound uniaxially along $c$. In the case of the magnetic exchange constants: $J_{1}$ and $J_{2}$, both the magnitude and sign of their refined values agree with their respective initial values. With refined values $J_{1}$ = 1.88~meV $> |J_{2}=-0.425|$~meV, mean field theory suggests that Ca$_{2}$RuO$_{4}$ would assume antiferromagnetic long range order at $T_{N}\sim\frac{2}{3}S(S+1)\{4J_{1}+4J_{2}\}=108(2)$~K, all consistent with both previously reported physical property and neutron diffraction measurements.\cite{braden98:58,nakatsuji00:62,nakatsuji00:84,Cao97:56,bertinshaw19:123} Intuitively, the presence of long range antiferromagnetic order in Ca$_{2}$RuO$_{4}$ at first appears somewhat perplexing considering the presence of a $j\rm{_{eff}}=0$ singlet ground state for $\hat{\mathcal{H}}_{SO}$. An explanation for the apparent contradiction is the presence of a tetragonal distortion that enables coupling and admixture between higher lying spin-orbit manifolds and the singlet ground state. Furthermore, with a value for the tetragonal distortion parameter $\Gamma=19(2)$~meV being smaller than the energy scales of interest defined by $\alpha'\lambda$, the spectrum would appear to originate from a system consisting of a non-distorted $j\rm{_{eff}}=0$. This seemingly non-magnetic singlet ground state would be consistent with the refined value of $H_{MF}$=0~meV instead of the initial mean field value listed in Tab.~\ref{tab:3} that was obtained from the Curie-Weiss temperature $\theta\rm{_{CW}}$.            

\indent It was this presence of a $j\rm{_{eff}}=0$ singlet ground state that made \caruo~such an attractive candidate in the search of a condensed-matter analog of the much-celebrated Higgs mode.~\cite{Higgs64:13} As was previously determined in the orignal study by Jain \emph{et al.},\cite{jain13:17} our spin-orbit exciton model produced a longitudinally polarized $L$ mode that remains well-defined throughout the Brillouin zone (Figs.~\ref{fig:fig4}(a,b)). Possessing 33(5)\% of the intensity of the corresponding transverse mode at $\mathbf{Q}=(0,0)$ (Fig.~\ref{fig:fig4}(c)), this $L$ mode corresponds to amplitude fluctuations of the magnetic moment of a system of interacting spins that is located near a quantum critical point, consistent with what one would expect for the Higgs mode. 

\indent The importance for the presence of the Higgs mode analog in a condensed matter system is that it provides a unique platform in the study of the decay processes of a particle (and its properties through inference) that has been postulated to play a key role in the determination of masses in the Standard model. According to earlier theoretical treatments, it has been postulated that the Higgs mode decays into a pair of Goldstone modes.~\cite{Munehisa15:5,Rose15:91} To pursue such a possibility, the kinematically accessible phase space permitted for such a decay process was calculated based on energy and momentum conservation given by~\cite{Huberman05:72,stock18:2,songvilay18:121} 

\begin{equation}
\begin{split}
&G(\mathbf{Q},E) = \\
&\sum\limits_{\mathbf{Q}_{1},\mathbf{Q}_{2}}\delta(\mathbf{Q}-\mathbf{Q}_{1}-\mathbf{Q}_{2})\delta(E-E_{\mathbf{Q}_{1}}-E_{\mathbf{Q}_{2}}),
\end{split}
\nonumber
\end{equation}

\noindent  where $E_{\mathbf{Q}_{1,2}}$ are the energies of transverse excitations at a given momentum transfer $\mathbf{Q}$. As illustrated in Fig.~\ref{fig:fig9}, the kinematically allowed region overlaps in both momentum and energy with the antiferromagnetic wave vector $(\pi,\pi)$. The predicted overlap naturally lends itself to the possibility of coupling between the Higgs $L$ mode and the ``multi-magnon" continuum. Such coupling was required in the previous theoretical treatment to address the broad scattering that was experimentally observed at $(\pi,\pi)$, yet clearly absent at $(0,0)$.     
\begin{figure}[htp!]
	\centering
	\includegraphics[width=1\linewidth]{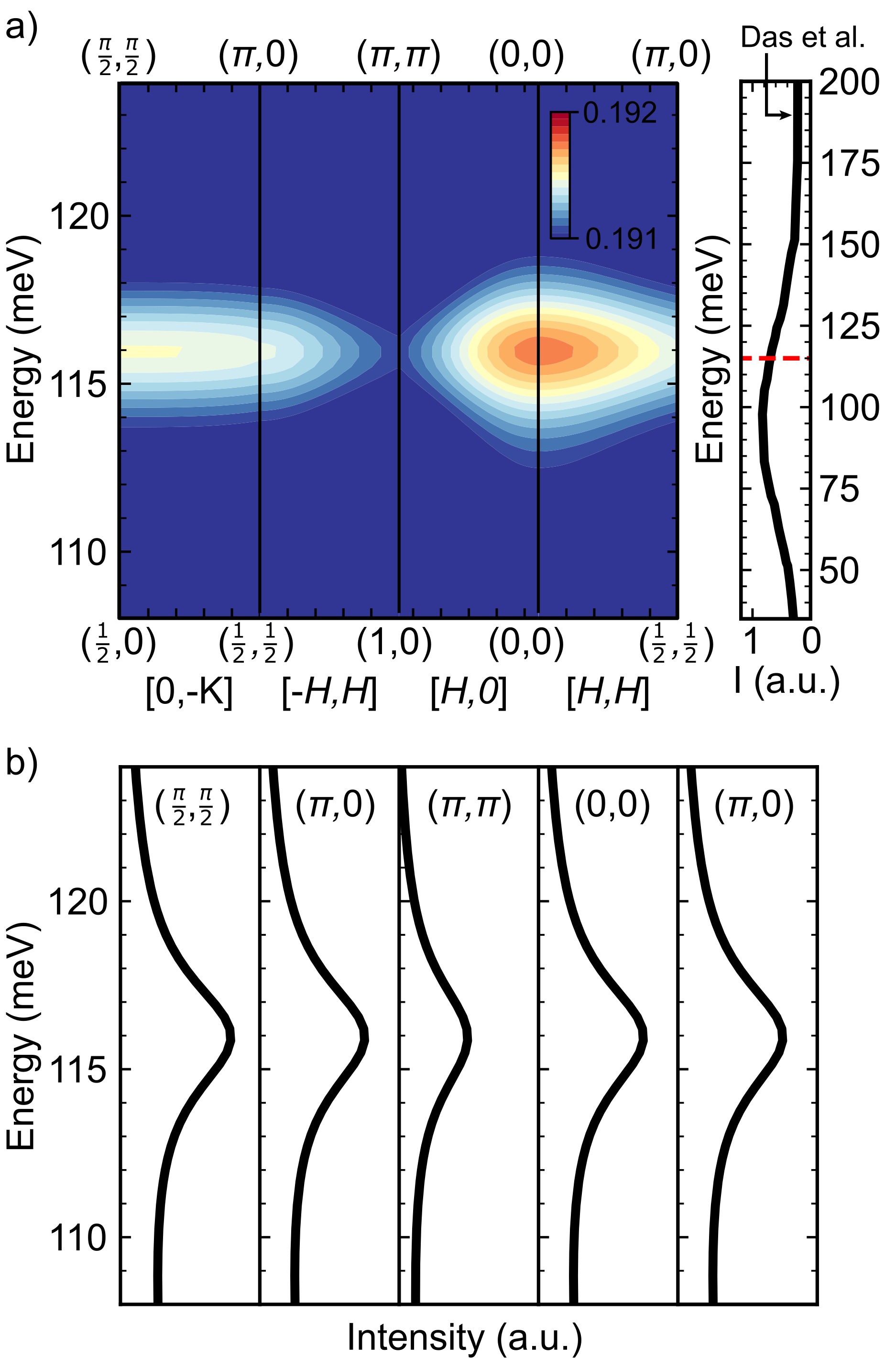}
	\caption{(a) Magnetic dynamic structure factor $S(\mathbf{Q},E)$ of Ca$_{2}$RuO$_{4}$ calculated using the spin-orbit exciton model in the $T\rightarrow0$~K limit exhibits a high energy mode ($E\sim$116~meV) associated with the dipolar forbidden $j{\rm{_{eff}}}=0\rightarrow2$ transition. (b) $\mathbf{Q}$-integrated cuts at various high symmetry points throughout the Brillouin zone reveal that the mode is minimally dispersive, (c) in close agreement with a spin-orbital excitation previously measured with RIXS.\cite{das18:8}}
	\label{fig:fig6}
\end{figure}

\indent As illustrated in Fig.~\ref{fig:fig6}, the extension of the spin-orbit exciton model to higher energy transfers ($E\gtrsim 0.1$~eV) identified an additional transverse mode centered at $E\sim 116$~meV. Corresponding to dipolar forbidden $j{\rm{_{eff}}}=0\rightarrow j\rm{_{eff}}=2$ transitions, the minimally dispersive mode exhibits an intensity three orders of magnitude lower than the corresponding $T$ mode at lower energy transfers. We note that while this mode is dipolar forbidden in the case of a perfect octahedral field, its corresponding transition is allowed in the case of \caruo~owing to the presence of a weak structural distortion.  Such a spin-orbital excitonic origin is consistent with previous theoretical approaches~\cite{das18:8,souliou17:119} addressing a minimally dispersive mode of magnetic origin that was measured at approximately the same energy transfer (Fig.~\ref{fig:fig6}(c)) with both RIXS and Raman spectroscopy.

\begin{figure*}[htp!]
	\centering
	\includegraphics[width=1\linewidth]{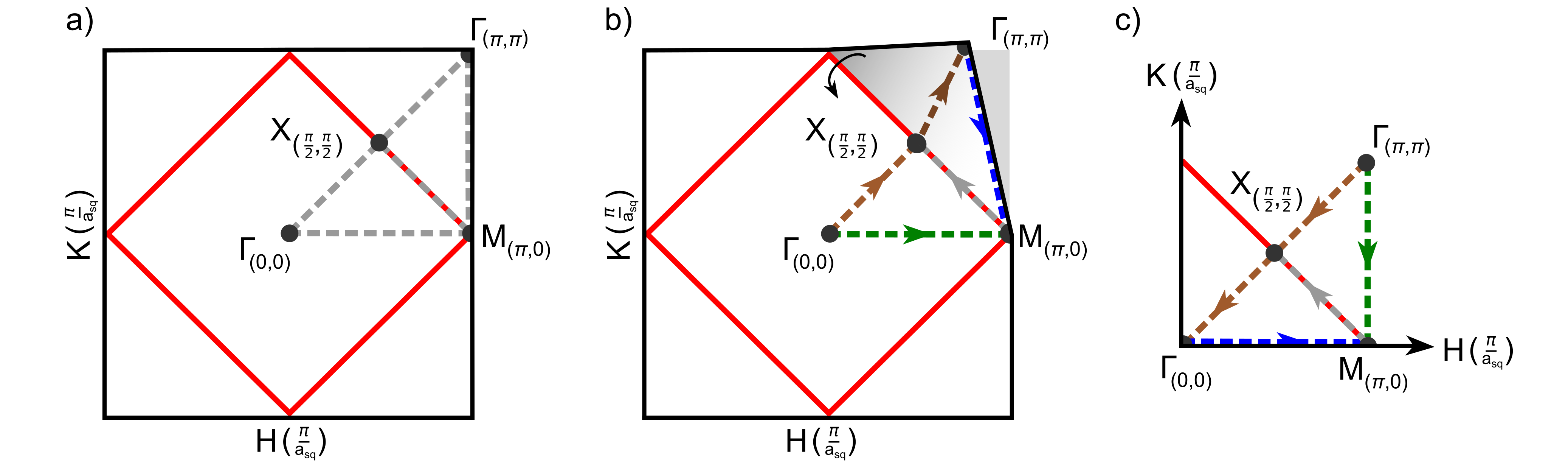}
	\caption{(a) Effective (second) Brillouin zone (b) backfolded onto the first Brillouin zone demonstrates that $\Gamma_{(\pi,\pi)} \rightarrow M_{(\pi,0)}$ (blue) in the second Brillouin zone is symmetrically equivalent to $\Gamma_{(0,0)} \rightarrow M_{(\pi,0)}$ (green) in the first Brillouin zone, $\Gamma_{(0,0)}\rightarrow\Gamma_{(\pi,\pi)}$ (brown) is reversed, while $M_{(\pi,0)} \rightarrow X_{\left(\frac{\pi}{2},\frac{\pi}{2}\right)}$ (gray) is unaffected. (c) New high symmetry directions between high symmetry points of the antiferromagnetic reciprocal unit cell (red) that has been symmetrized with respect to the square lattice (black) and whose zone center is now at $\Gamma_{(\pi,\pi)}$. Colors in (b) and (c) illustrate the relationship between specific dispersion relations for T and T$'$ modes.}
	\label{fig:fig5}
\end{figure*}


\indent We now shift the discussion to address the additional transverse $T'$ mode present in Fig. \ref{fig:fig4}.  Despite all the success of the spin-orbit exciton model to account for the experimental data presented so far (Figs.~\ref{fig:fig4}(a,b) and~\ref{fig:fig7}), constant-$\mathbf{Q}$ cuts (Fig.~\ref{fig:fig4}(c)) revealed that the combined intensities of the $L$ and $T$ modes alone could not account for all the scattering intensity that was observed experimentally.  Closer inspection of the unpolarized and polarized inelastic spectra previously measured on ARCS and PUMA,~\cite{jain13:17} respectively, revealed that the discrepancy in intensity was limited to transverse fluctuations and were extended throughout the Brillouin zone with energy transfers other than the $T$ and $L$ modes, including the presence of additional weak scattering present at the antiferromagnetic center $(\pi,\pi)$, confirming the need for an additional transverse mode, termed $T'$, that is absent in the spin-orbit exciton model. 

\indent The particular mode is weak in intensity, where its presence at both low energy and high energy transfers in Fig.~\ref{fig:fig7} was concealed by the dominant $T$ and $L$ modes, respectively. Possessing both an identical energy bandwidth and clear similarities to the dispersion relation exhibited by the $T$ mode throughout the entire Brillouin zone, the mode in question is consistent with one that originates from backfolding onto the first Brillouin zone, as originally proposed by Jain \emph{et al.}~\cite{jain13:17} Fig.~\ref{fig:fig5} illustrates that backfolding yields a mode whose dipsersion along $\Gamma_{(\pi,\pi)} \rightarrow M_{(\pi,0)}$ is replaced by $\Gamma_{(0,0)} \rightarrow M_{(\pi,0)}$ (and \emph{vice versa}), the dispersion along $\Gamma_{(0,0)}\rightarrow\Gamma_{(\pi,\pi)}$ is reversed, while the dispersion along $M_{(\pi,0)} \rightarrow X_{\left(\frac{\pi}{2},\frac{\pi}{2}\right)}$ is unaffected. Such a mode is illustrated in Fig.~\ref{fig:fig4}(b), demonstrating clear agreement with the dispersion relation reported in the experimental data, supporting the attribution of the $T'$ mode to backfolding. By considering the scattering intensity of the $T'$ mode as originating from a reciprocal unit cell that is identical to that illustrated in Fig.~\ref{fig:fig2}, but whose nuclear zone center coincides with $\Gamma_{(\pi,\pi)}$, the spin-orbit exciton model accounts for a significant portion of the intensity deficiency first identified in Fig.~\ref{fig:fig4}(c).   
 
\section{Discussion and Concluding Remarks} 

To summarize, by extending previous theoretical approaches,\cite{buyers11:75,sarte100:19} we have established the theoretical framework for a spin-orbit exciton model that accounts for the low energy magnetic excitations in the layered perovskite \caruo. In particular, the model successfully reproduces the longitudinal Higgs excitation. This mode has been theoretically predicted near quantum critical points in bilayer magnets,~\cite{Chubukov95:52} and has been experimentally identified in a plethora of systems spanning condensed matter physics including dimerized TlCuCl$_{3}$ near a pressure induced quantum critical point,~\cite{Ruegg08:100,merchant14:10} and in superconductors~\cite{Anderson15:1,Sherman15:11} including 2H-NbSe$_{2}$.~\cite{Measson14:89}  

In contrast to conventional pseudo-bosonic approaches that employ the Holstein-Primakoff transformation,\cite{majlis:book} the spin-orbit exciton approach presented here employs a minimalist Hamiltonian (Eq.~\ref{eq:5}), enabling not only the explicit, but also equal weighted incorporation of the individual contributions to the crystalline electric field with magnetic superexchange. As illustrated in Fig.~\ref{fig:fig8}, a clear advantage for such an approach is that it facilitates the quantization of the influences for each individual contribution. By understanding how each term in the Hamiltonian influences the excitation spectrum, the spin-orbit exciton model represents a tool that may be used to identify possible candidates possessing the appropriate conditions to host exotic excitations, such as the Higgs mode. 

In case of the longitudinally polarized Higgs mode, its existence is precluded when no orbital degree of freedom is present, such as is the case for $S=\frac{5}{2}$ in an undistorted octahedral weak ligand field. In this case, orthogonality ensures that matrix elements of the form $\langle m|J_{z}|0\rangle$ are zero, and thus by Eq.~\ref{eq:14} ensuring $g_{zz}$ is zero as well. In the case of \caruo, an orbital degree of freedom is introduced in the spin-orbit exciton model by spin-orbit coupling $\hat{\mathcal{H}}_{SO}$ (Eq.~\ref{eq:SO2}), enabling mixing of the crystalline electric field eigenstates. Such an effect can be rationalized by noting that the inclusion of a $\mathbf{l}\cdot\mathbf{S}$ term in the total magnetic Hamiltonian $\mathcal{H}$ results in a non-zero $[S_{z},\mathcal{H}]$, and by the Heisenberg equation-of-motion, $\langle S_{z} \rangle$ is not conserved, providing the possibility of a longitudinal mode to exist. Even in the case that $zz$ mode does exist, its dispersion is not necessarily unique compared to its transverse $+-$ and $-+$ counterparts, as is the case for an undistorted octahedral ligand field. As summarized in Fig.~\ref{fig:fig8}, it is the introduction of a uniaxial distortion $\hat{\mathcal{H}}_{dis}$ (Eq.~\ref{eq:distortion}) along $z$ that results in the separation of the $zz$ mode from the transverse modes, with the sign of $\Gamma$ determining the relative order in energy.  This term in the Hamiltonian also controls the magnetic anisotropy which has been suggested to be important for stabilizing the Higgs mode.~\cite{Ying20:102}

In addition to a structural distortion $\mathcal{H}_{dis}$, another contribution that has a particularly strong influence on the relative separation between the longitudinal and transverse branches is the molecular mean field. Having a disproportionately larger influence on the transverse $+-$ and $-+$ modes relative to their longitudinal counterpart, the value and sign of $\hat{\mathcal{H}}_{MF}$ (Eq.~\ref{eq:6_1}) determines the relative separation of the two transverse modes. An interesting result is that the least squares optimization of the spin-orbit exciton model yielded a zero molecular field $H_{MF}(i)$ for all sites $i$. This is consistent with the single-ion prediction of a $j\rm{_{eff}}$=0 ground state, but is not consistent with the presence of an ordered magnetic moment  characterized by a magnetic Bragg peak in the neutron scattering response.\cite{braden98:58,jain13:17,Pincini18:98} This discrepancy can be reconciled by noting that the value of molecular field was determined by fitting excitations with energies equal or greater than the distortion energy, and thus, at the energy transfers of interest, Ca$_{2}$RuO$_{4}$ could be approximated as a pure $j\rm{_{eff}}=0$ magnet.  One possible origin for the presence of an elastic Bragg peak in the neutron response despite a singlet ground state is the reported\cite{braden98:58} presence of a prominent distortion of the local octahedral coordination environment. However, other theoretical ideas have been proposed that may also be consistent with our refinement of the molecular field, including Hund's coupling~\cite{Svoboda17:95} and triplon condensation.~\cite{Chen11:84}  

In the discussion presented so far, key inferences concerning the properties of the magnetic excitation spectrum of \caruo~have been limited to the separate distinct influences for each of the spin-orbit exciton model's individual parameters. By considering a combination of these parameters, inferences with possibly large widespread applicability may be addressed, with one such example would be quantum criticality. A quantum critical point has been predicted to exist when the longitudinal response can be driven to zero energy~\cite{Rancon14:89} or when the longitudinal and transverse modes become degenerate.~\cite{oitmaa18:97} In such a situation the spectral weight in the longitudinal channel would be expected to diverge.  This can also be seen by applying the ``multi-magnon" formalism discussed in the case of classical two-dimensional magnets.~\cite{Huberman05:72} As illustrated in Fig.~\ref{fig:fig8}, the degeneracy of the longitudinal and transverse branches occurs in the absence of a distortion away from an ideal octahedra arrangement, while the shift of the $zz$ mode to lower energy transfers can be accomplished by an appropriate ratio of the exchange constants $J_{1}$, $J_{2}$ to the spin-orbit coupling constant $\lambda$. Possible mechanisms to achieve such conditions could involve strain or pressure, whose influence is not only restricted to the structural distortion of local coordination octahedra,\cite{ding06:74} but extends to the magnetic superexchange constants.\cite{blanco07:99,zhang06:74,rocquefelte12:2}    

In conclusion, we have established the theoretical framework for a spin-orbit exciton model where the use of a minimalist Hamiltonian enables for the direct and equal weighted incorporation of the individual contributions to the crystalline electric field with magnetic superexchange. Such an excitonic approach was then used to model and understand the magnetic excitations originating from the coupled spin and crystalline electric fields of a canonical Van Vleck $j\rm{_{eff}}=0$ ground state in \caruo. The anomalous longitudinally polarized Higgs mode,\cite{jain13:17} transverse spin-waves,\cite{Kunkem15:115} and orbital excitations\cite{das18:8} were successfully captured, and in good agreement with previously reported neutron and inelastic x-ray spectroscopic measurements. The framework established here illustrates how a crystalline electric field-induced singlet ground state can support coherent
longitudinal excitations, and transverse wavelike dynamics, while providing possible mechanisms for accessing a nearby quantum critical point.


\indent 

\section{Acknowledgements} 
We acknowledge useful conversations with W.J.L.~Buyers, R.A.~Cowley, H. Lane, K.J.~Camacho, C.~Schwenk, Y.~Wolde-Mariam, and A. Reyes. P.M.S. and B.R.O. acknowledge financial support from the University of California, Santa Barbara through the Elings Prize Fellowship. C.S. and K.H.H. would like to acknowledge the ERC, the EPSRC, the STFC, and the Carnegie Trust for the Universities of Scotland for financial support.  Finally, this material is based upon work supported by the National Science Foundation's Q-AMASE-i initiative under award DMR-1906325.


\begin{thebibliography}{102}%
\makeatletter
\providecommand \@ifxundefined [1]{%
 \@ifx{#1\undefined}
}%
\providecommand \@ifnum [1]{%
 \ifnum #1\expandafter \@firstoftwo
 \else \expandafter \@secondoftwo
 \fi
}%
\providecommand \@ifx [1]{%
 \ifx #1\expandafter \@firstoftwo
 \else \expandafter \@secondoftwo
 \fi
}%
\providecommand \natexlab [1]{#1}%
\providecommand \enquote  [1]{``#1''}%
\providecommand \bibnamefont  [1]{#1}%
\providecommand \bibfnamefont [1]{#1}%
\providecommand \citenamefont [1]{#1}%
\providecommand \href@noop [0]{\@secondoftwo}%
\providecommand \href [0]{\begingroup \@sanitize@url \@href}%
\providecommand \@href[1]{\@@startlink{#1}\@@href}%
\providecommand \@@href[1]{\endgroup#1\@@endlink}%
\providecommand \@sanitize@url [0]{\catcode `\\12\catcode `\$12\catcode
  `\&12\catcode `\#12\catcode `\^12\catcode `\_12\catcode `\%12\relax}%
\providecommand \@@startlink[1]{}%
\providecommand \@@endlink[0]{}%
\providecommand \url  [0]{\begingroup\@sanitize@url \@url }%
\providecommand \@url [1]{\endgroup\@href {#1}{\urlprefix }}%
\providecommand \urlprefix  [0]{URL }%
\providecommand \Eprint [0]{\href }%
\providecommand \doibase [0]{http://dx.doi.org/}%
\providecommand \selectlanguage [0]{\@gobble}%
\providecommand \bibinfo  [0]{\@secondoftwo}%
\providecommand \bibfield  [0]{\@secondoftwo}%
\providecommand \translation [1]{[#1]}%
\providecommand \BibitemOpen [0]{}%
\providecommand \bibitemStop [0]{}%
\providecommand \bibitemNoStop [0]{.\EOS\space}%
\providecommand \EOS [0]{\spacefactor3000\relax}%
\providecommand \BibitemShut  [1]{\csname bibitem#1\endcsname}%
\let\auto@bib@innerbib\@empty
\bibitem [{\citenamefont {Georges}\ \emph {et~al.}(2013)\citenamefont
  {Georges}, \citenamefont {de'Medici},\ and\ \citenamefont
  {Mravlje}}]{Georges17:4}%
  \BibitemOpen
  \bibfield  {author} {\bibinfo {author} {\bibfnamefont {A.}~\bibnamefont
  {Georges}}, \bibinfo {author} {\bibfnamefont {L.}~\bibnamefont {de'Medici}},
  \ and\ \bibinfo {author} {\bibfnamefont {J.}~\bibnamefont {Mravlje}},\ }\href
  {\doibase 10.1146/annurev-conmatphys-020911-125045} {\bibfield  {journal}
  {\bibinfo  {journal} {Annu. Rev. Condens. Matter Phys.}\ }\textbf {\bibinfo
  {volume} {4}},\ \bibinfo {pages} {137} (\bibinfo {year} {2013})}\BibitemShut
  {NoStop}%
\bibitem [{\citenamefont {Rau}\ \emph {et~al.}(2016)\citenamefont {Rau},
  \citenamefont {Lee},\ and\ \citenamefont {Kee}}]{Rau17:4}%
  \BibitemOpen
  \bibfield  {author} {\bibinfo {author} {\bibfnamefont {J.~G.}\ \bibnamefont
  {Rau}}, \bibinfo {author} {\bibfnamefont {E.}~\bibnamefont {Lee}}, \ and\
  \bibinfo {author} {\bibfnamefont {H.-Y.}\ \bibnamefont {Kee}},\ }\href
  {\doibase 10.1146/annurev-conmatphys-031115-011319} {\bibfield  {journal}
  {\bibinfo  {journal} {Annu. Rev. Condens. Matter Phys.}\ }\textbf {\bibinfo
  {volume} {7}},\ \bibinfo {pages} {195} (\bibinfo {year} {2016})}\BibitemShut
  {NoStop}%
\bibitem [{\citenamefont {Witczak-Krempa}\ \emph {et~al.}(2014)\citenamefont
  {Witczak-Krempa}, \citenamefont {Chen}, \citenamefont {Kim},\ and\
  \citenamefont {Balents}}]{Krempa14:5}%
  \BibitemOpen
  \bibfield  {author} {\bibinfo {author} {\bibfnamefont {W.}~\bibnamefont
  {Witczak-Krempa}}, \bibinfo {author} {\bibfnamefont {G.}~\bibnamefont
  {Chen}}, \bibinfo {author} {\bibfnamefont {Y.~B.}\ \bibnamefont {Kim}}, \
  and\ \bibinfo {author} {\bibfnamefont {L.}~\bibnamefont {Balents}},\ }\href
  {\doibase 10.1146/annurev-conmatphys-020911-125138} {\bibfield  {journal}
  {\bibinfo  {journal} {Annu. Rev. Condens. Matter Phys.}\ }\textbf {\bibinfo
  {volume} {5}},\ \bibinfo {pages} {57} (\bibinfo {year} {2014})}\BibitemShut
  {NoStop}%
\bibitem [{\citenamefont {Yin}\ \emph {et~al.}(2011)\citenamefont {Yin},
  \citenamefont {Haule},\ and\ \citenamefont {Kotliar}}]{Yin11:10}%
  \BibitemOpen
  \bibfield  {author} {\bibinfo {author} {\bibfnamefont {Z.~P.}\ \bibnamefont
  {Yin}}, \bibinfo {author} {\bibfnamefont {K.}~\bibnamefont {Haule}}, \ and\
  \bibinfo {author} {\bibfnamefont {G.}~\bibnamefont {Kotliar}},\ }\href
  {\doibase 10.1038/NMAT3120} {\bibfield  {journal} {\bibinfo  {journal} {Nat.
  Mater.}\ }\textbf {\bibinfo {volume} {10}},\ \bibinfo {pages} {932} (\bibinfo
  {year} {2011})}\BibitemShut {NoStop}%
\bibitem [{\citenamefont {Zhao}\ \emph {et~al.}(2019)\citenamefont {Zhao},
  \citenamefont {Hu}, \citenamefont {Ye}, \citenamefont {Hoffmann},
  \citenamefont {Kimchi},\ and\ \citenamefont {Cao}}]{Zhao19:100}%
  \BibitemOpen
  \bibfield  {author} {\bibinfo {author} {\bibfnamefont {H.}~\bibnamefont
  {Zhao}}, \bibinfo {author} {\bibfnamefont {B.}~\bibnamefont {Hu}}, \bibinfo
  {author} {\bibfnamefont {F.}~\bibnamefont {Ye}}, \bibinfo {author}
  {\bibfnamefont {C.}~\bibnamefont {Hoffmann}}, \bibinfo {author}
  {\bibfnamefont {I.}~\bibnamefont {Kimchi}}, \ and\ \bibinfo {author}
  {\bibfnamefont {G.}~\bibnamefont {Cao}},\ }\href {\doibase
  10.1103/PhysRevB.100.241104} {\bibfield  {journal} {\bibinfo  {journal}
  {Phys. Rev. B}\ }\textbf {\bibinfo {volume} {100}},\ \bibinfo {pages}
  {241104(R)} (\bibinfo {year} {2019})}\BibitemShut {NoStop}%
\bibitem [{\citenamefont {Cao}\ \emph {et~al.}(1997)\citenamefont {Cao},
  \citenamefont {McCall}, \citenamefont {Shepard}, \citenamefont {Crow},\ and\
  \citenamefont {Guertin}}]{Cao97:56}%
  \BibitemOpen
  \bibfield  {author} {\bibinfo {author} {\bibfnamefont {G.}~\bibnamefont
  {Cao}}, \bibinfo {author} {\bibfnamefont {S.}~\bibnamefont {McCall}},
  \bibinfo {author} {\bibfnamefont {M.}~\bibnamefont {Shepard}}, \bibinfo
  {author} {\bibfnamefont {J.~E.}\ \bibnamefont {Crow}}, \ and\ \bibinfo
  {author} {\bibfnamefont {R.~P.}\ \bibnamefont {Guertin}},\ }\href {\doibase
  10.1103/PhysRevB.56.R2916} {\bibfield  {journal} {\bibinfo  {journal} {Phys.
  Rev. B}\ }\textbf {\bibinfo {volume} {56}},\ \bibinfo {pages} {R2916}
  (\bibinfo {year} {1997})}\BibitemShut {NoStop}%
\bibitem [{\citenamefont {Nakatsuji}\ \emph
  {et~al.}(1997{\natexlab{a}})\citenamefont {Nakatsuji}, \citenamefont
  {Ikeda},\ and\ \citenamefont {Maeno}}]{nataksuji97:66}%
  \BibitemOpen
  \bibfield  {author} {\bibinfo {author} {\bibfnamefont {S.}~\bibnamefont
  {Nakatsuji}}, \bibinfo {author} {\bibfnamefont {S.~I.}\ \bibnamefont
  {Ikeda}}, \ and\ \bibinfo {author} {\bibfnamefont {Y.}~\bibnamefont
  {Maeno}},\ }\href {\doibase https://doi.org/10.1143/JPSJ.66.1868} {\bibfield
  {journal} {\bibinfo  {journal} {J. Phys. Soc. Jpn.}\ }\textbf {\bibinfo
  {volume} {66}},\ \bibinfo {pages} {1868} (\bibinfo {year}
  {1997}{\natexlab{a}})}\BibitemShut {NoStop}%
\bibitem [{\citenamefont {Nakatsuji}\ and\ \citenamefont
  {Maeno}(2000{\natexlab{a}})}]{nakatsuji00:84}%
  \BibitemOpen
  \bibfield  {author} {\bibinfo {author} {\bibfnamefont {S.}~\bibnamefont
  {Nakatsuji}}\ and\ \bibinfo {author} {\bibfnamefont {Y.}~\bibnamefont
  {Maeno}},\ }\href {\doibase 10.1103/PhysRevLett.84.2666} {\bibfield
  {journal} {\bibinfo  {journal} {Phys. Rev. Lett.}\ }\textbf {\bibinfo
  {volume} {84}},\ \bibinfo {pages} {2666} (\bibinfo {year}
  {2000}{\natexlab{a}})}\BibitemShut {NoStop}%
\bibitem [{\citenamefont {Nakatsuji}\ and\ \citenamefont
  {Maeno}(2000{\natexlab{b}})}]{nakatsuji00:62}%
  \BibitemOpen
  \bibfield  {author} {\bibinfo {author} {\bibfnamefont {S.}~\bibnamefont
  {Nakatsuji}}\ and\ \bibinfo {author} {\bibfnamefont {Y.}~\bibnamefont
  {Maeno}},\ }\href {\doibase 10.1103/PhysRevB.62.6458} {\bibfield  {journal}
  {\bibinfo  {journal} {Phys. Rev. B}\ }\textbf {\bibinfo {volume} {62}},\
  \bibinfo {pages} {6458} (\bibinfo {year} {2000}{\natexlab{b}})}\BibitemShut
  {NoStop}%
\bibitem [{\citenamefont {Nakatsuji}\ \emph {et~al.}(2003)\citenamefont
  {Nakatsuji}, \citenamefont {Hall}, \citenamefont {Balicas}, \citenamefont
  {Fisk}, \citenamefont {Sugahara}, \citenamefont {Yoshioka},\ and\
  \citenamefont {Maeno}}]{nataksuji03:90}%
  \BibitemOpen
  \bibfield  {author} {\bibinfo {author} {\bibfnamefont {S.}~\bibnamefont
  {Nakatsuji}}, \bibinfo {author} {\bibfnamefont {D.}~\bibnamefont {Hall}},
  \bibinfo {author} {\bibfnamefont {L.}~\bibnamefont {Balicas}}, \bibinfo
  {author} {\bibfnamefont {Z.}~\bibnamefont {Fisk}}, \bibinfo {author}
  {\bibfnamefont {K.}~\bibnamefont {Sugahara}}, \bibinfo {author}
  {\bibfnamefont {M.}~\bibnamefont {Yoshioka}}, \ and\ \bibinfo {author}
  {\bibfnamefont {Y.}~\bibnamefont {Maeno}},\ }\href {\doibase
  10.1103/PhysRevLett.90.137202} {\bibfield  {journal} {\bibinfo  {journal}
  {Phys. Rev. Lett.}\ }\textbf {\bibinfo {volume} {90}},\ \bibinfo {pages}
  {137202} (\bibinfo {year} {2003})}\BibitemShut {NoStop}%
\bibitem [{\citenamefont {Anisimov}\ \emph {et~al.}(2002)\citenamefont
  {Anisimov}, \citenamefont {Nekrasov}, \citenamefont {Kondakov}, \citenamefont
  {Rice},\ and\ \citenamefont {Sigrist}}]{Anisimov02:25}%
  \BibitemOpen
  \bibfield  {author} {\bibinfo {author} {\bibfnamefont {V.~I.}\ \bibnamefont
  {Anisimov}}, \bibinfo {author} {\bibfnamefont {I.~A.}\ \bibnamefont
  {Nekrasov}}, \bibinfo {author} {\bibfnamefont {D.~E.}\ \bibnamefont
  {Kondakov}}, \bibinfo {author} {\bibfnamefont {T.~M.}\ \bibnamefont {Rice}},
  \ and\ \bibinfo {author} {\bibfnamefont {M.}~\bibnamefont {Sigrist}},\ }\href
  {\doibase 10.1140/epjb/e20020021} {\bibfield  {journal} {\bibinfo  {journal}
  {Eur. Phys. J. B}\ }\textbf {\bibinfo {volume} {25}},\ \bibinfo {pages} {191}
  (\bibinfo {year} {2002})}\BibitemShut {NoStop}%
\bibitem [{\citenamefont {Nakamura}\ \emph {et~al.}(2002)\citenamefont
  {Nakamura}, \citenamefont {Goko}, \citenamefont {Ito}, \citenamefont
  {Fujita}, \citenamefont {Nakatsuji}, \citenamefont {Fukazawa}, \citenamefont
  {Maeno}, \citenamefont {Alireza}, \citenamefont {Forsythe},\ and\
  \citenamefont {Julian}}]{Nakamura02:65}%
  \BibitemOpen
  \bibfield  {author} {\bibinfo {author} {\bibfnamefont {F.}~\bibnamefont
  {Nakamura}}, \bibinfo {author} {\bibfnamefont {T.}~\bibnamefont {Goko}},
  \bibinfo {author} {\bibfnamefont {M.}~\bibnamefont {Ito}}, \bibinfo {author}
  {\bibfnamefont {T.}~\bibnamefont {Fujita}}, \bibinfo {author} {\bibfnamefont
  {S.}~\bibnamefont {Nakatsuji}}, \bibinfo {author} {\bibfnamefont
  {H.}~\bibnamefont {Fukazawa}}, \bibinfo {author} {\bibfnamefont
  {Y.}~\bibnamefont {Maeno}}, \bibinfo {author} {\bibfnamefont
  {P.}~\bibnamefont {Alireza}}, \bibinfo {author} {\bibfnamefont
  {D.}~\bibnamefont {Forsythe}}, \ and\ \bibinfo {author} {\bibfnamefont
  {S.~R.}\ \bibnamefont {Julian}},\ }\href {\doibase
  10.1103/PhysRevB.65.220402} {\bibfield  {journal} {\bibinfo  {journal} {Phys.
  Rev. B}\ }\textbf {\bibinfo {volume} {65}},\ \bibinfo {pages} {220402(R)}
  (\bibinfo {year} {2002})}\BibitemShut {NoStop}%
\bibitem [{\citenamefont {Porter}\ \emph {et~al.}(2018)\citenamefont {Porter},
  \citenamefont {Granata}, \citenamefont {Forte}, \citenamefont {Di~Matteo},
  \citenamefont {Cuoco}, \citenamefont {Fittipaldi}, \citenamefont
  {Vecchione},\ and\ \citenamefont {Bombardi}}]{Porter18:98}%
  \BibitemOpen
  \bibfield  {author} {\bibinfo {author} {\bibfnamefont {D.~G.}\ \bibnamefont
  {Porter}}, \bibinfo {author} {\bibfnamefont {V.}~\bibnamefont {Granata}},
  \bibinfo {author} {\bibfnamefont {F.}~\bibnamefont {Forte}}, \bibinfo
  {author} {\bibfnamefont {S.}~\bibnamefont {Di~Matteo}}, \bibinfo {author}
  {\bibfnamefont {M.}~\bibnamefont {Cuoco}}, \bibinfo {author} {\bibfnamefont
  {R.}~\bibnamefont {Fittipaldi}}, \bibinfo {author} {\bibfnamefont
  {A.}~\bibnamefont {Vecchione}}, \ and\ \bibinfo {author} {\bibfnamefont
  {A.}~\bibnamefont {Bombardi}},\ }\href {\doibase 10.1103/PhysRevB.98.125142}
  {\bibfield  {journal} {\bibinfo  {journal} {Phys. Rev. B}\ }\textbf {\bibinfo
  {volume} {98}},\ \bibinfo {pages} {125142} (\bibinfo {year}
  {2018})}\BibitemShut {NoStop}%
\bibitem [{\citenamefont {Braden}\ \emph {et~al.}(1998)\citenamefont {Braden},
  \citenamefont {Andr\'e}, \citenamefont {Nakatsuji},\ and\ \citenamefont
  {Maeno}}]{braden98:58}%
  \BibitemOpen
  \bibfield  {author} {\bibinfo {author} {\bibfnamefont {M.}~\bibnamefont
  {Braden}}, \bibinfo {author} {\bibfnamefont {G.}~\bibnamefont {Andr\'e}},
  \bibinfo {author} {\bibfnamefont {S.}~\bibnamefont {Nakatsuji}}, \ and\
  \bibinfo {author} {\bibfnamefont {Y.}~\bibnamefont {Maeno}},\ }\href
  {\doibase 10.1103/PhysRevB.58.847} {\bibfield  {journal} {\bibinfo  {journal}
  {Phys. Rev. B}\ }\textbf {\bibinfo {volume} {58}},\ \bibinfo {pages} {847}
  (\bibinfo {year} {1998})}\BibitemShut {NoStop}%
\bibitem [{\citenamefont {Steffens}\ \emph {et~al.}(2005)\citenamefont
  {Steffens}, \citenamefont {Friedt}, \citenamefont {Alireza}, \citenamefont
  {Marshall}, \citenamefont {Schmidt}, \citenamefont {Nakamura}, \citenamefont
  {Nakatsuji}, \citenamefont {Maeno}, \citenamefont {Lengsdorf}, \citenamefont
  {Abd-Elmeguid},\ and\ \citenamefont {Braden}}]{Steffens05:72}%
  \BibitemOpen
  \bibfield  {author} {\bibinfo {author} {\bibfnamefont {P.}~\bibnamefont
  {Steffens}}, \bibinfo {author} {\bibfnamefont {O.}~\bibnamefont {Friedt}},
  \bibinfo {author} {\bibfnamefont {P.}~\bibnamefont {Alireza}}, \bibinfo
  {author} {\bibfnamefont {W.~G.}\ \bibnamefont {Marshall}}, \bibinfo {author}
  {\bibfnamefont {W.}~\bibnamefont {Schmidt}}, \bibinfo {author} {\bibfnamefont
  {F.}~\bibnamefont {Nakamura}}, \bibinfo {author} {\bibfnamefont
  {S.}~\bibnamefont {Nakatsuji}}, \bibinfo {author} {\bibfnamefont
  {Y.}~\bibnamefont {Maeno}}, \bibinfo {author} {\bibfnamefont
  {R.}~\bibnamefont {Lengsdorf}}, \bibinfo {author} {\bibfnamefont {M.~M.}\
  \bibnamefont {Abd-Elmeguid}}, \ and\ \bibinfo {author} {\bibfnamefont
  {M.}~\bibnamefont {Braden}},\ }\href {\doibase 10.1103/PhysRevB.72.094104}
  {\bibfield  {journal} {\bibinfo  {journal} {Phys. Rev. B}\ }\textbf {\bibinfo
  {volume} {72}},\ \bibinfo {pages} {094104} (\bibinfo {year}
  {2005})}\BibitemShut {NoStop}%
\bibitem [{\citenamefont {Pincini}\ \emph {et~al.}(2018)\citenamefont
  {Pincini}, \citenamefont {Boseggia}, \citenamefont {Perry}, \citenamefont
  {Gutmann}, \citenamefont {Ricc\`o}, \citenamefont {Veiga}, \citenamefont
  {Dashwood}, \citenamefont {Collins}, \citenamefont {Nisbet}, \citenamefont
  {Bombardi}, \citenamefont {Porter}, \citenamefont {Baumberger}, \citenamefont
  {Boothroyd},\ and\ \citenamefont {McMorrow}}]{Pincini18:98}%
  \BibitemOpen
  \bibfield  {author} {\bibinfo {author} {\bibfnamefont {D.}~\bibnamefont
  {Pincini}}, \bibinfo {author} {\bibfnamefont {S.}~\bibnamefont {Boseggia}},
  \bibinfo {author} {\bibfnamefont {R.}~\bibnamefont {Perry}}, \bibinfo
  {author} {\bibfnamefont {M.~J.}\ \bibnamefont {Gutmann}}, \bibinfo {author}
  {\bibfnamefont {S.}~\bibnamefont {Ricc\`o}}, \bibinfo {author} {\bibfnamefont
  {L.~S.~I.}\ \bibnamefont {Veiga}}, \bibinfo {author} {\bibfnamefont {C.~D.}\
  \bibnamefont {Dashwood}}, \bibinfo {author} {\bibfnamefont {S.~P.}\
  \bibnamefont {Collins}}, \bibinfo {author} {\bibfnamefont {G.}~\bibnamefont
  {Nisbet}}, \bibinfo {author} {\bibfnamefont {A.}~\bibnamefont {Bombardi}},
  \bibinfo {author} {\bibfnamefont {D.~G.}\ \bibnamefont {Porter}}, \bibinfo
  {author} {\bibfnamefont {F.}~\bibnamefont {Baumberger}}, \bibinfo {author}
  {\bibfnamefont {A.~T.}\ \bibnamefont {Boothroyd}}, \ and\ \bibinfo {author}
  {\bibfnamefont {D.~F.}\ \bibnamefont {McMorrow}},\ }\href {\doibase
  10.1103/PhysRevB.98.014429} {\bibfield  {journal} {\bibinfo  {journal} {Phys.
  Rev. B}\ }\textbf {\bibinfo {volume} {98}},\ \bibinfo {pages} {014429}
  (\bibinfo {year} {2018})}\BibitemShut {NoStop}%
\bibitem [{\citenamefont {Friedt}\ \emph {et~al.}(2001)\citenamefont {Friedt},
  \citenamefont {Braden}, \citenamefont {Andr\'e}, \citenamefont {Adelmann},
  \citenamefont {Nakatsuji},\ and\ \citenamefont {Maeno}}]{Friedt01:63}%
  \BibitemOpen
  \bibfield  {author} {\bibinfo {author} {\bibfnamefont {O.}~\bibnamefont
  {Friedt}}, \bibinfo {author} {\bibfnamefont {M.}~\bibnamefont {Braden}},
  \bibinfo {author} {\bibfnamefont {G.}~\bibnamefont {Andr\'e}}, \bibinfo
  {author} {\bibfnamefont {P.}~\bibnamefont {Adelmann}}, \bibinfo {author}
  {\bibfnamefont {S.}~\bibnamefont {Nakatsuji}}, \ and\ \bibinfo {author}
  {\bibfnamefont {Y.}~\bibnamefont {Maeno}},\ }\href {\doibase
  10.1103/PhysRevB.63.174432} {\bibfield  {journal} {\bibinfo  {journal} {Phys.
  Rev. B}\ }\textbf {\bibinfo {volume} {63}},\ \bibinfo {pages} {174432}
  (\bibinfo {year} {2001})}\BibitemShut {NoStop}%
\bibitem [{\citenamefont {Feldmaier}\ \emph {et~al.}(2020)\citenamefont
  {Feldmaier}, \citenamefont {Strobel}, \citenamefont {Schmid}, \citenamefont
  {Hansmann},\ and\ \citenamefont {Daghofer}}]{Feldmaier19:xx}%
  \BibitemOpen
  \bibfield  {author} {\bibinfo {author} {\bibfnamefont {T.}~\bibnamefont
  {Feldmaier}}, \bibinfo {author} {\bibfnamefont {P.}~\bibnamefont {Strobel}},
  \bibinfo {author} {\bibfnamefont {M.}~\bibnamefont {Schmid}}, \bibinfo
  {author} {\bibfnamefont {P.}~\bibnamefont {Hansmann}}, \ and\ \bibinfo
  {author} {\bibfnamefont {M.}~\bibnamefont {Daghofer}},\ }\href {\doibase
  10.1103/PhysRevResearch.2.033201} {\bibfield  {journal} {\bibinfo  {journal}
  {Phys. Rev. Research}\ }\textbf {\bibinfo {volume} {2}},\ \bibinfo {pages}
  {033201} (\bibinfo {year} {2020})}\BibitemShut {NoStop}%
\bibitem [{\citenamefont {Khaliullin}(2013)}]{Khaliullin13:111}%
  \BibitemOpen
  \bibfield  {author} {\bibinfo {author} {\bibfnamefont {G.}~\bibnamefont
  {Khaliullin}},\ }\href {\doibase 10.1103/PhysRevLett.111.197201} {\bibfield
  {journal} {\bibinfo  {journal} {Phys. Rev. Lett.}\ }\textbf {\bibinfo
  {volume} {111}},\ \bibinfo {pages} {197201} (\bibinfo {year}
  {2013})}\BibitemShut {NoStop}%
\bibitem [{\citenamefont {P\'asztorov\'a}\ \emph {et~al.}(2019)\citenamefont
  {P\'asztorov\'a}, \citenamefont {Howell}, \citenamefont {Songvilay},
  \citenamefont {Sarte}, \citenamefont {Rodriguez-Rivera}, \citenamefont
  {Ar\'evalo-L\'opez}, \citenamefont {Schmalzl}, \citenamefont {Schneidewind},
  \citenamefont {Dunsiger}, \citenamefont {Singh}, \citenamefont {Petrovic},
  \citenamefont {Hu},\ and\ \citenamefont {Stock}}]{Pasztorova19:99}%
  \BibitemOpen
  \bibfield  {author} {\bibinfo {author} {\bibfnamefont {J.}~\bibnamefont
  {P\'asztorov\'a}}, \bibinfo {author} {\bibfnamefont {A.}~\bibnamefont
  {Howell}}, \bibinfo {author} {\bibfnamefont {M.}~\bibnamefont {Songvilay}},
  \bibinfo {author} {\bibfnamefont {P.~M.}\ \bibnamefont {Sarte}}, \bibinfo
  {author} {\bibfnamefont {J.~A.}\ \bibnamefont {Rodriguez-Rivera}}, \bibinfo
  {author} {\bibfnamefont {A.~M.}\ \bibnamefont {Ar\'evalo-L\'opez}}, \bibinfo
  {author} {\bibfnamefont {K.}~\bibnamefont {Schmalzl}}, \bibinfo {author}
  {\bibfnamefont {A.}~\bibnamefont {Schneidewind}}, \bibinfo {author}
  {\bibfnamefont {S.~R.}\ \bibnamefont {Dunsiger}}, \bibinfo {author}
  {\bibfnamefont {D.~K.}\ \bibnamefont {Singh}}, \bibinfo {author}
  {\bibfnamefont {C.}~\bibnamefont {Petrovic}}, \bibinfo {author}
  {\bibfnamefont {R.}~\bibnamefont {Hu}}, \ and\ \bibinfo {author}
  {\bibfnamefont {C.}~\bibnamefont {Stock}},\ }\href {\doibase
  10.1103/PhysRevB.99.125144} {\bibfield  {journal} {\bibinfo  {journal} {Phys.
  Rev. B}\ }\textbf {\bibinfo {volume} {99}},\ \bibinfo {pages} {125144}
  (\bibinfo {year} {2019})}\BibitemShut {NoStop}%
\bibitem [{\citenamefont {Stock}\ \emph {et~al.}(2012)\citenamefont {Stock},
  \citenamefont {Broholm}, \citenamefont {Demmel}, \citenamefont {Van~Duijn},
  \citenamefont {Taylor}, \citenamefont {Kang}, \citenamefont {Hu},\ and\
  \citenamefont {Petrovic}}]{Stock12:109}%
  \BibitemOpen
  \bibfield  {author} {\bibinfo {author} {\bibfnamefont {C.}~\bibnamefont
  {Stock}}, \bibinfo {author} {\bibfnamefont {C.}~\bibnamefont {Broholm}},
  \bibinfo {author} {\bibfnamefont {F.}~\bibnamefont {Demmel}}, \bibinfo
  {author} {\bibfnamefont {J.}~\bibnamefont {Van~Duijn}}, \bibinfo {author}
  {\bibfnamefont {J.~W.}\ \bibnamefont {Taylor}}, \bibinfo {author}
  {\bibfnamefont {H.~J.}\ \bibnamefont {Kang}}, \bibinfo {author}
  {\bibfnamefont {R.}~\bibnamefont {Hu}}, \ and\ \bibinfo {author}
  {\bibfnamefont {C.}~\bibnamefont {Petrovic}},\ }\href {\doibase
  10.1103/PhysRevLett.109.127201} {\bibfield  {journal} {\bibinfo  {journal}
  {Phys. Rev. Lett.}\ }\textbf {\bibinfo {volume} {109}},\ \bibinfo {pages}
  {127201} (\bibinfo {year} {2012})}\BibitemShut {NoStop}%
\bibitem [{\citenamefont {Gretarsson}\ \emph {et~al.}(2019)\citenamefont
  {Gretarsson}, \citenamefont {Suzuki}, \citenamefont {Kim}, \citenamefont
  {Ueda}, \citenamefont {Krautloher}, \citenamefont {Kim}, \citenamefont
  {Yavas}, \citenamefont {Khaliullin},\ and\ \citenamefont
  {Keimer}}]{Gretarsson19:100}%
  \BibitemOpen
  \bibfield  {author} {\bibinfo {author} {\bibfnamefont {H.}~\bibnamefont
  {Gretarsson}}, \bibinfo {author} {\bibfnamefont {H.}~\bibnamefont {Suzuki}},
  \bibinfo {author} {\bibfnamefont {H.}~\bibnamefont {Kim}}, \bibinfo {author}
  {\bibfnamefont {K.}~\bibnamefont {Ueda}}, \bibinfo {author} {\bibfnamefont
  {M.}~\bibnamefont {Krautloher}}, \bibinfo {author} {\bibfnamefont {B.~J.}\
  \bibnamefont {Kim}}, \bibinfo {author} {\bibfnamefont {H.}~\bibnamefont
  {Yavas}}, \bibinfo {author} {\bibfnamefont {G.}~\bibnamefont {Khaliullin}}, \
  and\ \bibinfo {author} {\bibfnamefont {B.}~\bibnamefont {Keimer}},\ }\href
  {\doibase 10.1103/PhysRevB.100.045123} {\bibfield  {journal} {\bibinfo
  {journal} {Phys. Rev. B}\ }\textbf {\bibinfo {volume} {100}},\ \bibinfo
  {pages} {045123} (\bibinfo {year} {2019})}\BibitemShut {NoStop}%
\bibitem [{\citenamefont {Wang}\ and\ \citenamefont
  {Cooper}(1968)}]{Wang68:172}%
  \BibitemOpen
  \bibfield  {author} {\bibinfo {author} {\bibfnamefont {Y.-L.}\ \bibnamefont
  {Wang}}\ and\ \bibinfo {author} {\bibfnamefont {B.~R.}\ \bibnamefont
  {Cooper}},\ }\href {\doibase 10.1103/PhysRev.172.539} {\bibfield  {journal}
  {\bibinfo  {journal} {Phys. Rev.}\ }\textbf {\bibinfo {volume} {172}},\
  \bibinfo {pages} {539} (\bibinfo {year} {1968})}\BibitemShut {NoStop}%
\bibitem [{\citenamefont {Birgeneau}\ \emph {et~al.}(1971)\citenamefont
  {Birgeneau}, \citenamefont {Als-Nielsen},\ and\ \citenamefont
  {Bucher}}]{Birgeneau71:27}%
  \BibitemOpen
  \bibfield  {author} {\bibinfo {author} {\bibfnamefont {R.~J.}\ \bibnamefont
  {Birgeneau}}, \bibinfo {author} {\bibfnamefont {J.}~\bibnamefont
  {Als-Nielsen}}, \ and\ \bibinfo {author} {\bibfnamefont {E.}~\bibnamefont
  {Bucher}},\ }\href {\doibase 10.1103/PhysRevLett.27.1530} {\bibfield
  {journal} {\bibinfo  {journal} {Phys. Rev. Lett.}\ }\textbf {\bibinfo
  {volume} {27}},\ \bibinfo {pages} {1530} (\bibinfo {year}
  {1971})}\BibitemShut {NoStop}%
\bibitem [{\citenamefont {Holden}\ and\ \citenamefont
  {Buyers}(1974)}]{Holden74:9}%
  \BibitemOpen
  \bibfield  {author} {\bibinfo {author} {\bibfnamefont {T.~M.}\ \bibnamefont
  {Holden}}\ and\ \bibinfo {author} {\bibfnamefont {W.~J.~L.}\ \bibnamefont
  {Buyers}},\ }\href {\doibase 10.1103/PhysRevB.9.3797} {\bibfield  {journal}
  {\bibinfo  {journal} {Phys. Rev. B}\ }\textbf {\bibinfo {volume} {9}},\
  \bibinfo {pages} {3797} (\bibinfo {year} {1974})}\BibitemShut {NoStop}%
\bibitem [{\citenamefont {Hsieh}\ and\ \citenamefont
  {Blume}(1972)}]{hsieh72:6}%
  \BibitemOpen
  \bibfield  {author} {\bibinfo {author} {\bibfnamefont {Y.~Y.}\ \bibnamefont
  {Hsieh}}\ and\ \bibinfo {author} {\bibfnamefont {M.}~\bibnamefont {Blume}},\
  }\href {\doibase 10.1103/PhysRevB.6.2684} {\bibfield  {journal} {\bibinfo
  {journal} {Phys. Rev. B}\ }\textbf {\bibinfo {volume} {6}},\ \bibinfo {pages}
  {2684} (\bibinfo {year} {1972})}\BibitemShut {NoStop}%
\bibitem [{\citenamefont {Birgeneau}\ \emph {et~al.}(1972)\citenamefont
  {Birgeneau}, \citenamefont {Als-Nielsen},\ and\ \citenamefont
  {Bucher}}]{Birgeneau72:6}%
  \BibitemOpen
  \bibfield  {author} {\bibinfo {author} {\bibfnamefont {R.~J.}\ \bibnamefont
  {Birgeneau}}, \bibinfo {author} {\bibfnamefont {J.}~\bibnamefont
  {Als-Nielsen}}, \ and\ \bibinfo {author} {\bibfnamefont {E.}~\bibnamefont
  {Bucher}},\ }\href {\doibase 10.1103/PhysRevB.6.2724} {\bibfield  {journal}
  {\bibinfo  {journal} {Phys. Rev. B}\ }\textbf {\bibinfo {volume} {6}},\
  \bibinfo {pages} {2724} (\bibinfo {year} {1972})}\BibitemShut {NoStop}%
\bibitem [{\citenamefont {Cooper}\ and\ \citenamefont
  {Vogt}(1971)}]{Cooper71:59}%
  \BibitemOpen
  \bibfield  {author} {\bibinfo {author} {\bibfnamefont {B.}~\bibnamefont
  {Cooper}}\ and\ \bibinfo {author} {\bibfnamefont {O.}~\bibnamefont {Vogt}},\
  }\href {\doibase 10.1051/jphyscol:19711343} {\bibfield  {journal} {\bibinfo
  {journal} {J. Phys. Colloq.}\ }\textbf {\bibinfo {volume} {32}},\ \bibinfo
  {pages} {C1} (\bibinfo {year} {1971})}\BibitemShut {NoStop}%
\bibitem [{\citenamefont {Pink}(1968)}]{pink68:1}%
  \BibitemOpen
  \bibfield  {author} {\bibinfo {author} {\bibfnamefont {D.~A.}\ \bibnamefont
  {Pink}},\ }\href {\doibase 0.1088/0022-3719/1/5/313} {\bibfield  {journal}
  {\bibinfo  {journal} {J. Phys. C: Solid State Phys.}\ }\textbf {\bibinfo
  {volume} {1}},\ \bibinfo {pages} {1246} (\bibinfo {year} {1968})}\BibitemShut
  {NoStop}%
\bibitem [{\citenamefont {Fang}\ \emph {et~al.}(2004)\citenamefont {Fang},
  \citenamefont {Nagaosa},\ and\ \citenamefont {Terakura}}]{Fang04:69}%
  \BibitemOpen
  \bibfield  {author} {\bibinfo {author} {\bibfnamefont {Z.}~\bibnamefont
  {Fang}}, \bibinfo {author} {\bibfnamefont {N.}~\bibnamefont {Nagaosa}}, \
  and\ \bibinfo {author} {\bibfnamefont {K.}~\bibnamefont {Terakura}},\ }\href
  {\doibase 10.1103/PhysRevB.69.045116} {\bibfield  {journal} {\bibinfo
  {journal} {Phys. Rev. B}\ }\textbf {\bibinfo {volume} {69}},\ \bibinfo
  {pages} {045116} (\bibinfo {year} {2004})}\BibitemShut {NoStop}%
\bibitem [{\citenamefont {Kunkem\"oller}\ \emph {et~al.}(2015)\citenamefont
  {Kunkem\"oller}, \citenamefont {Khomskii}, \citenamefont {Steffens},
  \citenamefont {Piovano}, \citenamefont {Nugroho},\ and\ \citenamefont
  {Braden}}]{Kunkem15:115}%
  \BibitemOpen
  \bibfield  {author} {\bibinfo {author} {\bibfnamefont {S.}~\bibnamefont
  {Kunkem\"oller}}, \bibinfo {author} {\bibfnamefont {D.}~\bibnamefont
  {Khomskii}}, \bibinfo {author} {\bibfnamefont {P.}~\bibnamefont {Steffens}},
  \bibinfo {author} {\bibfnamefont {A.}~\bibnamefont {Piovano}}, \bibinfo
  {author} {\bibfnamefont {A.~A.}\ \bibnamefont {Nugroho}}, \ and\ \bibinfo
  {author} {\bibfnamefont {M.}~\bibnamefont {Braden}},\ }\href {\doibase
  10.1103/PhysRevLett.115.247201} {\bibfield  {journal} {\bibinfo  {journal}
  {Phys. Rev. Lett.}\ }\textbf {\bibinfo {volume} {115}},\ \bibinfo {pages}
  {247201} (\bibinfo {year} {2015})}\BibitemShut {NoStop}%
\bibitem [{\citenamefont {Jain}\ \emph {et~al.}(2017)\citenamefont {Jain},
  \citenamefont {Krautloher}, \citenamefont {Porras}, \citenamefont {Ryu},
  \citenamefont {Chen}, \citenamefont {Abernathy}, \citenamefont {Park},
  \citenamefont {Ivanov}, \citenamefont {Chaloupka}, \citenamefont
  {Khaliullin}, \citenamefont {Keimer},\ and\ \citenamefont {Kim}}]{jain13:17}%
  \BibitemOpen
  \bibfield  {author} {\bibinfo {author} {\bibfnamefont {A.}~\bibnamefont
  {Jain}}, \bibinfo {author} {\bibfnamefont {M.}~\bibnamefont {Krautloher}},
  \bibinfo {author} {\bibfnamefont {J.}~\bibnamefont {Porras}}, \bibinfo
  {author} {\bibfnamefont {G.~H.}\ \bibnamefont {Ryu}}, \bibinfo {author}
  {\bibfnamefont {D.~P.}\ \bibnamefont {Chen}}, \bibinfo {author}
  {\bibfnamefont {D.~L.}\ \bibnamefont {Abernathy}}, \bibinfo {author}
  {\bibfnamefont {J.~T.}\ \bibnamefont {Park}}, \bibinfo {author}
  {\bibfnamefont {A.}~\bibnamefont {Ivanov}}, \bibinfo {author} {\bibfnamefont
  {J.}~\bibnamefont {Chaloupka}}, \bibinfo {author} {\bibfnamefont
  {G.}~\bibnamefont {Khaliullin}}, \bibinfo {author} {\bibfnamefont
  {B.}~\bibnamefont {Keimer}}, \ and\ \bibinfo {author} {\bibfnamefont {B.~J.}\
  \bibnamefont {Kim}},\ }\href {\doibase https://doi.org/10.1038/nphys4077}
  {\bibfield  {journal} {\bibinfo  {journal} {Nat. Phys.}\ }\textbf {\bibinfo
  {volume} {13}},\ \bibinfo {pages} {633} (\bibinfo {year} {2017})}\BibitemShut
  {NoStop}%
\bibitem [{\citenamefont {Peschel}\ \emph {et~al.}(1972)\citenamefont
  {Peschel}, \citenamefont {Klenin},\ and\ \citenamefont
  {Fulde}}]{Peschel71:5}%
  \BibitemOpen
  \bibfield  {author} {\bibinfo {author} {\bibfnamefont {I.}~\bibnamefont
  {Peschel}}, \bibinfo {author} {\bibfnamefont {M.}~\bibnamefont {Klenin}}, \
  and\ \bibinfo {author} {\bibfnamefont {P.}~\bibnamefont {Fulde}},\ }\href
  {\doibase 10.1088/0022-3719/5/15/004} {\bibfield  {journal} {\bibinfo
  {journal} {J. Phys. C: Solid State Phys.}\ }\textbf {\bibinfo {volume} {5}},\
  \bibinfo {pages} {L194} (\bibinfo {year} {1972})}\BibitemShut {NoStop}%
\bibitem [{\citenamefont {Kunkem\"oller}\ \emph {et~al.}(2017)\citenamefont
  {Kunkem\"oller}, \citenamefont {Komleva}, \citenamefont {Streltsov},
  \citenamefont {Hoffmann}, \citenamefont {Khomskii}, \citenamefont {Steffens},
  \citenamefont {Sidis}, \citenamefont {Schmalzl},\ and\ \citenamefont
  {Braden}}]{Kunkem17:95}%
  \BibitemOpen
  \bibfield  {author} {\bibinfo {author} {\bibfnamefont {S.}~\bibnamefont
  {Kunkem\"oller}}, \bibinfo {author} {\bibfnamefont {E.}~\bibnamefont
  {Komleva}}, \bibinfo {author} {\bibfnamefont {S.~V.}\ \bibnamefont
  {Streltsov}}, \bibinfo {author} {\bibfnamefont {S.}~\bibnamefont {Hoffmann}},
  \bibinfo {author} {\bibfnamefont {D.~I.}\ \bibnamefont {Khomskii}}, \bibinfo
  {author} {\bibfnamefont {P.}~\bibnamefont {Steffens}}, \bibinfo {author}
  {\bibfnamefont {Y.}~\bibnamefont {Sidis}}, \bibinfo {author} {\bibfnamefont
  {K.}~\bibnamefont {Schmalzl}}, \ and\ \bibinfo {author} {\bibfnamefont
  {M.}~\bibnamefont {Braden}},\ }\href {\doibase 10.1103/PhysRevB.95.214408}
  {\bibfield  {journal} {\bibinfo  {journal} {Phys. Rev. B}\ }\textbf {\bibinfo
  {volume} {95}},\ \bibinfo {pages} {214408} (\bibinfo {year}
  {2017})}\BibitemShut {NoStop}%
\bibitem [{\citenamefont {Das}\ \emph {et~al.}(2018)\citenamefont {Das},
  \citenamefont {Forte}, \citenamefont {Fittipaldi}, \citenamefont {Fatuzzo},
  \citenamefont {Granata}, \citenamefont {Ivashko}, \citenamefont {Horio},
  \citenamefont {Schindler}, \citenamefont {Dantz}, \citenamefont {Tseng},
  \citenamefont {McNally}, \citenamefont {R\o{}nnow}, \citenamefont {Wan},
  \citenamefont {Christensen}, \citenamefont {Pelliciari}, \citenamefont
  {Olalde-Velasco}, \citenamefont {Kikugawa}, \citenamefont {Neupert},
  \citenamefont {Vecchione}, \citenamefont {Schmitt}, \citenamefont {Cuoco},\
  and\ \citenamefont {Chang}}]{das18:8}%
  \BibitemOpen
  \bibfield  {author} {\bibinfo {author} {\bibfnamefont {L.}~\bibnamefont
  {Das}}, \bibinfo {author} {\bibfnamefont {F.}~\bibnamefont {Forte}}, \bibinfo
  {author} {\bibfnamefont {R.}~\bibnamefont {Fittipaldi}}, \bibinfo {author}
  {\bibfnamefont {C.~G.}\ \bibnamefont {Fatuzzo}}, \bibinfo {author}
  {\bibfnamefont {V.}~\bibnamefont {Granata}}, \bibinfo {author} {\bibfnamefont
  {O.}~\bibnamefont {Ivashko}}, \bibinfo {author} {\bibfnamefont
  {M.}~\bibnamefont {Horio}}, \bibinfo {author} {\bibfnamefont
  {F.}~\bibnamefont {Schindler}}, \bibinfo {author} {\bibfnamefont
  {M.}~\bibnamefont {Dantz}}, \bibinfo {author} {\bibfnamefont
  {Y.}~\bibnamefont {Tseng}}, \bibinfo {author} {\bibfnamefont {D.~E.}\
  \bibnamefont {McNally}}, \bibinfo {author} {\bibfnamefont {H.~M.}\
  \bibnamefont {R\o{}nnow}}, \bibinfo {author} {\bibfnamefont {W.}~\bibnamefont
  {Wan}}, \bibinfo {author} {\bibfnamefont {N.~B.}\ \bibnamefont
  {Christensen}}, \bibinfo {author} {\bibfnamefont {J.}~\bibnamefont
  {Pelliciari}}, \bibinfo {author} {\bibfnamefont {P.}~\bibnamefont
  {Olalde-Velasco}}, \bibinfo {author} {\bibfnamefont {N.}~\bibnamefont
  {Kikugawa}}, \bibinfo {author} {\bibfnamefont {T.}~\bibnamefont {Neupert}},
  \bibinfo {author} {\bibfnamefont {A.}~\bibnamefont {Vecchione}}, \bibinfo
  {author} {\bibfnamefont {T.}~\bibnamefont {Schmitt}}, \bibinfo {author}
  {\bibfnamefont {M.}~\bibnamefont {Cuoco}}, \ and\ \bibinfo {author}
  {\bibfnamefont {J.}~\bibnamefont {Chang}},\ }\href {\doibase
  10.1103/PhysRevX.8.011048} {\bibfield  {journal} {\bibinfo  {journal} {Phys.
  Rev. X}\ }\textbf {\bibinfo {volume} {8}},\ \bibinfo {pages} {011048}
  (\bibinfo {year} {2018})}\BibitemShut {NoStop}%
\bibitem [{\citenamefont {Buyers}\ \emph {et~al.}(1975)\citenamefont {Buyers},
  \citenamefont {Holden},\ and\ \citenamefont {Perreault}}]{buyers11:75}%
  \BibitemOpen
  \bibfield  {author} {\bibinfo {author} {\bibfnamefont {W.~J.~L.}\
  \bibnamefont {Buyers}}, \bibinfo {author} {\bibfnamefont {T.~M.}\
  \bibnamefont {Holden}}, \ and\ \bibinfo {author} {\bibfnamefont
  {A.}~\bibnamefont {Perreault}},\ }\href {\doibase 10.1103/PhysRevB.11.266}
  {\bibfield  {journal} {\bibinfo  {journal} {Phys. Rev. B}\ }\textbf {\bibinfo
  {volume} {11}},\ \bibinfo {pages} {266} (\bibinfo {year} {1975})}\BibitemShut
  {NoStop}%
\bibitem [{\citenamefont {Sarte}\ \emph {et~al.}(2019)\citenamefont {Sarte},
  \citenamefont {Songvilay}, \citenamefont {Pachoud}, \citenamefont {Ewings},
  \citenamefont {Frost}, \citenamefont {Prabhakaran}, \citenamefont {Hong},
  \citenamefont {Browne}, \citenamefont {Yamani}, \citenamefont {Attfield},
  \citenamefont {Rodriguez}, \citenamefont {Wilson},\ and\ \citenamefont
  {Stock}}]{sarte100:19}%
  \BibitemOpen
  \bibfield  {author} {\bibinfo {author} {\bibfnamefont {P.~M.}\ \bibnamefont
  {Sarte}}, \bibinfo {author} {\bibfnamefont {M.}~\bibnamefont {Songvilay}},
  \bibinfo {author} {\bibfnamefont {E.}~\bibnamefont {Pachoud}}, \bibinfo
  {author} {\bibfnamefont {R.~A.}\ \bibnamefont {Ewings}}, \bibinfo {author}
  {\bibfnamefont {C.~D.}\ \bibnamefont {Frost}}, \bibinfo {author}
  {\bibfnamefont {D.}~\bibnamefont {Prabhakaran}}, \bibinfo {author}
  {\bibfnamefont {K.~H.}\ \bibnamefont {Hong}}, \bibinfo {author}
  {\bibfnamefont {A.~J.}\ \bibnamefont {Browne}}, \bibinfo {author}
  {\bibfnamefont {Z.}~\bibnamefont {Yamani}}, \bibinfo {author} {\bibfnamefont
  {J.~P.}\ \bibnamefont {Attfield}}, \bibinfo {author} {\bibfnamefont {E.~E.}\
  \bibnamefont {Rodriguez}}, \bibinfo {author} {\bibfnamefont {S.~D.}\
  \bibnamefont {Wilson}}, \ and\ \bibinfo {author} {\bibfnamefont
  {C.}~\bibnamefont {Stock}},\ }\href {\doibase 10.1103/PhysRevB.100.075143}
  {\bibfield  {journal} {\bibinfo  {journal} {Phys. Rev. B}\ }\textbf {\bibinfo
  {volume} {100}},\ \bibinfo {pages} {075143} (\bibinfo {year}
  {2019})}\BibitemShut {NoStop}%
\bibitem [{\citenamefont {Yosida}(1996)}]{Yosida:book}%
  \BibitemOpen
  \bibfield  {author} {\bibinfo {author} {\bibfnamefont {K.}~\bibnamefont
  {Yosida}},\ }\href@noop {} {\emph {\bibinfo {title} {Theory of Magnetism}}}\
  (\bibinfo  {publisher} {Springer},\ \bibinfo {address} {New York},\ \bibinfo
  {year} {1996})\BibitemShut {NoStop}%
\bibitem [{\citenamefont {Zubarev}(1960)}]{Zubarev60:3}%
  \BibitemOpen
  \bibfield  {author} {\bibinfo {author} {\bibfnamefont {D.~N.}\ \bibnamefont
  {Zubarev}},\ }\href {\doibase 10.3367/UFNr.0071.196005c.0071} {\bibfield
  {journal} {\bibinfo  {journal} {Phys.-Uspekhi}\ }\textbf {\bibinfo {volume}
  {3}},\ \bibinfo {pages} {320} (\bibinfo {year} {1960})}\BibitemShut {NoStop}%
\bibitem [{\citenamefont {Dong}\ \emph {et~al.}(2018)\citenamefont {Dong},
  \citenamefont {Wang},\ and\ \citenamefont {Li}}]{Dong18:97}%
  \BibitemOpen
  \bibfield  {author} {\bibinfo {author} {\bibfnamefont {Z.-Y.}\ \bibnamefont
  {Dong}}, \bibinfo {author} {\bibfnamefont {W.}~\bibnamefont {Wang}}, \ and\
  \bibinfo {author} {\bibfnamefont {J.-X.}\ \bibnamefont {Li}},\ }\href
  {\doibase 10.1103/PhysRevB.97.205106} {\bibfield  {journal} {\bibinfo
  {journal} {Phys. Rev. B}\ }\textbf {\bibinfo {volume} {97}},\ \bibinfo
  {pages} {205106} (\bibinfo {year} {2018})}\BibitemShut {NoStop}%
\bibitem [{\citenamefont {Muniz}\ \emph {et~al.}(2014)\citenamefont {Muniz},
  \citenamefont {Kato},\ and\ \citenamefont {Batista}}]{Muniz04:14}%
  \BibitemOpen
  \bibfield  {author} {\bibinfo {author} {\bibfnamefont {R.~A.}\ \bibnamefont
  {Muniz}}, \bibinfo {author} {\bibfnamefont {Y.}~\bibnamefont {Kato}}, \ and\
  \bibinfo {author} {\bibfnamefont {C.~D.}\ \bibnamefont {Batista}},\ }\href
  {\doibase 10.1093/ptep/ptu109} {\bibfield  {journal} {\bibinfo  {journal}
  {Prog. Theor. Exp. Phys.}\ }\textbf {\bibinfo {volume} {2014}},\ \bibinfo
  {pages} {83101} (\bibinfo {year} {2014})}\BibitemShut {NoStop}%
\bibitem [{\citenamefont {Sarte}\ \emph {et~al.}(2020)\citenamefont {Sarte},
  \citenamefont {Wilson}, \citenamefont {Attfield},\ and\ \citenamefont
  {Stock}}]{Sarte20:32}%
  \BibitemOpen
  \bibfield  {author} {\bibinfo {author} {\bibfnamefont {P.~M.}\ \bibnamefont
  {Sarte}}, \bibinfo {author} {\bibfnamefont {S.~D.}\ \bibnamefont {Wilson}},
  \bibinfo {author} {\bibfnamefont {J.~P.}\ \bibnamefont {Attfield}}, \ and\
  \bibinfo {author} {\bibfnamefont {C.}~\bibnamefont {Stock}},\ }\href
  {\doibase 10.1088/1361-648X/ab8498} {\bibfield  {journal} {\bibinfo
  {journal} {J. Phys.: Condens. Matter}\ }\textbf {\bibinfo {volume} {32}},\
  \bibinfo {pages} {374011} (\bibinfo {year} {2020})}\BibitemShut {NoStop}%
\bibitem [{\citenamefont {Wolff}(1960)}]{wolff60:120}%
  \BibitemOpen
  \bibfield  {author} {\bibinfo {author} {\bibfnamefont {P.~A.}\ \bibnamefont
  {Wolff}},\ }\href {\doibase 10.1103/PhysRev.120.814} {\bibfield  {journal}
  {\bibinfo  {journal} {Phys. Rev.}\ }\textbf {\bibinfo {volume} {120}},\
  \bibinfo {pages} {814} (\bibinfo {year} {1960})}\BibitemShut {NoStop}%
\bibitem [{\citenamefont {Cooke}(1973)}]{cooke73:7}%
  \BibitemOpen
  \bibfield  {author} {\bibinfo {author} {\bibfnamefont {J.~F.}\ \bibnamefont
  {Cooke}},\ }\href {\doibase 10.1103/PhysRevB.7.1108} {\bibfield  {journal}
  {\bibinfo  {journal} {Phys. Rev. B}\ }\textbf {\bibinfo {volume} {7}},\
  \bibinfo {pages} {1108} (\bibinfo {year} {1973})}\BibitemShut {NoStop}%
\bibitem [{\citenamefont {Yamada}\ and\ \citenamefont
  {Shimizu}(1966)}]{yamada66:21}%
  \BibitemOpen
  \bibfield  {author} {\bibinfo {author} {\bibfnamefont {H.}~\bibnamefont
  {Yamada}}\ and\ \bibinfo {author} {\bibfnamefont {M.}~\bibnamefont
  {Shimizu}},\ }\href {\doibase 10.1143/JPSJ.21.1517} {\bibfield  {journal}
  {\bibinfo  {journal} {J. Phys. Soc. Jpn}\ }\textbf {\bibinfo {volume} {21}},\
  \bibinfo {pages} {1517} (\bibinfo {year} {1966})}\BibitemShut {NoStop}%
\bibitem [{\citenamefont {Yamada}\ and\ \citenamefont
  {Shimizu}(1967)}]{yamada67:22}%
  \BibitemOpen
  \bibfield  {author} {\bibinfo {author} {\bibfnamefont {H.}~\bibnamefont
  {Yamada}}\ and\ \bibinfo {author} {\bibfnamefont {M.}~\bibnamefont
  {Shimizu}},\ }\href {\doibase 10.1143/JPSJ.22.1404} {\bibfield  {journal}
  {\bibinfo  {journal} {J. Phys. Soc. Jpn.}\ }\textbf {\bibinfo {volume}
  {22}},\ \bibinfo {pages} {1404} (\bibinfo {year} {1967})}\BibitemShut
  {NoStop}%
\bibitem [{\citenamefont {Cao}\ and\ \citenamefont {DeLong}(2013)}]{Cao:book}%
  \BibitemOpen
  \bibfield  {author} {\bibinfo {author} {\bibfnamefont {G.}~\bibnamefont
  {Cao}}\ and\ \bibinfo {author} {\bibfnamefont {L.}~\bibnamefont {DeLong}},\
  }\href@noop {} {\emph {\bibinfo {title} {Frontiers of 4d and 5d transition
  metal oxides}}}\ (\bibinfo  {publisher} {World Scientific},\ \bibinfo
  {address} {Singapore},\ \bibinfo {year} {2013})\BibitemShut {NoStop}%
\bibitem [{\citenamefont {Chatterji}\ and\ \citenamefont
  {Schneider}(2009)}]{Chatterji09:79}%
  \BibitemOpen
  \bibfield  {author} {\bibinfo {author} {\bibfnamefont {T.}~\bibnamefont
  {Chatterji}}\ and\ \bibinfo {author} {\bibfnamefont {G.~J.}\ \bibnamefont
  {Schneider}},\ }\href {\doibase 10.1103/PhysRevB.79.212409} {\bibfield
  {journal} {\bibinfo  {journal} {Phys. Rev. B}\ }\textbf {\bibinfo {volume}
  {79}},\ \bibinfo {pages} {212409} (\bibinfo {year} {2009})}\BibitemShut
  {NoStop}%
\bibitem [{\citenamefont {Cowley}\ \emph {et~al.}(2013)\citenamefont {Cowley},
  \citenamefont {Buyers}, \citenamefont {Stock}, \citenamefont {Yamani},
  \citenamefont {Frost}, \citenamefont {Taylor},\ and\ \citenamefont
  {Prabhakaran}}]{Cowley13:88}%
  \BibitemOpen
  \bibfield  {author} {\bibinfo {author} {\bibfnamefont {R.~A.}\ \bibnamefont
  {Cowley}}, \bibinfo {author} {\bibfnamefont {W.~J.~L.}\ \bibnamefont
  {Buyers}}, \bibinfo {author} {\bibfnamefont {C.}~\bibnamefont {Stock}},
  \bibinfo {author} {\bibfnamefont {Z.}~\bibnamefont {Yamani}}, \bibinfo
  {author} {\bibfnamefont {C.}~\bibnamefont {Frost}}, \bibinfo {author}
  {\bibfnamefont {J.~W.}\ \bibnamefont {Taylor}}, \ and\ \bibinfo {author}
  {\bibfnamefont {D.}~\bibnamefont {Prabhakaran}},\ }\href {\doibase
  10.1103/PhysRevB.88.205117} {\bibfield  {journal} {\bibinfo  {journal} {Phys.
  Rev. B}\ }\textbf {\bibinfo {volume} {88}},\ \bibinfo {pages} {205117}
  (\bibinfo {year} {2013})}\BibitemShut {NoStop}%
\bibitem [{\citenamefont {McClure}(1959)}]{McClure59:9}%
  \BibitemOpen
  \bibfield  {author} {\bibinfo {author} {\bibfnamefont {D.~S.}\ \bibnamefont
  {McClure}},\ }\href {\doibase 10.1016/S0081-1947(08)60569-X} {\bibfield
  {journal} {\bibinfo  {journal} {Solid State Phys.}\ }\textbf {\bibinfo
  {volume} {9}},\ \bibinfo {pages} {399} (\bibinfo {year} {1959})}\BibitemShut
  {NoStop}%
\bibitem [{\citenamefont {Tanabe}\ and\ \citenamefont
  {Sugano}(1954{\natexlab{a}})}]{Tanabe54:9}%
  \BibitemOpen
  \bibfield  {author} {\bibinfo {author} {\bibfnamefont {Y.}~\bibnamefont
  {Tanabe}}\ and\ \bibinfo {author} {\bibfnamefont {S.}~\bibnamefont
  {Sugano}},\ }\href {\doibase 10.1143/JPSJ.9.753} {\bibfield  {journal}
  {\bibinfo  {journal} {J. Phys. Soc. Jpn.}\ }\textbf {\bibinfo {volume} {9}},\
  \bibinfo {pages} {753} (\bibinfo {year} {1954}{\natexlab{a}})}\BibitemShut
  {NoStop}%
\bibitem [{\citenamefont {Tanabe}\ and\ \citenamefont
  {Sugano}(1954{\natexlab{b}})}]{Tanabe54:9_2}%
  \BibitemOpen
  \bibfield  {author} {\bibinfo {author} {\bibfnamefont {Y.}~\bibnamefont
  {Tanabe}}\ and\ \bibinfo {author} {\bibfnamefont {S.}~\bibnamefont
  {Sugano}},\ }\href {\doibase 10.1143/JPSJ.9.766} {\bibfield  {journal}
  {\bibinfo  {journal} {J. Phys. Soc. Jpn.}\ }\textbf {\bibinfo {volume} {9}},\
  \bibinfo {pages} {766} (\bibinfo {year} {1954}{\natexlab{b}})}\BibitemShut
  {NoStop}%
\bibitem [{\citenamefont {Landau}\ and\ \citenamefont
  {Lifshitz}(1977)}]{Landau:book}%
  \BibitemOpen
  \bibfield  {author} {\bibinfo {author} {\bibfnamefont {L.~D.}\ \bibnamefont
  {Landau}}\ and\ \bibinfo {author} {\bibfnamefont {E.~M.}\ \bibnamefont
  {Lifshitz}},\ }\href@noop {} {\emph {\bibinfo {title} {Quantum Mechanics
  (Non-relativistic Theory)}}}\ (\bibinfo  {publisher} {Elsevier Science},\
  \bibinfo {address} {Oxford},\ \bibinfo {year} {1977})\BibitemShut {NoStop}%
\bibitem [{\citenamefont {Pekker}\ and\ \citenamefont
  {Varma}(2015)}]{pekker15:36}%
  \BibitemOpen
  \bibfield  {author} {\bibinfo {author} {\bibfnamefont {D.}~\bibnamefont
  {Pekker}}\ and\ \bibinfo {author} {\bibfnamefont {C.~M.}\ \bibnamefont
  {Varma}},\ }\href {\doibase 10.1146/annurev-conmatphys-031214-014350}
  {\bibfield  {journal} {\bibinfo  {journal} {Annu. Rev. Condens. Matter
  Phys.}\ }\textbf {\bibinfo {volume} {6}},\ \bibinfo {pages} {269} (\bibinfo
  {year} {2015})}\BibitemShut {NoStop}%
\bibitem [{\citenamefont {Mizokawa}\ \emph {et~al.}(2001)\citenamefont
  {Mizokawa}, \citenamefont {Tjeng}, \citenamefont {Sawatzky}, \citenamefont
  {Ghiringhelli}, \citenamefont {Tjernberg}, \citenamefont {Brookes},
  \citenamefont {Fukazawa}, \citenamefont {Nakatsuji},\ and\ \citenamefont
  {Maeno}}]{mizokawa01:87}%
  \BibitemOpen
  \bibfield  {author} {\bibinfo {author} {\bibfnamefont {T.}~\bibnamefont
  {Mizokawa}}, \bibinfo {author} {\bibfnamefont {L.~H.}\ \bibnamefont {Tjeng}},
  \bibinfo {author} {\bibfnamefont {G.~A.}\ \bibnamefont {Sawatzky}}, \bibinfo
  {author} {\bibfnamefont {G.}~\bibnamefont {Ghiringhelli}}, \bibinfo {author}
  {\bibfnamefont {O.}~\bibnamefont {Tjernberg}}, \bibinfo {author}
  {\bibfnamefont {N.~B.}\ \bibnamefont {Brookes}}, \bibinfo {author}
  {\bibfnamefont {H.}~\bibnamefont {Fukazawa}}, \bibinfo {author}
  {\bibfnamefont {S.}~\bibnamefont {Nakatsuji}}, \ and\ \bibinfo {author}
  {\bibfnamefont {Y.}~\bibnamefont {Maeno}},\ }\href {\doibase
  10.1103/PhysRevLett.87.077202} {\bibfield  {journal} {\bibinfo  {journal}
  {Phys. Rev. Lett.}\ }\textbf {\bibinfo {volume} {87}},\ \bibinfo {pages}
  {077202} (\bibinfo {year} {2001})}\BibitemShut {NoStop}%
\bibitem [{\citenamefont {Fatuzzo}\ \emph {et~al.}(2015)\citenamefont
  {Fatuzzo}, \citenamefont {Dantz}, \citenamefont {Fatale}, \citenamefont
  {Olalde-Velasco}, \citenamefont {Shaik}, \citenamefont {Dalla~Piazza},
  \citenamefont {Toth}, \citenamefont {Pelliciari}, \citenamefont {Fittipaldi},
  \citenamefont {Vecchione}, \citenamefont {Kikugawa}, \citenamefont {Brooks},
  \citenamefont {R\o{}nnow}, \citenamefont {Grioni}, \citenamefont {R\"uegg},
  \citenamefont {Schmitt},\ and\ \citenamefont {Chang}}]{fatuzzo15:91}%
  \BibitemOpen
  \bibfield  {author} {\bibinfo {author} {\bibfnamefont {C.~G.}\ \bibnamefont
  {Fatuzzo}}, \bibinfo {author} {\bibfnamefont {M.}~\bibnamefont {Dantz}},
  \bibinfo {author} {\bibfnamefont {S.}~\bibnamefont {Fatale}}, \bibinfo
  {author} {\bibfnamefont {P.}~\bibnamefont {Olalde-Velasco}}, \bibinfo
  {author} {\bibfnamefont {N.~E.}\ \bibnamefont {Shaik}}, \bibinfo {author}
  {\bibfnamefont {B.}~\bibnamefont {Dalla~Piazza}}, \bibinfo {author}
  {\bibfnamefont {S.}~\bibnamefont {Toth}}, \bibinfo {author} {\bibfnamefont
  {J.}~\bibnamefont {Pelliciari}}, \bibinfo {author} {\bibfnamefont
  {R.}~\bibnamefont {Fittipaldi}}, \bibinfo {author} {\bibfnamefont
  {A.}~\bibnamefont {Vecchione}}, \bibinfo {author} {\bibfnamefont
  {N.}~\bibnamefont {Kikugawa}}, \bibinfo {author} {\bibfnamefont {J.~S.}\
  \bibnamefont {Brooks}}, \bibinfo {author} {\bibfnamefont {H.~M.}\
  \bibnamefont {R\o{}nnow}}, \bibinfo {author} {\bibfnamefont {M.}~\bibnamefont
  {Grioni}}, \bibinfo {author} {\bibfnamefont {C.}~\bibnamefont {R\"uegg}},
  \bibinfo {author} {\bibfnamefont {T.}~\bibnamefont {Schmitt}}, \ and\
  \bibinfo {author} {\bibfnamefont {J.}~\bibnamefont {Chang}},\ }\href
  {\doibase 10.1103/PhysRevB.91.155104} {\bibfield  {journal} {\bibinfo
  {journal} {Phys. Rev. B}\ }\textbf {\bibinfo {volume} {91}},\ \bibinfo
  {pages} {155104} (\bibinfo {year} {2015})}\BibitemShut {NoStop}%
\bibitem [{\citenamefont {Griffith}(1960)}]{Griffith60:56}%
  \BibitemOpen
  \bibfield  {author} {\bibinfo {author} {\bibfnamefont {J.~S.}\ \bibnamefont
  {Griffith}},\ }\href {\doibase 10.1039/TF9605600193} {\bibfield  {journal}
  {\bibinfo  {journal} {Trans. Faraday Soc.}\ }\textbf {\bibinfo {volume}
  {56}},\ \bibinfo {pages} {193} (\bibinfo {year} {1960})}\BibitemShut
  {NoStop}%
\bibitem [{\citenamefont {Moffitt}\ \emph {et~al.}(1959)\citenamefont
  {Moffitt}, \citenamefont {Goodman}, \citenamefont {Fred},\ and\ \citenamefont
  {Weinstock}}]{moffitt59:2}%
  \BibitemOpen
  \bibfield  {author} {\bibinfo {author} {\bibfnamefont {W.}~\bibnamefont
  {Moffitt}}, \bibinfo {author} {\bibfnamefont {G.~L.}\ \bibnamefont
  {Goodman}}, \bibinfo {author} {\bibfnamefont {M.}~\bibnamefont {Fred}}, \
  and\ \bibinfo {author} {\bibfnamefont {B.}~\bibnamefont {Weinstock}},\ }\href
  {\doibase 10.1080/00268975900100101} {\bibfield  {journal} {\bibinfo
  {journal} {Mol. Phys.}\ }\textbf {\bibinfo {volume} {2}},\ \bibinfo {pages}
  {109} (\bibinfo {year} {1959})}\BibitemShut {NoStop}%
\bibitem [{\citenamefont {Alexander}\ \emph {et~al.}(1999)\citenamefont
  {Alexander}, \citenamefont {Cao}, \citenamefont {Dobrosavljevic},
  \citenamefont {McCall}, \citenamefont {Crow}, \citenamefont {Lochner},\ and\
  \citenamefont {Guertin}}]{Alexander99:60}%
  \BibitemOpen
  \bibfield  {author} {\bibinfo {author} {\bibfnamefont {C.~S.}\ \bibnamefont
  {Alexander}}, \bibinfo {author} {\bibfnamefont {G.}~\bibnamefont {Cao}},
  \bibinfo {author} {\bibfnamefont {V.}~\bibnamefont {Dobrosavljevic}},
  \bibinfo {author} {\bibfnamefont {S.}~\bibnamefont {McCall}}, \bibinfo
  {author} {\bibfnamefont {J.~E.}\ \bibnamefont {Crow}}, \bibinfo {author}
  {\bibfnamefont {E.}~\bibnamefont {Lochner}}, \ and\ \bibinfo {author}
  {\bibfnamefont {R.~P.}\ \bibnamefont {Guertin}},\ }\href {\doibase
  10.1103/PhysRevB.60.R8422} {\bibfield  {journal} {\bibinfo  {journal} {Phys.
  Rev. B}\ }\textbf {\bibinfo {volume} {60}},\ \bibinfo {pages} {R8422}
  (\bibinfo {year} {1999})}\BibitemShut {NoStop}%
\bibitem [{\citenamefont {Zegkinoglou}\ \emph {et~al.}(2005)\citenamefont
  {Zegkinoglou}, \citenamefont {Strempfer}, \citenamefont {Nelson},
  \citenamefont {Hill}, \citenamefont {Chakhalian}, \citenamefont {Bernhard},
  \citenamefont {Lang}, \citenamefont {Srajer}, \citenamefont {Fukazawa},
  \citenamefont {Nakatsuji}, \citenamefont {Maeno},\ and\ \citenamefont
  {Keimer}}]{Zegkinoglou05:95}%
  \BibitemOpen
  \bibfield  {author} {\bibinfo {author} {\bibfnamefont {I.}~\bibnamefont
  {Zegkinoglou}}, \bibinfo {author} {\bibfnamefont {J.}~\bibnamefont
  {Strempfer}}, \bibinfo {author} {\bibfnamefont {C.~S.}\ \bibnamefont
  {Nelson}}, \bibinfo {author} {\bibfnamefont {J.~P.}\ \bibnamefont {Hill}},
  \bibinfo {author} {\bibfnamefont {J.}~\bibnamefont {Chakhalian}}, \bibinfo
  {author} {\bibfnamefont {C.}~\bibnamefont {Bernhard}}, \bibinfo {author}
  {\bibfnamefont {J.~C.}\ \bibnamefont {Lang}}, \bibinfo {author}
  {\bibfnamefont {G.}~\bibnamefont {Srajer}}, \bibinfo {author} {\bibfnamefont
  {H.}~\bibnamefont {Fukazawa}}, \bibinfo {author} {\bibfnamefont
  {S.}~\bibnamefont {Nakatsuji}}, \bibinfo {author} {\bibfnamefont
  {Y.}~\bibnamefont {Maeno}}, \ and\ \bibinfo {author} {\bibfnamefont
  {B.}~\bibnamefont {Keimer}},\ }\href {\doibase 10.1103/PhysRevLett.95.136401}
  {\bibfield  {journal} {\bibinfo  {journal} {Phys. Rev. Lett.}\ }\textbf
  {\bibinfo {volume} {95}},\ \bibinfo {pages} {136401} (\bibinfo {year}
  {2005})}\BibitemShut {NoStop}%
\bibitem [{\citenamefont {Qi}\ \emph {et~al.}(2010)\citenamefont {Qi},
  \citenamefont {Korneta}, \citenamefont {Parkin}, \citenamefont {De~Long},
  \citenamefont {Schlottmann},\ and\ \citenamefont {Cao}}]{Qi10:105}%
  \BibitemOpen
  \bibfield  {author} {\bibinfo {author} {\bibfnamefont {T.~F.}\ \bibnamefont
  {Qi}}, \bibinfo {author} {\bibfnamefont {O.~B.}\ \bibnamefont {Korneta}},
  \bibinfo {author} {\bibfnamefont {S.}~\bibnamefont {Parkin}}, \bibinfo
  {author} {\bibfnamefont {L.~E.}\ \bibnamefont {De~Long}}, \bibinfo {author}
  {\bibfnamefont {P.}~\bibnamefont {Schlottmann}}, \ and\ \bibinfo {author}
  {\bibfnamefont {G.}~\bibnamefont {Cao}},\ }\href {\doibase
  10.1103/PhysRevLett.105.177203} {\bibfield  {journal} {\bibinfo  {journal}
  {Phys. Rev. Lett.}\ }\textbf {\bibinfo {volume} {105}},\ \bibinfo {pages}
  {177203} (\bibinfo {year} {2010})}\BibitemShut {NoStop}%
\bibitem [{\citenamefont {Lee}\ \emph {et~al.}(2002)\citenamefont {Lee},
  \citenamefont {Lee}, \citenamefont {Noh}, \citenamefont {Oh}, \citenamefont
  {Yu}, \citenamefont {Nakatsuji}, \citenamefont {Fukazawa},\ and\
  \citenamefont {Maeno}}]{Lee02:89}%
  \BibitemOpen
  \bibfield  {author} {\bibinfo {author} {\bibfnamefont {J.~S.}\ \bibnamefont
  {Lee}}, \bibinfo {author} {\bibfnamefont {Y.~S.}\ \bibnamefont {Lee}},
  \bibinfo {author} {\bibfnamefont {T.~W.}\ \bibnamefont {Noh}}, \bibinfo
  {author} {\bibfnamefont {S.-J.}\ \bibnamefont {Oh}}, \bibinfo {author}
  {\bibfnamefont {J.}~\bibnamefont {Yu}}, \bibinfo {author} {\bibfnamefont
  {S.}~\bibnamefont {Nakatsuji}}, \bibinfo {author} {\bibfnamefont
  {H.}~\bibnamefont {Fukazawa}}, \ and\ \bibinfo {author} {\bibfnamefont
  {Y.}~\bibnamefont {Maeno}},\ }\href {\doibase 10.1103/PhysRevLett.89.257402}
  {\bibfield  {journal} {\bibinfo  {journal} {Phys. Rev. Lett.}\ }\textbf
  {\bibinfo {volume} {89}},\ \bibinfo {pages} {257402} (\bibinfo {year}
  {2002})}\BibitemShut {NoStop}%
\bibitem [{\citenamefont {Zhang}\ and\ \citenamefont
  {Pavarini}(2020)}]{Zhang20:101}%
  \BibitemOpen
  \bibfield  {author} {\bibinfo {author} {\bibfnamefont {G.}~\bibnamefont
  {Zhang}}\ and\ \bibinfo {author} {\bibfnamefont {E.}~\bibnamefont
  {Pavarini}},\ }\href {\doibase 10.1103/PhysRevB.101.205128} {\bibfield
  {journal} {\bibinfo  {journal} {Phys. Rev. B}\ }\textbf {\bibinfo {volume}
  {101}},\ \bibinfo {pages} {205128} (\bibinfo {year} {2020})}\BibitemShut
  {NoStop}%
\bibitem [{\citenamefont {Hutchings}(1964)}]{Hutchings64:16}%
  \BibitemOpen
  \bibfield  {author} {\bibinfo {author} {\bibfnamefont {M.~T.}\ \bibnamefont
  {Hutchings}},\ }\href {\doibase
  https://doi.org/10.1016/S0081-1947(08)60517-2} {\bibfield  {journal}
  {\bibinfo  {journal} {Solid State Phys.}\ }\textbf {\bibinfo {volume} {16}},\
  \bibinfo {pages} {227} (\bibinfo {year} {1964})}\BibitemShut {NoStop}%
\bibitem [{\citenamefont {Bauer}\ and\ \citenamefont
  {Rotter}(2010)}]{Bauer:book}%
  \BibitemOpen
  \bibfield  {author} {\bibinfo {author} {\bibfnamefont {E.}~\bibnamefont
  {Bauer}}\ and\ \bibinfo {author} {\bibfnamefont {M.}~\bibnamefont {Rotter}},\
  }\href@noop {} {\emph {\bibinfo {title} {Magnetism of complex metallic
  alloys: Crystalline electric field effects}}}\ (\bibinfo  {publisher} {World
  Scientific},\ \bibinfo {address} {Singapore},\ \bibinfo {year}
  {2010})\BibitemShut {NoStop}%
\bibitem [{\citenamefont {Abragam}\ and\ \citenamefont
  {Bleaney}(1986)}]{Abragam:book}%
  \BibitemOpen
  \bibfield  {author} {\bibinfo {author} {\bibfnamefont {A.}~\bibnamefont
  {Abragam}}\ and\ \bibinfo {author} {\bibfnamefont {B.}~\bibnamefont
  {Bleaney}},\ }\href@noop {} {\emph {\bibinfo {title} {Electron paramagnetic
  resonance of transition ions}}}\ (\bibinfo  {publisher} {Dover
  Publications},\ \bibinfo {address} {New York},\ \bibinfo {year}
  {1986})\BibitemShut {NoStop}%
\bibitem [{\citenamefont {Hotta}(2006)}]{Hotta06:69}%
  \BibitemOpen
  \bibfield  {author} {\bibinfo {author} {\bibfnamefont {H.}~\bibnamefont
  {Hotta}},\ }\href {\doibase 10.1088/0034-4885/69/7/R02} {\bibfield  {journal}
  {\bibinfo  {journal} {Rep. Prog. Phys.}\ }\textbf {\bibinfo {volume} {69}},\
  \bibinfo {pages} {2061} (\bibinfo {year} {2006})}\BibitemShut {NoStop}%
\bibitem [{\citenamefont {Walter}(1960)}]{Walter87:59}%
  \BibitemOpen
  \bibfield  {author} {\bibinfo {author} {\bibfnamefont {U.}~\bibnamefont
  {Walter}},\ }\href {\doibase https://doi.org/10.1016/0022-3697(84)90147-1}
  {\bibfield  {journal} {\bibinfo  {journal} {J. Phys. Chem. Solids}\ }\textbf
  {\bibinfo {volume} {45}},\ \bibinfo {pages} {401} (\bibinfo {year}
  {1960})}\BibitemShut {NoStop}%
\bibitem [{\citenamefont {Sarte}\ \emph
  {et~al.}(2018{\natexlab{a}})\citenamefont {Sarte}, \citenamefont {Cowley},
  \citenamefont {Rodriguez}, \citenamefont {Pachoud}, \citenamefont {Le},
  \citenamefont {Garc\'{\i}a-Sakai}, \citenamefont {Taylor}, \citenamefont
  {Frost}, \citenamefont {Prabhakaran}, \citenamefont {MacEwen}, \citenamefont
  {Kitada}, \citenamefont {Browne}, \citenamefont {Songvilay}, \citenamefont
  {Yamani}, \citenamefont {Buyers}, \citenamefont {Attfield},\ and\
  \citenamefont {Stock}}]{Sarte18:98_2}%
  \BibitemOpen
  \bibfield  {author} {\bibinfo {author} {\bibfnamefont {P.~M.}\ \bibnamefont
  {Sarte}}, \bibinfo {author} {\bibfnamefont {R.~A.}\ \bibnamefont {Cowley}},
  \bibinfo {author} {\bibfnamefont {E.~E.}\ \bibnamefont {Rodriguez}}, \bibinfo
  {author} {\bibfnamefont {E.}~\bibnamefont {Pachoud}}, \bibinfo {author}
  {\bibfnamefont {D.}~\bibnamefont {Le}}, \bibinfo {author} {\bibfnamefont
  {V.}~\bibnamefont {Garc\'{\i}a-Sakai}}, \bibinfo {author} {\bibfnamefont
  {J.~W.}\ \bibnamefont {Taylor}}, \bibinfo {author} {\bibfnamefont {C.~D.}\
  \bibnamefont {Frost}}, \bibinfo {author} {\bibfnamefont {D.}~\bibnamefont
  {Prabhakaran}}, \bibinfo {author} {\bibfnamefont {C.}~\bibnamefont
  {MacEwen}}, \bibinfo {author} {\bibfnamefont {A.}~\bibnamefont {Kitada}},
  \bibinfo {author} {\bibfnamefont {A.~J.}\ \bibnamefont {Browne}}, \bibinfo
  {author} {\bibfnamefont {M.}~\bibnamefont {Songvilay}}, \bibinfo {author}
  {\bibfnamefont {Z.}~\bibnamefont {Yamani}}, \bibinfo {author} {\bibfnamefont
  {W.~J.~L.}\ \bibnamefont {Buyers}}, \bibinfo {author} {\bibfnamefont {J.~P.}\
  \bibnamefont {Attfield}}, \ and\ \bibinfo {author} {\bibfnamefont
  {C.}~\bibnamefont {Stock}},\ }\href {\doibase 10.1103/PhysRevB.98.024415}
  {\bibfield  {journal} {\bibinfo  {journal} {Phys. Rev. B}\ }\textbf {\bibinfo
  {volume} {98}},\ \bibinfo {pages} {024415} (\bibinfo {year}
  {2018}{\natexlab{a}})}\BibitemShut {NoStop}%
\bibitem [{\citenamefont {Sarte}\ \emph
  {et~al.}(2018{\natexlab{b}})\citenamefont {Sarte}, \citenamefont
  {Ar\'evalo-L\'opez}, \citenamefont {Songvilay}, \citenamefont {Le},
  \citenamefont {Guidi}, \citenamefont {Garc\'{\i}a-Sakai}, \citenamefont
  {Mukhopadhyay}, \citenamefont {Capelli}, \citenamefont {Ratcliff},
  \citenamefont {Hong}, \citenamefont {McNally}, \citenamefont {Pachoud},
  \citenamefont {Attfield},\ and\ \citenamefont {Stock}}]{sarte98:18}%
  \BibitemOpen
  \bibfield  {author} {\bibinfo {author} {\bibfnamefont {P.~M.}\ \bibnamefont
  {Sarte}}, \bibinfo {author} {\bibfnamefont {A.~M.}\ \bibnamefont
  {Ar\'evalo-L\'opez}}, \bibinfo {author} {\bibfnamefont {M.}~\bibnamefont
  {Songvilay}}, \bibinfo {author} {\bibfnamefont {D.}~\bibnamefont {Le}},
  \bibinfo {author} {\bibfnamefont {T.}~\bibnamefont {Guidi}}, \bibinfo
  {author} {\bibfnamefont {V.}~\bibnamefont {Garc\'{\i}a-Sakai}}, \bibinfo
  {author} {\bibfnamefont {S.}~\bibnamefont {Mukhopadhyay}}, \bibinfo {author}
  {\bibfnamefont {S.~C.}\ \bibnamefont {Capelli}}, \bibinfo {author}
  {\bibfnamefont {W.~D.}\ \bibnamefont {Ratcliff}}, \bibinfo {author}
  {\bibfnamefont {K.~H.}\ \bibnamefont {Hong}}, \bibinfo {author}
  {\bibfnamefont {G.~M.}\ \bibnamefont {McNally}}, \bibinfo {author}
  {\bibfnamefont {E.}~\bibnamefont {Pachoud}}, \bibinfo {author} {\bibfnamefont
  {J.~P.}\ \bibnamefont {Attfield}}, \ and\ \bibinfo {author} {\bibfnamefont
  {C.}~\bibnamefont {Stock}},\ }\href {\doibase 10.1103/PhysRevB.98.224410}
  {\bibfield  {journal} {\bibinfo  {journal} {Phys. Rev. B}\ }\textbf {\bibinfo
  {volume} {98}},\ \bibinfo {pages} {224410} (\bibinfo {year}
  {2018}{\natexlab{b}})}\BibitemShut {NoStop}%
\bibitem [{\citenamefont {Momma}\ and\ \citenamefont
  {Izumi}(2011)}]{Momma11:44}%
  \BibitemOpen
  \bibfield  {author} {\bibinfo {author} {\bibfnamefont {K.}~\bibnamefont
  {Momma}}\ and\ \bibinfo {author} {\bibfnamefont {F.}~\bibnamefont {Izumi}},\
  }\href {\doibase 10.1107/s0021889811038970} {\bibfield  {journal} {\bibinfo
  {journal} {J. Appl. Crystallogr.}\ }\textbf {\bibinfo {volume} {44}},\
  \bibinfo {pages} {1272} (\bibinfo {year} {2011})}\BibitemShut {NoStop}%
\bibitem [{\citenamefont {Nakatsuji}\ \emph
  {et~al.}(1997{\natexlab{b}})\citenamefont {Nakatsuji}, \citenamefont {ichi
  Ikeda},\ and\ \citenamefont {Maeno}}]{nakatsuji97:282}%
  \BibitemOpen
  \bibfield  {author} {\bibinfo {author} {\bibfnamefont {S.}~\bibnamefont
  {Nakatsuji}}, \bibinfo {author} {\bibfnamefont {S.}~\bibnamefont {ichi
  Ikeda}}, \ and\ \bibinfo {author} {\bibfnamefont {Y.}~\bibnamefont {Maeno}},\
  }\href {\doibase https://doi.org/10.1016/S0921-4534(97)00545-5} {\bibfield
  {journal} {\bibinfo  {journal} {Physica C}\ }\textbf {\bibinfo {volume}
  {282-287}},\ \bibinfo {pages} {729 } (\bibinfo {year}
  {1997}{\natexlab{b}})}\BibitemShut {NoStop}%
\bibitem [{\citenamefont {Zhou}\ \emph {et~al.}(2018)\citenamefont {Zhou},
  \citenamefont {Sarte}, \citenamefont {Conner}, \citenamefont {Balicas},
  \citenamefont {Wiebe}, \citenamefont {Chen}, \citenamefont {Wu},
  \citenamefont {Wu}, \citenamefont {Liu}, \citenamefont {Chen},\ and\
  \citenamefont {Fang}}]{Zhou18:30}%
  \BibitemOpen
  \bibfield  {author} {\bibinfo {author} {\bibfnamefont {H.~D.}\ \bibnamefont
  {Zhou}}, \bibinfo {author} {\bibfnamefont {P.~M.}\ \bibnamefont {Sarte}},
  \bibinfo {author} {\bibfnamefont {B.~S.}\ \bibnamefont {Conner}}, \bibinfo
  {author} {\bibfnamefont {L.}~\bibnamefont {Balicas}}, \bibinfo {author}
  {\bibfnamefont {C.~R.}\ \bibnamefont {Wiebe}}, \bibinfo {author}
  {\bibfnamefont {X.~H.}\ \bibnamefont {Chen}}, \bibinfo {author}
  {\bibfnamefont {T.}~\bibnamefont {Wu}}, \bibinfo {author} {\bibfnamefont
  {G.}~\bibnamefont {Wu}}, \bibinfo {author} {\bibfnamefont {R.~H.}\
  \bibnamefont {Liu}}, \bibinfo {author} {\bibfnamefont {H.}~\bibnamefont
  {Chen}}, \ and\ \bibinfo {author} {\bibfnamefont {D.~F.}\ \bibnamefont
  {Fang}},\ }\href {\doibase 10.1088/1361-648x/aaa3b0} {\bibfield  {journal}
  {\bibinfo  {journal} {J. Phys. Condens. Matter}\ }\textbf {\bibinfo {volume}
  {30}},\ \bibinfo {pages} {095601} (\bibinfo {year} {2018})}\BibitemShut
  {NoStop}%
\bibitem [{\citenamefont {Veenstra}\ \emph {et~al.}(2014)\citenamefont
  {Veenstra}, \citenamefont {Zhu}, \citenamefont {Raichle}, \citenamefont
  {Ludbrook}, \citenamefont {Nicolaou}, \citenamefont {Slomski}, \citenamefont
  {Landolt}, \citenamefont {Kittaka}, \citenamefont {Maeno}, \citenamefont
  {Dil}, \citenamefont {Elfimov}, \citenamefont {Haverkort},\ and\
  \citenamefont {Damascelli}}]{Veenstra14:112}%
  \BibitemOpen
  \bibfield  {author} {\bibinfo {author} {\bibfnamefont {C.~N.}\ \bibnamefont
  {Veenstra}}, \bibinfo {author} {\bibfnamefont {Z.-H.}\ \bibnamefont {Zhu}},
  \bibinfo {author} {\bibfnamefont {M.}~\bibnamefont {Raichle}}, \bibinfo
  {author} {\bibfnamefont {B.~M.}\ \bibnamefont {Ludbrook}}, \bibinfo {author}
  {\bibfnamefont {A.}~\bibnamefont {Nicolaou}}, \bibinfo {author}
  {\bibfnamefont {B.}~\bibnamefont {Slomski}}, \bibinfo {author} {\bibfnamefont
  {G.}~\bibnamefont {Landolt}}, \bibinfo {author} {\bibfnamefont
  {S.}~\bibnamefont {Kittaka}}, \bibinfo {author} {\bibfnamefont
  {Y.}~\bibnamefont {Maeno}}, \bibinfo {author} {\bibfnamefont {J.~H.}\
  \bibnamefont {Dil}}, \bibinfo {author} {\bibfnamefont {I.~S.}\ \bibnamefont
  {Elfimov}}, \bibinfo {author} {\bibfnamefont {M.~W.}\ \bibnamefont
  {Haverkort}}, \ and\ \bibinfo {author} {\bibfnamefont {A.}~\bibnamefont
  {Damascelli}},\ }\href {\doibase 10.1103/PhysRevLett.112.127002} {\bibfield
  {journal} {\bibinfo  {journal} {Phys. Rev. Lett.}\ }\textbf {\bibinfo
  {volume} {112}},\ \bibinfo {pages} {127002} (\bibinfo {year}
  {2014})}\BibitemShut {NoStop}%
\bibitem [{\citenamefont {Rho}\ \emph {et~al.}(2003)\citenamefont {Rho},
  \citenamefont {Cooper}, \citenamefont {Nakatsuji}, \citenamefont {Fukazawa},\
  and\ \citenamefont {Maeno}}]{Rho03:68}%
  \BibitemOpen
  \bibfield  {author} {\bibinfo {author} {\bibfnamefont {H.}~\bibnamefont
  {Rho}}, \bibinfo {author} {\bibfnamefont {S.~L.}\ \bibnamefont {Cooper}},
  \bibinfo {author} {\bibfnamefont {S.}~\bibnamefont {Nakatsuji}}, \bibinfo
  {author} {\bibfnamefont {H.}~\bibnamefont {Fukazawa}}, \ and\ \bibinfo
  {author} {\bibfnamefont {Y.}~\bibnamefont {Maeno}},\ }\href {\doibase
  10.1103/PhysRevB.68.100404} {\bibfield  {journal} {\bibinfo  {journal} {Phys.
  Rev. B}\ }\textbf {\bibinfo {volume} {68}},\ \bibinfo {pages} {100404(R)}
  (\bibinfo {year} {2003})}\BibitemShut {NoStop}%
\bibitem [{\citenamefont {Rho}\ \emph {et~al.}(2005)\citenamefont {Rho},
  \citenamefont {Cooper}, \citenamefont {Nakatsuji}, \citenamefont {Fukazawa},\
  and\ \citenamefont {Maeno}}]{Rho05:71}%
  \BibitemOpen
  \bibfield  {author} {\bibinfo {author} {\bibfnamefont {H.}~\bibnamefont
  {Rho}}, \bibinfo {author} {\bibfnamefont {S.~L.}\ \bibnamefont {Cooper}},
  \bibinfo {author} {\bibfnamefont {S.}~\bibnamefont {Nakatsuji}}, \bibinfo
  {author} {\bibfnamefont {H.}~\bibnamefont {Fukazawa}}, \ and\ \bibinfo
  {author} {\bibfnamefont {Y.}~\bibnamefont {Maeno}},\ }\href {\doibase
  10.1103/PhysRevB.71.245121} {\bibfield  {journal} {\bibinfo  {journal} {Phys.
  Rev. B}\ }\textbf {\bibinfo {volume} {71}},\ \bibinfo {pages} {245121}
  (\bibinfo {year} {2005})}\BibitemShut {NoStop}%
\bibitem [{\citenamefont {Sugai}\ \emph {et~al.}(1990)\citenamefont {Sugai},
  \citenamefont {Sato}, \citenamefont {Kobayashi}, \citenamefont {Akimitsu},
  \citenamefont {Ito}, \citenamefont {Takagi}, \citenamefont {Uchida},
  \citenamefont {Hosoya}, \citenamefont {Kajitani},\ and\ \citenamefont
  {Fukuda}}]{Sugai90:42}%
  \BibitemOpen
  \bibfield  {author} {\bibinfo {author} {\bibfnamefont {S.}~\bibnamefont
  {Sugai}}, \bibinfo {author} {\bibfnamefont {M.}~\bibnamefont {Sato}},
  \bibinfo {author} {\bibfnamefont {T.}~\bibnamefont {Kobayashi}}, \bibinfo
  {author} {\bibfnamefont {J.}~\bibnamefont {Akimitsu}}, \bibinfo {author}
  {\bibfnamefont {T.}~\bibnamefont {Ito}}, \bibinfo {author} {\bibfnamefont
  {H.}~\bibnamefont {Takagi}}, \bibinfo {author} {\bibfnamefont
  {S.}~\bibnamefont {Uchida}}, \bibinfo {author} {\bibfnamefont
  {S.}~\bibnamefont {Hosoya}}, \bibinfo {author} {\bibfnamefont
  {T.}~\bibnamefont {Kajitani}}, \ and\ \bibinfo {author} {\bibfnamefont
  {T.}~\bibnamefont {Fukuda}},\ }\href {\doibase 10.1103/PhysRevB.42.1045}
  {\bibfield  {journal} {\bibinfo  {journal} {Phys. Rev. B}\ }\textbf {\bibinfo
  {volume} {42}},\ \bibinfo {pages} {1045} (\bibinfo {year}
  {1990})}\BibitemShut {NoStop}%
\bibitem [{\citenamefont {Buyers}\ \emph {et~al.}(1971)\citenamefont {Buyers},
  \citenamefont {Holden}, \citenamefont {Svensson}, \citenamefont {Cowley},\
  and\ \citenamefont {Hutchings}}]{Buyers71:4}%
  \BibitemOpen
  \bibfield  {author} {\bibinfo {author} {\bibfnamefont {W.~J.~L.}\
  \bibnamefont {Buyers}}, \bibinfo {author} {\bibfnamefont {T.~M.}\
  \bibnamefont {Holden}}, \bibinfo {author} {\bibfnamefont {E.~C.}\
  \bibnamefont {Svensson}}, \bibinfo {author} {\bibfnamefont {R.~A.}\
  \bibnamefont {Cowley}}, \ and\ \bibinfo {author} {\bibfnamefont {M.~T.}\
  \bibnamefont {Hutchings}},\ }\href {\doibase 10.1088/0022-3719/4/14/028}
  {\bibfield  {journal} {\bibinfo  {journal} {J. Phys. C: Solid St. Phys.}\
  }\textbf {\bibinfo {volume} {1971}},\ \bibinfo {pages} {2139} (\bibinfo
  {year} {1971})}\BibitemShut {NoStop}%
\bibitem [{\citenamefont {Bertinshaw}\ \emph {et~al.}(2019)\citenamefont
  {Bertinshaw}, \citenamefont {Gurung}, \citenamefont {Jorba}, \citenamefont
  {Liu}, \citenamefont {Schmid}, \citenamefont {Mantadakis}, \citenamefont
  {Daghofer}, \citenamefont {Krautloher}, \citenamefont {Jain}, \citenamefont
  {Ryu}, \citenamefont {Fabelo}, \citenamefont {Hansmann}, \citenamefont
  {Khaliullin}, \citenamefont {Pfleiderer}, \citenamefont {Keimer},\ and\
  \citenamefont {Kim}}]{bertinshaw19:123}%
  \BibitemOpen
  \bibfield  {author} {\bibinfo {author} {\bibfnamefont {J.}~\bibnamefont
  {Bertinshaw}}, \bibinfo {author} {\bibfnamefont {N.}~\bibnamefont {Gurung}},
  \bibinfo {author} {\bibfnamefont {P.}~\bibnamefont {Jorba}}, \bibinfo
  {author} {\bibfnamefont {H.}~\bibnamefont {Liu}}, \bibinfo {author}
  {\bibfnamefont {M.}~\bibnamefont {Schmid}}, \bibinfo {author} {\bibfnamefont
  {D.~T.}\ \bibnamefont {Mantadakis}}, \bibinfo {author} {\bibfnamefont
  {M.}~\bibnamefont {Daghofer}}, \bibinfo {author} {\bibfnamefont
  {M.}~\bibnamefont {Krautloher}}, \bibinfo {author} {\bibfnamefont
  {A.}~\bibnamefont {Jain}}, \bibinfo {author} {\bibfnamefont {G.~H.}\
  \bibnamefont {Ryu}}, \bibinfo {author} {\bibfnamefont {O.}~\bibnamefont
  {Fabelo}}, \bibinfo {author} {\bibfnamefont {P.}~\bibnamefont {Hansmann}},
  \bibinfo {author} {\bibfnamefont {G.}~\bibnamefont {Khaliullin}}, \bibinfo
  {author} {\bibfnamefont {C.}~\bibnamefont {Pfleiderer}}, \bibinfo {author}
  {\bibfnamefont {B.}~\bibnamefont {Keimer}}, \ and\ \bibinfo {author}
  {\bibfnamefont {B.~J.}\ \bibnamefont {Kim}},\ }\href {\doibase
  10.1103/PhysRevLett.123.137204} {\bibfield  {journal} {\bibinfo  {journal}
  {Phys. Rev. Lett.}\ }\textbf {\bibinfo {volume} {123}},\ \bibinfo {pages}
  {137204} (\bibinfo {year} {2019})}\BibitemShut {NoStop}%
\bibitem [{\citenamefont {Higgs}(1964)}]{Higgs64:13}%
  \BibitemOpen
  \bibfield  {author} {\bibinfo {author} {\bibfnamefont {P.~W.}\ \bibnamefont
  {Higgs}},\ }\href {\doibase 10.1103/PhysRevLett.13.508} {\bibfield  {journal}
  {\bibinfo  {journal} {Phys. Rev. Lett.}\ }\textbf {\bibinfo {volume} {13}},\
  \bibinfo {pages} {508} (\bibinfo {year} {1964})}\BibitemShut {NoStop}%
\bibitem [{\citenamefont {Munehisa}(2015)}]{Munehisa15:5}%
  \BibitemOpen
  \bibfield  {author} {\bibinfo {author} {\bibfnamefont {T.}~\bibnamefont
  {Munehisa}},\ }\href {\doibase 10.4236/wjcmp.2015.54027} {\bibfield
  {journal} {\bibinfo  {journal} {World J. Condens. Matter Phys.}\ }\textbf
  {\bibinfo {volume} {87}},\ \bibinfo {pages} {077202} (\bibinfo {year}
  {2015})}\BibitemShut {NoStop}%
\bibitem [{\citenamefont {Rose}\ \emph {et~al.}(2015)\citenamefont {Rose},
  \citenamefont {L\'eonard},\ and\ \citenamefont {Dupuis}}]{Rose15:91}%
  \BibitemOpen
  \bibfield  {author} {\bibinfo {author} {\bibfnamefont {F.}~\bibnamefont
  {Rose}}, \bibinfo {author} {\bibfnamefont {F.}~\bibnamefont {L\'eonard}}, \
  and\ \bibinfo {author} {\bibfnamefont {N.}~\bibnamefont {Dupuis}},\ }\href
  {\doibase 10.1103/PhysRevB.91.224501} {\bibfield  {journal} {\bibinfo
  {journal} {Phys. Rev. B}\ }\textbf {\bibinfo {volume} {91}},\ \bibinfo
  {pages} {224501} (\bibinfo {year} {2015})}\BibitemShut {NoStop}%
\bibitem [{\citenamefont {Huberman}\ \emph {et~al.}(2005)\citenamefont
  {Huberman}, \citenamefont {Coldea}, \citenamefont {Cowley}, \citenamefont
  {Tennant}, \citenamefont {Leheny}, \citenamefont {Christianson},\ and\
  \citenamefont {Frost}}]{Huberman05:72}%
  \BibitemOpen
  \bibfield  {author} {\bibinfo {author} {\bibfnamefont {T.}~\bibnamefont
  {Huberman}}, \bibinfo {author} {\bibfnamefont {R.}~\bibnamefont {Coldea}},
  \bibinfo {author} {\bibfnamefont {R.~A.}\ \bibnamefont {Cowley}}, \bibinfo
  {author} {\bibfnamefont {D.~A.}\ \bibnamefont {Tennant}}, \bibinfo {author}
  {\bibfnamefont {R.~L.}\ \bibnamefont {Leheny}}, \bibinfo {author}
  {\bibfnamefont {R.~J.}\ \bibnamefont {Christianson}}, \ and\ \bibinfo
  {author} {\bibfnamefont {C.~D.}\ \bibnamefont {Frost}},\ }\href {\doibase
  10.1103/PhysRevB.72.014413} {\bibfield  {journal} {\bibinfo  {journal} {Phys.
  Rev. B}\ }\textbf {\bibinfo {volume} {72}},\ \bibinfo {pages} {014413}
  (\bibinfo {year} {2005})}\BibitemShut {NoStop}%
\bibitem [{\citenamefont {Stock}\ \emph {et~al.}(2018)\citenamefont {Stock},
  \citenamefont {Gehring}, \citenamefont {Ewings}, \citenamefont {Xu},
  \citenamefont {Li}, \citenamefont {Viehland},\ and\ \citenamefont
  {Luo}}]{stock18:2}%
  \BibitemOpen
  \bibfield  {author} {\bibinfo {author} {\bibfnamefont {C.}~\bibnamefont
  {Stock}}, \bibinfo {author} {\bibfnamefont {P.~M.}\ \bibnamefont {Gehring}},
  \bibinfo {author} {\bibfnamefont {R.~A.}\ \bibnamefont {Ewings}}, \bibinfo
  {author} {\bibfnamefont {G.}~\bibnamefont {Xu}}, \bibinfo {author}
  {\bibfnamefont {J.}~\bibnamefont {Li}}, \bibinfo {author} {\bibfnamefont
  {D.}~\bibnamefont {Viehland}}, \ and\ \bibinfo {author} {\bibfnamefont
  {H.}~\bibnamefont {Luo}},\ }\href {\doibase
  10.1103/PhysRevMaterials.2.024404} {\bibfield  {journal} {\bibinfo  {journal}
  {Phys. Rev. Materials}\ }\textbf {\bibinfo {volume} {2}},\ \bibinfo {pages}
  {024404} (\bibinfo {year} {2018})}\BibitemShut {NoStop}%
\bibitem [{\citenamefont {Songvilay}\ \emph {et~al.}(2018)\citenamefont
  {Songvilay}, \citenamefont {Rodriguez}, \citenamefont {Lindsay},
  \citenamefont {Green}, \citenamefont {Walker}, \citenamefont
  {Rodriguez-Rivera},\ and\ \citenamefont {Stock}}]{songvilay18:121}%
  \BibitemOpen
  \bibfield  {author} {\bibinfo {author} {\bibfnamefont {M.}~\bibnamefont
  {Songvilay}}, \bibinfo {author} {\bibfnamefont {E.~E.}\ \bibnamefont
  {Rodriguez}}, \bibinfo {author} {\bibfnamefont {R.}~\bibnamefont {Lindsay}},
  \bibinfo {author} {\bibfnamefont {M.~A.}\ \bibnamefont {Green}}, \bibinfo
  {author} {\bibfnamefont {H.~C.}\ \bibnamefont {Walker}}, \bibinfo {author}
  {\bibfnamefont {J.~A.}\ \bibnamefont {Rodriguez-Rivera}}, \ and\ \bibinfo
  {author} {\bibfnamefont {C.}~\bibnamefont {Stock}},\ }\href {\doibase
  10.1103/PhysRevLett.121.087201} {\bibfield  {journal} {\bibinfo  {journal}
  {Phys. Rev. Lett.}\ }\textbf {\bibinfo {volume} {121}},\ \bibinfo {pages}
  {087201} (\bibinfo {year} {2018})}\BibitemShut {NoStop}%
\bibitem [{\citenamefont {Souliou}\ \emph {et~al.}(2017)\citenamefont
  {Souliou}, \citenamefont {Chaloupka}, \citenamefont {Khaliullin},
  \citenamefont {Ryu}, \citenamefont {Jain}, \citenamefont {Kim}, \citenamefont
  {Le~Tacon},\ and\ \citenamefont {Keimer}}]{souliou17:119}%
  \BibitemOpen
  \bibfield  {author} {\bibinfo {author} {\bibfnamefont {S.-M.}\ \bibnamefont
  {Souliou}}, \bibinfo {author} {\bibfnamefont {J.}~\bibnamefont {Chaloupka}},
  \bibinfo {author} {\bibfnamefont {G.}~\bibnamefont {Khaliullin}}, \bibinfo
  {author} {\bibfnamefont {G.}~\bibnamefont {Ryu}}, \bibinfo {author}
  {\bibfnamefont {A.}~\bibnamefont {Jain}}, \bibinfo {author} {\bibfnamefont
  {B.~J.}\ \bibnamefont {Kim}}, \bibinfo {author} {\bibfnamefont
  {M.}~\bibnamefont {Le~Tacon}}, \ and\ \bibinfo {author} {\bibfnamefont
  {B.}~\bibnamefont {Keimer}},\ }\href {\doibase
  10.1103/PhysRevLett.119.067201} {\bibfield  {journal} {\bibinfo  {journal}
  {Phys. Rev. Lett.}\ }\textbf {\bibinfo {volume} {119}},\ \bibinfo {pages}
  {067201} (\bibinfo {year} {2017})}\BibitemShut {NoStop}%
\bibitem [{\citenamefont {Chubukov}\ and\ \citenamefont
  {Morr}(1995)}]{Chubukov95:52}%
  \BibitemOpen
  \bibfield  {author} {\bibinfo {author} {\bibfnamefont {A.~V.}\ \bibnamefont
  {Chubukov}}\ and\ \bibinfo {author} {\bibfnamefont {D.~K.}\ \bibnamefont
  {Morr}},\ }\href {\doibase 10.1103/PhysRevB.52.3521} {\bibfield  {journal}
  {\bibinfo  {journal} {Phys. Rev. B}\ }\textbf {\bibinfo {volume} {52}},\
  \bibinfo {pages} {3521} (\bibinfo {year} {1995})}\BibitemShut {NoStop}%
\bibitem [{\citenamefont {R\"uegg}\ \emph {et~al.}(2008)\citenamefont
  {R\"uegg}, \citenamefont {Normand}, \citenamefont {Matsumoto}, \citenamefont
  {Furrer}, \citenamefont {McMorrow}, \citenamefont {Kr\"amer}, \citenamefont
  {G\"udel}, \citenamefont {Gvasaliya}, \citenamefont {Mutka},\ and\
  \citenamefont {Boehm}}]{Ruegg08:100}%
  \BibitemOpen
  \bibfield  {author} {\bibinfo {author} {\bibfnamefont {C.}~\bibnamefont
  {R\"uegg}}, \bibinfo {author} {\bibfnamefont {B.}~\bibnamefont {Normand}},
  \bibinfo {author} {\bibfnamefont {M.}~\bibnamefont {Matsumoto}}, \bibinfo
  {author} {\bibfnamefont {A.}~\bibnamefont {Furrer}}, \bibinfo {author}
  {\bibfnamefont {D.~F.}\ \bibnamefont {McMorrow}}, \bibinfo {author}
  {\bibfnamefont {K.~W.}\ \bibnamefont {Kr\"amer}}, \bibinfo {author}
  {\bibfnamefont {H.~U.}\ \bibnamefont {G\"udel}}, \bibinfo {author}
  {\bibfnamefont {S.~N.}\ \bibnamefont {Gvasaliya}}, \bibinfo {author}
  {\bibfnamefont {H.}~\bibnamefont {Mutka}}, \ and\ \bibinfo {author}
  {\bibfnamefont {M.}~\bibnamefont {Boehm}},\ }\href {\doibase
  10.1103/PhysRevLett.100.205701} {\bibfield  {journal} {\bibinfo  {journal}
  {Phys. Rev. Lett.}\ }\textbf {\bibinfo {volume} {100}},\ \bibinfo {pages}
  {205701} (\bibinfo {year} {2008})}\BibitemShut {NoStop}%
\bibitem [{\citenamefont {Merchant}\ \emph {et~al.}(2014)\citenamefont
  {Merchant}, \citenamefont {Normand}, \citenamefont {Kramer}, \citenamefont
  {Boehm}, \citenamefont {McMorrow},\ and\ \citenamefont
  {Ruegg}}]{merchant14:10}%
  \BibitemOpen
  \bibfield  {author} {\bibinfo {author} {\bibfnamefont {P.}~\bibnamefont
  {Merchant}}, \bibinfo {author} {\bibfnamefont {B.}~\bibnamefont {Normand}},
  \bibinfo {author} {\bibfnamefont {K.~W.}\ \bibnamefont {Kramer}}, \bibinfo
  {author} {\bibfnamefont {M.}~\bibnamefont {Boehm}}, \bibinfo {author}
  {\bibfnamefont {D.~F.}\ \bibnamefont {McMorrow}}, \ and\ \bibinfo {author}
  {\bibfnamefont {C.}~\bibnamefont {Ruegg}},\ }\href {\doibase
  10.1038/NPHYS2902} {\bibfield  {journal} {\bibinfo  {journal} {Nat. Phys.}\
  }\textbf {\bibinfo {volume} {10}},\ \bibinfo {pages} {373} (\bibinfo {year}
  {2014})}\BibitemShut {NoStop}%
\bibitem [{\citenamefont {Anderson}(2015)}]{Anderson15:1}%
  \BibitemOpen
  \bibfield  {author} {\bibinfo {author} {\bibfnamefont {P.}~\bibnamefont
  {Anderson}},\ }\href {\doibase 10.1038/nphys3247} {\bibfield  {journal}
  {\bibinfo  {journal} {Nat. Phys.}\ }\textbf {\bibinfo {volume} {11}},\
  \bibinfo {pages} {93} (\bibinfo {year} {2015})}\BibitemShut {NoStop}%
\bibitem [{\citenamefont {Sherman}\ \emph {et~al.}(2015)\citenamefont
  {Sherman}, \citenamefont {Pracht}, \citenamefont {Gorshunov}, \citenamefont
  {Poran}, \citenamefont {Jesudasan}, \citenamefont {Chand}, \citenamefont
  {Raychaudhuri}, \citenamefont {Swanson}, \citenamefont {Trivedi},
  \citenamefont {Auerbach}, \citenamefont {Scheffler}, \citenamefont
  {Frydman},\ and\ \citenamefont {Dressell}}]{Sherman15:11}%
  \BibitemOpen
  \bibfield  {author} {\bibinfo {author} {\bibfnamefont {D.}~\bibnamefont
  {Sherman}}, \bibinfo {author} {\bibfnamefont {U.~S.}\ \bibnamefont {Pracht}},
  \bibinfo {author} {\bibfnamefont {B.}~\bibnamefont {Gorshunov}}, \bibinfo
  {author} {\bibfnamefont {S.}~\bibnamefont {Poran}}, \bibinfo {author}
  {\bibfnamefont {J.}~\bibnamefont {Jesudasan}}, \bibinfo {author}
  {\bibfnamefont {M.}~\bibnamefont {Chand}}, \bibinfo {author} {\bibfnamefont
  {P.}~\bibnamefont {Raychaudhuri}}, \bibinfo {author} {\bibfnamefont
  {M.}~\bibnamefont {Swanson}}, \bibinfo {author} {\bibfnamefont
  {N.}~\bibnamefont {Trivedi}}, \bibinfo {author} {\bibfnamefont
  {A.}~\bibnamefont {Auerbach}}, \bibinfo {author} {\bibfnamefont
  {M.}~\bibnamefont {Scheffler}}, \bibinfo {author} {\bibfnamefont
  {A.}~\bibnamefont {Frydman}}, \ and\ \bibinfo {author} {\bibfnamefont
  {M.}~\bibnamefont {Dressell}},\ }\href {\doibase 10.1038/nphys3227}
  {\bibfield  {journal} {\bibinfo  {journal} {Nat. Phys.}\ }\textbf {\bibinfo
  {volume} {11}},\ \bibinfo {pages} {188} (\bibinfo {year} {2015})}\BibitemShut
  {NoStop}%
\bibitem [{\citenamefont {M\'easson}\ \emph {et~al.}(2014)\citenamefont
  {M\'easson}, \citenamefont {Gallais}, \citenamefont {Cazayous}, \citenamefont
  {Clair}, \citenamefont {Rodi\`ere}, \citenamefont {Cario},\ and\
  \citenamefont {Sacuto}}]{Measson14:89}%
  \BibitemOpen
  \bibfield  {author} {\bibinfo {author} {\bibfnamefont {M.-A.}\ \bibnamefont
  {M\'easson}}, \bibinfo {author} {\bibfnamefont {Y.}~\bibnamefont {Gallais}},
  \bibinfo {author} {\bibfnamefont {M.}~\bibnamefont {Cazayous}}, \bibinfo
  {author} {\bibfnamefont {B.}~\bibnamefont {Clair}}, \bibinfo {author}
  {\bibfnamefont {P.}~\bibnamefont {Rodi\`ere}}, \bibinfo {author}
  {\bibfnamefont {L.}~\bibnamefont {Cario}}, \ and\ \bibinfo {author}
  {\bibfnamefont {A.}~\bibnamefont {Sacuto}},\ }\href {\doibase
  10.1103/PhysRevB.89.060503} {\bibfield  {journal} {\bibinfo  {journal} {Phys.
  Rev. B}\ }\textbf {\bibinfo {volume} {89}},\ \bibinfo {pages} {060503(R)}
  (\bibinfo {year} {2014})}\BibitemShut {NoStop}%
\bibitem [{\citenamefont {Majlis}(2007)}]{majlis:book}%
  \BibitemOpen
  \bibfield  {author} {\bibinfo {author} {\bibfnamefont {N.}~\bibnamefont
  {Majlis}},\ }\href@noop {} {\emph {\bibinfo {title} {Quantum Theory of
  Magnetism}}}\ (\bibinfo  {publisher} {World Scientific},\ \bibinfo {address}
  {Hackensack},\ \bibinfo {year} {2007})\BibitemShut {NoStop}%
\bibitem [{\citenamefont {Su}\ \emph {et~al.}(2020)\citenamefont {Su},
  \citenamefont {Masaki-Kato}, \citenamefont {Zhu}, \citenamefont {Zhu},
  \citenamefont {Kamiya},\ and\ \citenamefont {Lin}}]{Ying20:102}%
  \BibitemOpen
  \bibfield  {author} {\bibinfo {author} {\bibfnamefont {Y.}~\bibnamefont
  {Su}}, \bibinfo {author} {\bibfnamefont {A.}~\bibnamefont {Masaki-Kato}},
  \bibinfo {author} {\bibfnamefont {W.}~\bibnamefont {Zhu}}, \bibinfo {author}
  {\bibfnamefont {J.-X.}\ \bibnamefont {Zhu}}, \bibinfo {author} {\bibfnamefont
  {Y.}~\bibnamefont {Kamiya}}, \ and\ \bibinfo {author} {\bibfnamefont {S.-Z.}\
  \bibnamefont {Lin}},\ }\href {\doibase 10.1103/PhysRevB.102.125102}
  {\bibfield  {journal} {\bibinfo  {journal} {Phys. Rev. B}\ }\textbf {\bibinfo
  {volume} {102}},\ \bibinfo {pages} {125102} (\bibinfo {year}
  {2020})}\BibitemShut {NoStop}%
\bibitem [{\citenamefont {Svoboda}\ \emph {et~al.}(2017)\citenamefont
  {Svoboda}, \citenamefont {Randeria},\ and\ \citenamefont
  {Trivedi}}]{Svoboda17:95}%
  \BibitemOpen
  \bibfield  {author} {\bibinfo {author} {\bibfnamefont {C.}~\bibnamefont
  {Svoboda}}, \bibinfo {author} {\bibfnamefont {M.}~\bibnamefont {Randeria}}, \
  and\ \bibinfo {author} {\bibfnamefont {N.}~\bibnamefont {Trivedi}},\ }\href
  {\doibase 10.1103/PhysRevB.95.014409} {\bibfield  {journal} {\bibinfo
  {journal} {Phys. Rev. B}\ }\textbf {\bibinfo {volume} {95}},\ \bibinfo
  {pages} {014409} (\bibinfo {year} {2017})}\BibitemShut {NoStop}%
\bibitem [{\citenamefont {Chen}\ and\ \citenamefont
  {Balents}(2011)}]{Chen11:84}%
  \BibitemOpen
  \bibfield  {author} {\bibinfo {author} {\bibfnamefont {G.}~\bibnamefont
  {Chen}}\ and\ \bibinfo {author} {\bibfnamefont {L.}~\bibnamefont {Balents}},\
  }\href {\doibase 10.1103/PhysRevB.84.094420} {\bibfield  {journal} {\bibinfo
  {journal} {Phys. Rev. B}\ }\textbf {\bibinfo {volume} {84}},\ \bibinfo
  {pages} {094420} (\bibinfo {year} {2011})}\BibitemShut {NoStop}%
\bibitem [{\citenamefont {Rancon}\ and\ \citenamefont
  {Dupuis}(2014)}]{Rancon14:89}%
  \BibitemOpen
  \bibfield  {author} {\bibinfo {author} {\bibfnamefont {A.}~\bibnamefont
  {Rancon}}\ and\ \bibinfo {author} {\bibfnamefont {N.}~\bibnamefont
  {Dupuis}},\ }\href {\doibase 10.1103/PhysRevB.89.180501} {\bibfield
  {journal} {\bibinfo  {journal} {Phys. Rev. B}\ }\textbf {\bibinfo {volume}
  {89}},\ \bibinfo {pages} {180501(R)} (\bibinfo {year} {2014})}\BibitemShut
  {NoStop}%
\bibitem [{\citenamefont {Oitmaa}(2018)}]{oitmaa18:97}%
  \BibitemOpen
  \bibfield  {author} {\bibinfo {author} {\bibfnamefont {J.}~\bibnamefont
  {Oitmaa}},\ }\href {\doibase 10.1103/PhysRevB.97.174421} {\bibfield
  {journal} {\bibinfo  {journal} {Phys. Rev. B}\ }\textbf {\bibinfo {volume}
  {97}},\ \bibinfo {pages} {174421} (\bibinfo {year} {2018})}\BibitemShut
  {NoStop}%
\bibitem [{\citenamefont {Ding}\ \emph {et~al.}(2006)\citenamefont {Ding},
  \citenamefont {Ren}, \citenamefont {Chow}, \citenamefont {Zhang},
  \citenamefont {Vogel}, \citenamefont {Winkler}, \citenamefont {Xu},
  \citenamefont {Zhao},\ and\ \citenamefont {Mao}}]{ding06:74}%
  \BibitemOpen
  \bibfield  {author} {\bibinfo {author} {\bibfnamefont {Y.}~\bibnamefont
  {Ding}}, \bibinfo {author} {\bibfnamefont {Y.}~\bibnamefont {Ren}}, \bibinfo
  {author} {\bibfnamefont {P.}~\bibnamefont {Chow}}, \bibinfo {author}
  {\bibfnamefont {J.}~\bibnamefont {Zhang}}, \bibinfo {author} {\bibfnamefont
  {S.~C.}\ \bibnamefont {Vogel}}, \bibinfo {author} {\bibfnamefont
  {B.}~\bibnamefont {Winkler}}, \bibinfo {author} {\bibfnamefont
  {J.}~\bibnamefont {Xu}}, \bibinfo {author} {\bibfnamefont {Y.}~\bibnamefont
  {Zhao}}, \ and\ \bibinfo {author} {\bibfnamefont {H.-k.}\ \bibnamefont
  {Mao}},\ }\href {\doibase 10.1103/PhysRevB.74.144101} {\bibfield  {journal}
  {\bibinfo  {journal} {Phys. Rev. B}\ }\textbf {\bibinfo {volume} {74}},\
  \bibinfo {pages} {144101} (\bibinfo {year} {2006})}\BibitemShut {NoStop}%
\bibitem [{\citenamefont {Blanco-Canosa}\ \emph {et~al.}(2007)\citenamefont
  {Blanco-Canosa}, \citenamefont {Rivadulla}, \citenamefont {Pardo},
  \citenamefont {Baldomir}, \citenamefont {Zhou}, \citenamefont
  {Garc\'{\i}a-Hern\'andez}, \citenamefont {L\'opez-Quintela}, \citenamefont
  {Rivas},\ and\ \citenamefont {Goodenough}}]{blanco07:99}%
  \BibitemOpen
  \bibfield  {author} {\bibinfo {author} {\bibfnamefont {S.}~\bibnamefont
  {Blanco-Canosa}}, \bibinfo {author} {\bibfnamefont {F.}~\bibnamefont
  {Rivadulla}}, \bibinfo {author} {\bibfnamefont {V.}~\bibnamefont {Pardo}},
  \bibinfo {author} {\bibfnamefont {D.}~\bibnamefont {Baldomir}}, \bibinfo
  {author} {\bibfnamefont {J.-S.}\ \bibnamefont {Zhou}}, \bibinfo {author}
  {\bibfnamefont {M.}~\bibnamefont {Garc\'{\i}a-Hern\'andez}}, \bibinfo
  {author} {\bibfnamefont {M.~A.}\ \bibnamefont {L\'opez-Quintela}}, \bibinfo
  {author} {\bibfnamefont {J.}~\bibnamefont {Rivas}}, \ and\ \bibinfo {author}
  {\bibfnamefont {J.~B.}\ \bibnamefont {Goodenough}},\ }\href {\doibase
  10.1103/PhysRevLett.99.187201} {\bibfield  {journal} {\bibinfo  {journal}
  {Phys. Rev. Lett.}\ }\textbf {\bibinfo {volume} {99}},\ \bibinfo {pages}
  {187201} (\bibinfo {year} {2007})}\BibitemShut {NoStop}%
\bibitem [{\citenamefont {Zhang}\ \emph {et~al.}(2006)\citenamefont {Zhang},
  \citenamefont {Hu}, \citenamefont {Han},\ and\ \citenamefont
  {Tang}}]{zhang06:74}%
  \BibitemOpen
  \bibfield  {author} {\bibinfo {author} {\bibfnamefont {W.-B.}\ \bibnamefont
  {Zhang}}, \bibinfo {author} {\bibfnamefont {Y.-L.}\ \bibnamefont {Hu}},
  \bibinfo {author} {\bibfnamefont {K.-L.}\ \bibnamefont {Han}}, \ and\
  \bibinfo {author} {\bibfnamefont {B.-Y.}\ \bibnamefont {Tang}},\ }\href
  {\doibase 10.1103/PhysRevB.74.054421} {\bibfield  {journal} {\bibinfo
  {journal} {Phys. Rev. B}\ }\textbf {\bibinfo {volume} {74}},\ \bibinfo
  {pages} {054421} (\bibinfo {year} {2006})}\BibitemShut {NoStop}%
\bibitem [{\citenamefont {Rocquefelte}\ \emph {et~al.}(2012)\citenamefont
  {Rocquefelte}, \citenamefont {Schwarz},\ and\ \citenamefont
  {Blaha}}]{rocquefelte12:2}%
  \BibitemOpen
  \bibfield  {author} {\bibinfo {author} {\bibfnamefont {X.}~\bibnamefont
  {Rocquefelte}}, \bibinfo {author} {\bibfnamefont {K.}~\bibnamefont
  {Schwarz}}, \ and\ \bibinfo {author} {\bibfnamefont {P.}~\bibnamefont
  {Blaha}},\ }\href {\doibase https://doi.org/10.1038/srep00759} {\bibfield
  {journal} {\bibinfo  {journal} {Sci. Rep.}\ }\textbf {\bibinfo {volume}
  {2}},\ \bibinfo {pages} {1} (\bibinfo {year} {2012})}\BibitemShut {NoStop}%
\end{thebibliography}

%

\end{document}